\documentclass[useAMS,usenatbib,usegraphicx]{mn2e}
\usepackage{amsmath,amssymb}
\usepackage{epsfig,subfigure,color}

\newcommand{\veck}{\ensuremath{\bmath{k}}}
\newcommand{\vecB}{\ensuremath{\bmath{B}}}
\newcommand{\vecv}{\ensuremath{\bmath{v}}}
\newcommand{\vecg}{\ensuremath{\bmath{g}}}
\newcommand{\vecnab}{\ensuremath{\bmath{\nabla}}}
\newcommand{\DD}{\ensuremath{\mathrm{D}}}
\newcommand{\dd}{\ensuremath{\mathrm{d}}}

\title[Local support against gravity in magneto-turbulent fluids]{
	Local support against gravity in magneto-turbulent fluids}
\author[W. Schmidt, D. C. Collins, and A. G. Kritsuk]{W. Schmidt$^{1}$\thanks{E-mail:
schmidt@astro.physik.uni-goettingen.de}, D. C. Collins$^{2,3}$, and A. G. Kritsuk$^{3}$\\
$^{1}$Institut f\"ur Astrophysik, Universit\"at G\"ottingen, Friedrich-Hund Platz 1, D-37077 G\"ottingen, Germany\\
$^{2}$Theoretical Division, Los Alamos National Lab, Los Alamos, NM, 87545, USA\\
$^{3}$Physics Department and Center for Astrophysics \& Space Sciences, University of California, San Diego, 9500 Gilman 
Drive,\\ La Jolla, CA 92093-0424, USA}

\begin{document}

\date{xx/xx/2012}


\maketitle

\label{firstpage}

\begin{abstract}
Comparisons of the integrated thermal pressure support of gas against its gravitational potential energy lead to critical mass scales 
for gravitational instability such as the Jeans and the Bonnor-Ebert masses, which play an important
role in analysis of many physical systems, including the heuristics of
numerical simulations. In a strict theoretical sense, however, neither the Jeans nor the 
Bonnor-Ebert mass are meaningful when applied locally to substructure in a self-gravitating turbulent medium.
For this reason, we investigate the local support by thermal pressure, turbulence,
and magnetic fields against gravitational compression through an approach that is independent of these concepts.
At the centre of our approach is the dynamical equation for the divergence of the velocity field.
We carry out a statistical analysis of the source terms of the local compression rate (the negative time derivative
of the divergence) for simulations of forced self-gravitating turbulence in periodic boxes
with zero, weak, and moderately strong mean magnetic fields (measured by the averages of the magnetic and thermal
pressures). We also consider the amplification of the magnetic field energy by shear and by compression. 
Thereby, we are able to demonstrate that the support against gravity is dominated by thermal pressure fluctuations,
although magnetic pressure also yields a significant contribution. The net effect of turbulence 
in the highly supersonic regime, however,
is to enhance compression rather than supporting overdense gas even if the vorticity is very high. This is
incommensurate with the support of the highly dynamical substructures in magneto-turbulent fluids
being determined by local virial equilibria of volume energies without surface stresses. 
\end{abstract}

\section{Introduction}
\label{sc:intro}

Self-gravitating turbulent fluids cover an enormous range of length scales, from molecular clouds all the way up to 
cosmological scales. Traditionally, the linear perturbation analysis of an extended homogeneous medium by \citet{Jeans02} is 
applied to infer the stability against gravitational collapse. Apart from a geometrical factor, the resulting critical mass 
and length scales also apply to isolated systems (``clouds'') in  dynamical equilibrium. This is a consequence of the viral 
theorem. The critical mass of a spherical isothermal cloud that is solely supported by its thermal pressure is known as 
Bonnor-Ebert mass \citep{Ebert55,Bonnor56}. As pointed out by \cite{Ballest06}, however, the generalized virial theorem also
includes the integrated energies of non-thermal motions and magnetic fields, as well as contributions from stresses at the
boundaries of non-isolated systems \citep[see also][]{Lequeux}. The former can be interpreted as turbulent and magnetic pressures, 
respectively. While these mean quantities characterize the global state, no conclusions about local effects can be drawn. Nevertheless,
it is commonly assumed that Jeans-like scales can be applied locally if the density is scaled from the mean value to peak values 
in overdense structures. However, in self-gravitating turbulent gas, 
such structures are highly non-linear and generally not in equilibrium. Consequently, an extrapolation to this regime is 
questionable on theoretical grounds. \citet{Chandra51} incorporated turbulence into the Jeans criterion by introducing an 
effective pressure that depends on the internal velocity dispersion of density enhancements \citep[see also][]{BonaHey87}. 
This heuristic approach was later put onto a firmer basis by means of renormalization group theory \citep{BonaPer92}. 
However, a fundamental limitation 
remains even with this method: The Jeans mass with an effective pressure in the fashion of Chandrasekhar can be derived
only for length scales above the energy injection scale of turbulence. To a certain degree, this might be the case for cosmological
structure formation, where turbulence is typically driven by gravity on length scales smaller than the size of halos 
\citep[e.~g.][]{SurSchl10,FederSur11,TurkOishi12,LatSchl13}. Overdense structures in star-forming clouds, on the other hand,  
collapse on scales below the integral scale of turbulence, which can be much larger than the size of the whole cloud 
\citep{ElmeScal04,ElmeBurk10,KlessHenne10}. Of course, it is still plausible that gravity becomes dominant on length scales 
comparable to the local, density-dependent Jeans length, but a strict derivation is not possible if
gravitational instabilities develop on time scales smaller than the dynamical time scale of turbulence.

In this article, we attempt to overcome these limitations by considering the fully non-linear dynamical equations of self-gravitating gas. 
This approach necessarily involves statistical analysis. Apart from the adopted simulation scenario and methodology, however,
the results we obtain are independent of any theoretical assumptions. 
The dynamics of gas compression is characterized by a partial differential equation for the divergence of the velocity. By applying the Poisson equation, the Laplacian of the gravitational potential can be substituted
by a term that is proportional to the local gas density. This term is amenable to a comparison with the other sources
of gas compression, which are derived from the local thermal, turbulent, and magnetic pressures of the gas. 
If these source terms are positive, they drive expansion of the gas or slow its collapse.  Otherwise, they
aid to its compression.
In a collapsing region, the rate of change is mainly driven by the gravitational compression rate and, consequently, the divergence becomes ever more negative \citep{Schmidt09}. The relevance of turbulent support relative to thermal support of the gas in the intracluster medium
was already investigated by \citet{ZhuFeng10} and \citet{IapiSchm11}. We elaborate on this approach by generalizing the divergence equation to magnetohydrodynamics and by analyzing numerical data from simulations of strongly self-gravitating turbulence in the
interstellar medium. 

The production of dense clumps in the turbulent two-phase medium resulting from colliding flows were investigated, for example, by \citet{BanVaz09}. Turbulence produced by homogeneous and isotropic forcing in periodic boxes, on the other hand, has
been demonstrated to be particularly suitable for calculating statistics on molecular-cloud scales \citep[e.~g., ][]{KritNor06,KritNor07,SchmFeder09,KritUsty09,FederRom10,FederChab11}. 
To avoid a globally inhomogeneous flow structure, we thus utilize data from adaptive mesh refinement (AMR) simulations of 
self-gravitating hydrodynamical (HD) turbulence \citep{KritNor11} and magnetohydrodynamical (MHD) turbulence \citep{CollKrit12}.
These simulations were initialized by forced isotropic turbulence without self-gravity and then AMR was used to compute the
contraction of the gas under the action of gravity. Since the statistically homogeneous forcing completely randomizes the
turbulent flow, also the dependence on initial conditions is minimized. Particularly the HD simulation offers an extremely 
large dynamical range, which resolves collapsing objects very well. An important question regarding the simulation of self-gravitating gas
is the  utilization of sink particles. If excess mass is dumped into sink particles, the gas density can be kept below a given threshold 
even at the maximum refinement level. Such a threshold is typically set to a multiple of the Jeans length \citep{Truelove97}.
By following this approach, \citet{FederBan10} demonstrated that gravitationally collapsing cores in a turbulent cloud can be 
identified with a set of heuristic rules, which test the divergence of the flow, the local gravitational potential, the local 
Jeans length, etc. However, two uncertainties remain. Firstly, the removal of mass from grid cells inevitably produces small-
scale perturbations in the solution of the equations of gas dynamics. This is potentially problematic for the calculations presented in this 
article because they depend on higher-order derivatives of gas-dynamical variables, especially, in the high-density regions that 
would produce sink particles. Secondly, existing sink particle prescriptions do not absorb magnetic fields. While
magnetic field lines are at least partially dragged into collapsing objects, the gas absorbed into sink particles is
effectively decoupled from the magnetic field. In this work, we avoid these uncertainties by letting the gas contract to 
arbitrarily high densities (limited only by the dynamical range and the robustness of the numerical simulations), although
this comes at the cost of eventually violating the resolution limit set by a Truelove-like criterion. As a consequence, 
results for extremely high densities are tentative. 

This article is structured as follows. In Section~\ref{sc:support}, we define support against gravity
by the source terms in the partial differential equation for the divergence of the velocity field. The
simulation methods and parameters are briefly summarized in Section~\ref{sc:sim}.
In Sections~\ref{sc:hd_stat} 
and~\ref{sc:mhd_stat}, we carry out a statistical analysis of the different source terms for purely hydrodynamical as well as 
magnetohydrodynamical turbulence simulations with weak and strong magnetic fields. Thereby, we are able to infer turbulent, thermal, 
and magnetic support depending on various gas-dynamical properties. In addition, turbulent dynamo action and the compressive magnetic field amplification is analyzed. In the last section, we summarize our results and discuss their implications.

\section{The local support function}
\label{sc:support}

The momentum equation for a perfectly conducting ideal fluid subject to the gravitational
potential $\phi$ and the magnetic field \vecB\ can be written as
\begin{equation}
  \label{eq:momt}
  \frac{\partial}{\partial t}(\rho\vecv) + \vecnab\cdot(\rho\vecv\otimes\vecv) =
  - \vecnab P + \frac{1}{c}(\bmath{J}\times\vecB) - \rho\vecnab\phi.
\end{equation}
In ideal magnetohydrodynamics, the current density $\bmath{J}$ is related to the curl of the magnetic field 
via Ampere's law,
\begin{equation}
  	\label{eq:ampere}
	\vecnab\times\vecB = \frac{4\pi}{c}\bmath{J},
\end{equation}
and the magnetic field is given by the compressible induction equation
\begin{equation}
	\label{eq:induction}
	\frac{\DD\vecB}{\DD t} =
	(\vecB\cdot\vecnab)\vecv-\vecB d,
\end{equation}
where the substantial time derivative is defined by
\begin{equation}
    \frac{\DD}{\DD t} =
    \frac{\partial}{\partial t} + \vecv\cdot\vecnab,
\end{equation}
and $d=\vecnab\cdot\vecv$ is the divergence of the velocity.

The gravitational field $\vecg=-\vecnab\phi$ directly produces divergence $d$, but not vorticity $\omega=\vecnab\times\vecv$
(the rotation operator applied to $\vecg$ is identical to zero). Because of 
the non-linear turbulent interactions of turbulent velocity fluctuations, however, rotational motions (vortices)
affect the divergence of the velocity. For magnetized fluids, the following equation for the rate of change of the divergence 
$d$ is obtained by applying the divergence operator to the momentum equation \citep{Schmidt09,ZhuFeng10} and by substituting 
Ampere's law~(\ref{eq:ampere}):
\begin{equation}
  \label{eq:div}
  \begin{split}
  \frac{\DD d}{\DD t}
  = & \frac{1}{2}\left(\omega^{2}-|S|^{2}\right) -
      \frac{1}{\rho}\left[\nabla^{2}P - \vecnab\cdot\left(\vecnab\cdot\bmath{\tau}_{\mathrm{m}}\right)\right]\\
    & + \frac{1}{\rho^{2}}\vecnab\rho\cdot
	  \left(\vecnab P - \vecnab\cdot\bmath{\tau}_{\mathrm{m}}\right) -\nabla^2\phi.
  \end{split}
\end{equation}
The vorticity of the flow, $\bmath{\omega}=\vecnab\times\vecv$, which is related to the antisymmetric
part of the velocity derivative, tends to increase the divergence. The rate of strain, $|S|=(2S_{ij}S_{ij})^{1/2}$, where
\begin{equation}
	S_{ij}=\frac{1}{2}\left(\frac{\partial v_i}{\partial x_j}+\frac{\partial v_i}{\partial x_j}\right)
\end{equation}
is the symmetric part of the velocity derivative, has the opposite effect. The Maxwell stresses of the magnetic field are 
given by
\begin{equation}
	(\tau_{ij})_{\mathrm{m}} = \frac{1}{4\pi}\left(B_i B_j - \frac{1}{2}B^2\delta_{ij}\right).
\end{equation}
The last term on the right-hand side corresponds to the magnetic pressure $P_{\rm m}=B^2/8\pi$, while the former term 
specifies the anisotropic tension of magnetic field lines.

In a gravitationally collapsing region, flux conservation squeezes the magnetic field lines as long as the ideal MHD 
approximation holds. The rate of change of the magnetic pressure is easily obtained from the induction 
equation~(\ref{eq:induction}):
\begin{equation}
	\label{eq:dynamo}
	\frac{\DD}{\DD t}\left(\frac{B^2}{8\pi}\right) =
	\frac{1}{4\pi}\left(B_i B_j S_{ij}^{\ast} - \frac{2}{3}B^2 d\right),
\end{equation}
where $S_{ij}^{\ast}=S_{ij}-\frac{1}{3}d\delta_{ij}$ is the trace-free rate-of-strain tensor (the
contribution from the antisymmetric part $W_{ij}=v_{i,j}-S_{ij}$ of the velocity derivative $v_{i,j}$ vanishes because
$B_i B_j W_{ij} =0$). 
The first term on the right-hand side results from the action of the turbulent shear on the magnetic
field. This term can be either positive or negative. Turbulent dynamo action amplifies the magnetic field.
The rate of increase of the magnetic pressure due to gas contraction ($d<0$), on the other hand, is 
$-\frac{2}{3}B^2 d$.

For given boundary conditions, the gravitational potential $\phi$ is determined by the mass density of
all gravitating matter (baryonic gas, stars, dark matter) through the Poisson equation. For a compact mass 
distribution surrounded by vacuum, the Poisson equation takes the simple form
\begin{equation}
  \label{eq:poisson_vac}
  \nabla^{2}\phi = 4\pi G\rho.
\end{equation}
For periodic boundary conditions, on the other hand, the mean density $\langle\rho\rangle$ is conserved and the gas 
contraction must be zero for uniform mass density in the absence of any flow and magnetic fields. Consistency with the 
boundary conditions thus requires
\begin{equation}
  \label{eq:poisson_box}
  \nabla^{2}\phi = 4\pi G(\rho-\rho_0),
\end{equation}
so that $d=0$ for $\rho=\rho_0=\langle\rho\rangle$, $\vecv=0$, and $\vecB=0$. 

By using the notation of \citet{ZhuFeng10}, we can generically write the Poisson equation in the form
\begin{equation}
  \label{eq:poisson}
  \nabla^{2}\phi = 4\pi G\rho_0\delta.
\end{equation}
Here, $\rho_0$ denotes a suitable reference density, and it is understood that the dimensionless density variation $\delta$ 
is consistent with the source term of the Poisson equation for any given system. For example, $\delta=\rho/
\rho_0 - 1$ for periodic boundary conditions.

We can now write an equation for the \emph{rate of compression} of advected fluid elements:
\begin{equation}
  \label{eq:compr}
  -\frac{\DD d}{\DD t}
  = 4\pi G\rho_0\delta - \Lambda.
\end{equation}
Each of the terms in this equation has the dimension of inverse time squared. By applying the 
product rule and $\vecnab\cdot\vecB=0$ to the magnetohydrodynamical terms in Eq.~(\ref{eq:div}), it follows 
that the \emph{local} support of the gas against gravity is given by the function
\begin{equation}
  \begin{split}
  \label{eq:support}
  \Lambda
  = & \frac{1}{2}\left(\omega^{2}-|S|^{2}\right) \\
    & - \frac{1}{\rho}\left[\frac{\partial^2}{\partial x_i\partial x_i}\left(P + \frac{B^2}{8\pi}\right) 
      - \frac{1}{4\pi}\frac{\partial B_i}{\partial x_j}\frac{\partial B_j}{\partial x_i}\right]\\
    & + \frac{1}{\rho^{2}}\frac{\partial \rho}{\partial x_i}\left[
    	    \frac{\partial}{\partial x_i}\left(P + \frac{B^2}{8\pi}\right)
	  - \frac{1}{4\pi}\left(B_j\frac{\partial B_i}{\partial x_j}\right)\right],
  \end{split}
\end{equation}
The contributions of turbulence, thermal pressure, and magnetic fields to the support of the gas are given by
\begin{align}
  \label{eq:support_turb}
	\Lambda_{\rm turb} =& 
	\frac{1}{2}\left(\omega^{2}-|S|^{2}\right),\\
  \label{eq:support_therm}
	\Lambda_{\rm therm} =& 
	-\frac{1}{\rho}\frac{\partial^2 P}{\partial x_i\partial x_i} 
    + \frac{1}{\rho^{2}}\frac{\partial \rho}{\partial x_i}\frac{\partial P}{\partial x_i},
\end{align}
and	
\begin{equation}
  \begin{split}
  \label{eq:support_magn}
    \Lambda_{\rm magn} =&
    \frac{1}{4\pi\rho}\left[-\frac{\partial^2}{\partial x_i\partial x_i}\left(\frac{1}{2}B^2\right)
	+ \frac{\partial B_i}{\partial x_j}\frac{\partial B_j}{\partial x_i}\right],\\
    &+ \frac{1}{4\pi\rho^2}\frac{\partial\rho}{\partial x_i}
    \left[\frac{\partial}{\partial x_i}\left(\frac{1}{2}B^2\right) - B_j\frac{\partial B_i}{\partial x_j}\right],
  \end{split}
\end{equation}
respectively. The net effect is to resist gravitational collapse if $\Lambda=\Lambda_{\rm therm}+\Lambda_{\rm magn}+
\Lambda_{\rm magn}>0$. Negative support ($\Lambda<0$) can result, for instance, from strong shock compression (large rate of 
strain). For isothermal gas, $P=c_{0}^2\rho$ and $c_{0}= \mathrm{const}$. The thermal support is then given by
\begin{equation}
  \label{eq:support_isoth}
  \Lambda_{\rm isoth} = -c_{0}^2\nabla^2\ln\rho.
\end{equation}

The classical Jeans length can be obtained from a linear perturbation analysis of  
rate of compression (Eq.~\ref{eq:compr}) for small density perturbations $\delta\ll 1$. 
Let the gas initially be at rest and the magnetic field be zero.
Linearization of the continuity and induction equations then implies
\[
	\frac{\partial\delta}{\partial t}\simeq -d' \quad\mbox{and}\quad 
	\frac{\partial B'}{\partial t}\simeq 0,
\]
where $d'$ is the divergence of the flow and $B'$ the magnetic field induced by the density perturbation.  With the usual 
plane-wave ansatz $\delta\propto\exp[i{\tilde{\omega} t+\bmath{k}\cdot\bmath{x}}]$, where $\tilde{\omega}$ is the
angular frequency and $k$ the wavenumber,
we have $d'= -i\tilde{\omega} \delta$ and a dispersion relation is obtained by linearizing Eq.~(\ref{eq:compr})
for the compression rate:
\begin{equation}
	\tilde{\omega}^2 \simeq -4\pi G\rho_0 + k^2 c_s^2.
\end{equation}
Here, $c_s$ is the speed of sound for the unperturbed state. Hence, the perturbation is unstable if
\begin{equation}
	k<k_{\rm J}=\left(\frac{4\pi G\rho_0}{c_s^2}\right)^{1/2},
\end{equation}
which is the result found by \citet{Jeans02}. The Jeans criterion 
corresponds to an equilibrium between the thermal pressure support for a single wave mode and the gravity term in 
Eq.~(\ref{eq:compr}):
\[
	k_{\rm J}^2c_s^2 \simeq 4\pi G\rho_0.
\] 
An \emph{ad hoc} modification of the Jeans criterion for a uniform turbulent velocity dispersion was introduced by 
\citet{Chandra51} and later derived in a rigorous manner by \citet{BonaPer92}. However, their theory is valid only if 
turbulence is produced on length scales \emph{smaller than the Jeans length}, which excludes most applications in 
astrophysics.  

Equation~(\ref{eq:compr}), on the other hand applies to density variations of any magnitude in magnetized turbulent gas. 
We can distinguish the following different regimes:
\begin{enumerate}
\item For $\Lambda>4\pi G\rho_0\delta> 0$, the gas is supported 
	against gravity. Since $\DD d/\DD t>0$, gas contraction ($d<0$) is slowed down.
\item Fluid elements undergo transient phases of vanishing support ($\Lambda \ll 4\pi G\rho_0\delta$)
	or non-gravitational compression ($\Lambda < 0$), e.~g., by shocks, as long as 
	turbulent fluctuations of the pressure, the velocity, and the magnetic field push the gas
	back into regions with positive support. However, as the overdensity $\delta$ grows, 
	it becomes increasingly improbable that a fluid element can escape the
	pull of gravity and collapse may ensue.
\item For a fluid element in a collapsing region, $d<0$, $4\pi G\rho_0\delta\gg|\Lambda|$, and
	the free-fall time scale $\sim (4\pi G\rho_0\delta)^{-1/2}$ associated with the gravity
	term becomes the dynamically dominant time scale. 
\item At the end of collapse, the gas reaches an equilibrium state, in which
	the volume-averaged support is balanced by gravity, i.~e.,
	$\langle\Lambda\rangle\sim\langle 4\pi G\rho_0\delta\rangle$. For an isolated object,
	this corresponds to the simple virial equilibrium (possibly, with non-thermal kinetic and 
	magnetic contributions).
\end{enumerate}
Since the terms in the local support function $\Lambda$ are second derivatives or products of derivatives, this
function has a factor $k^2$ in Fourier space, which is analogous to the above Jeans equilibrium condition. Generally,
however, there is whole spectrum of perturbations, ranging from the smallest to largest wave numbers of the system.

\begin{figure*}
  \begin{center}
    \mbox{\subfigure[$\Lambda_{\rm turb}/4\pi G\rho_0$]{\includegraphics[width=0.445\linewidth]{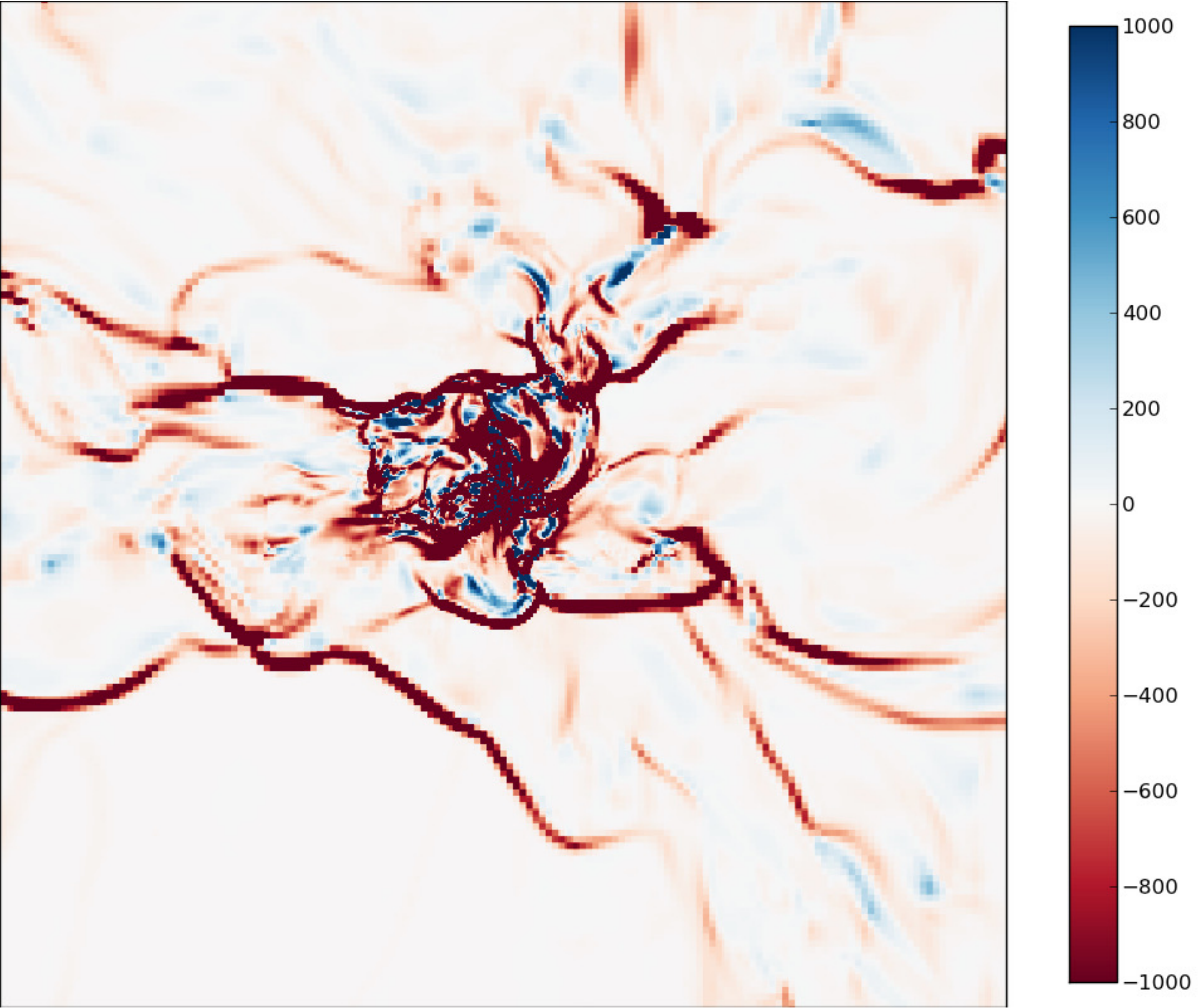}}\qquad
          \subfigure[$\Omega_{1/2}=\frac{1}{2}|\vecnab\times(\rho^{1/2}\vecv)|^2$]{\includegraphics[width=0.445\linewidth]{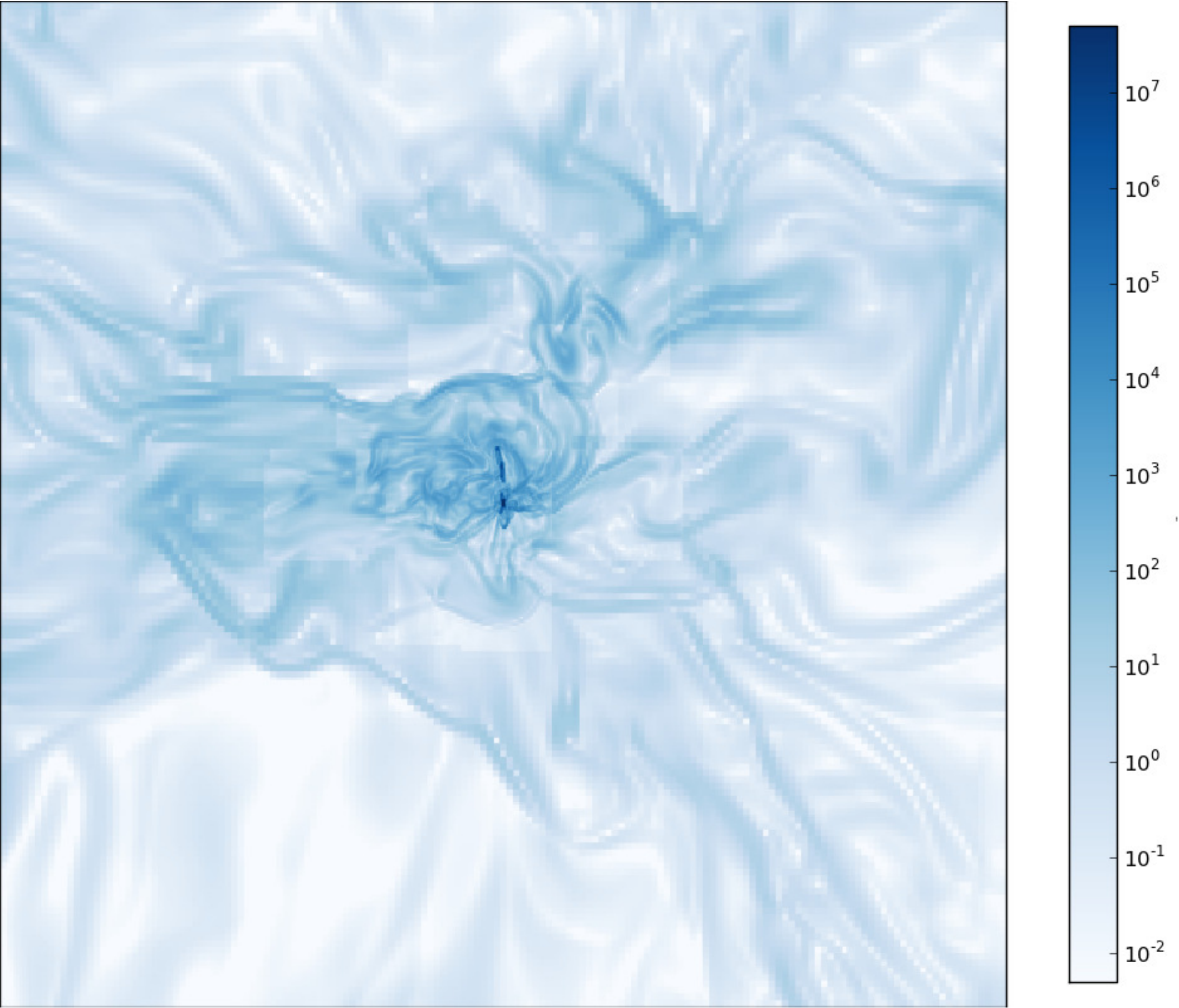}}}
    \mbox{\subfigure[$\Lambda_{\rm therm}/4\pi G\rho_0$]{\includegraphics[width=0.445\linewidth]{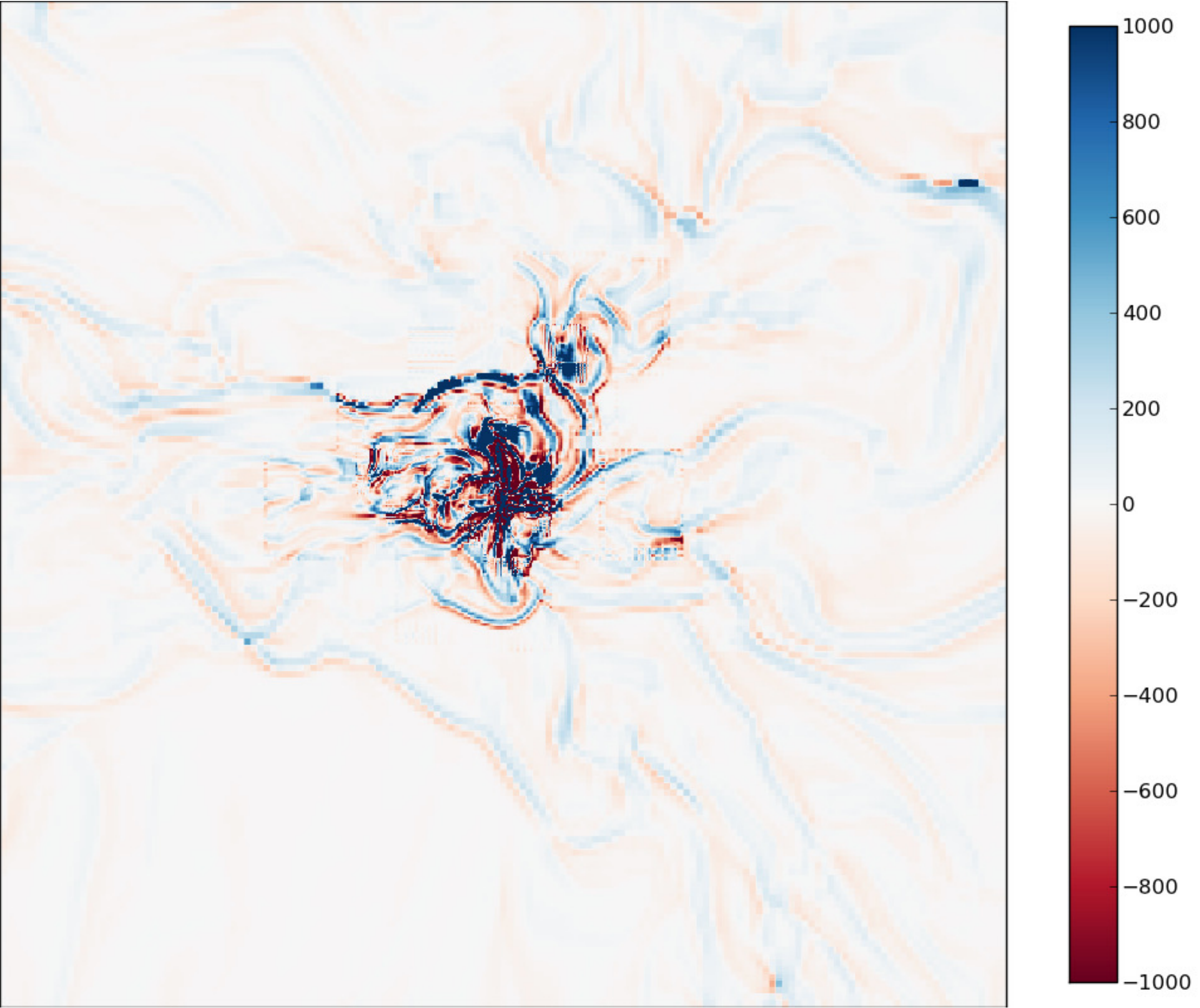}}\qquad
          \subfigure[$P$]{\includegraphics[width=0.445\linewidth]{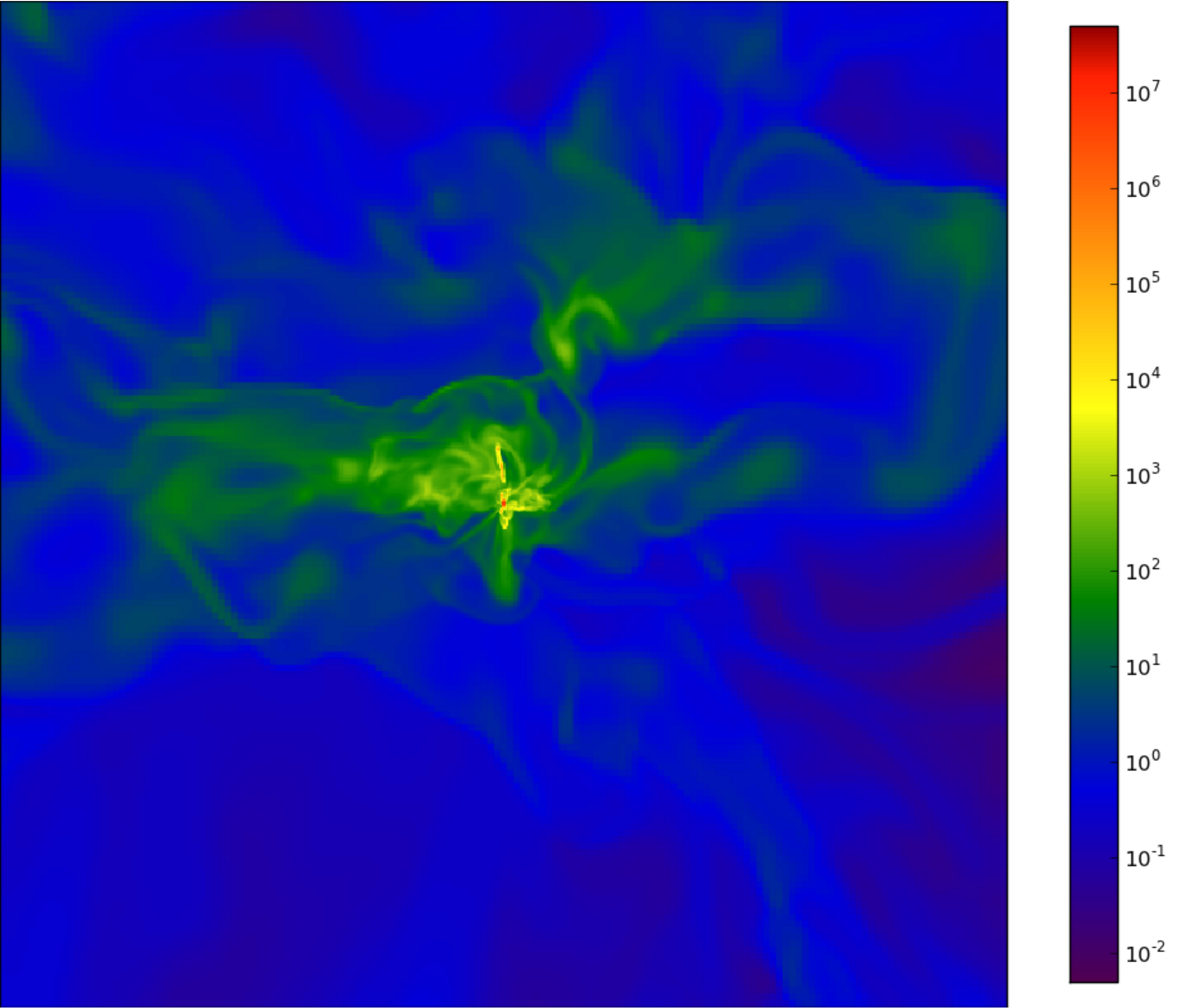}}}
    \mbox{\subfigure[$\Lambda_{\rm magn}/4\pi G\rho_0$]{\includegraphics[width=0.445\linewidth]{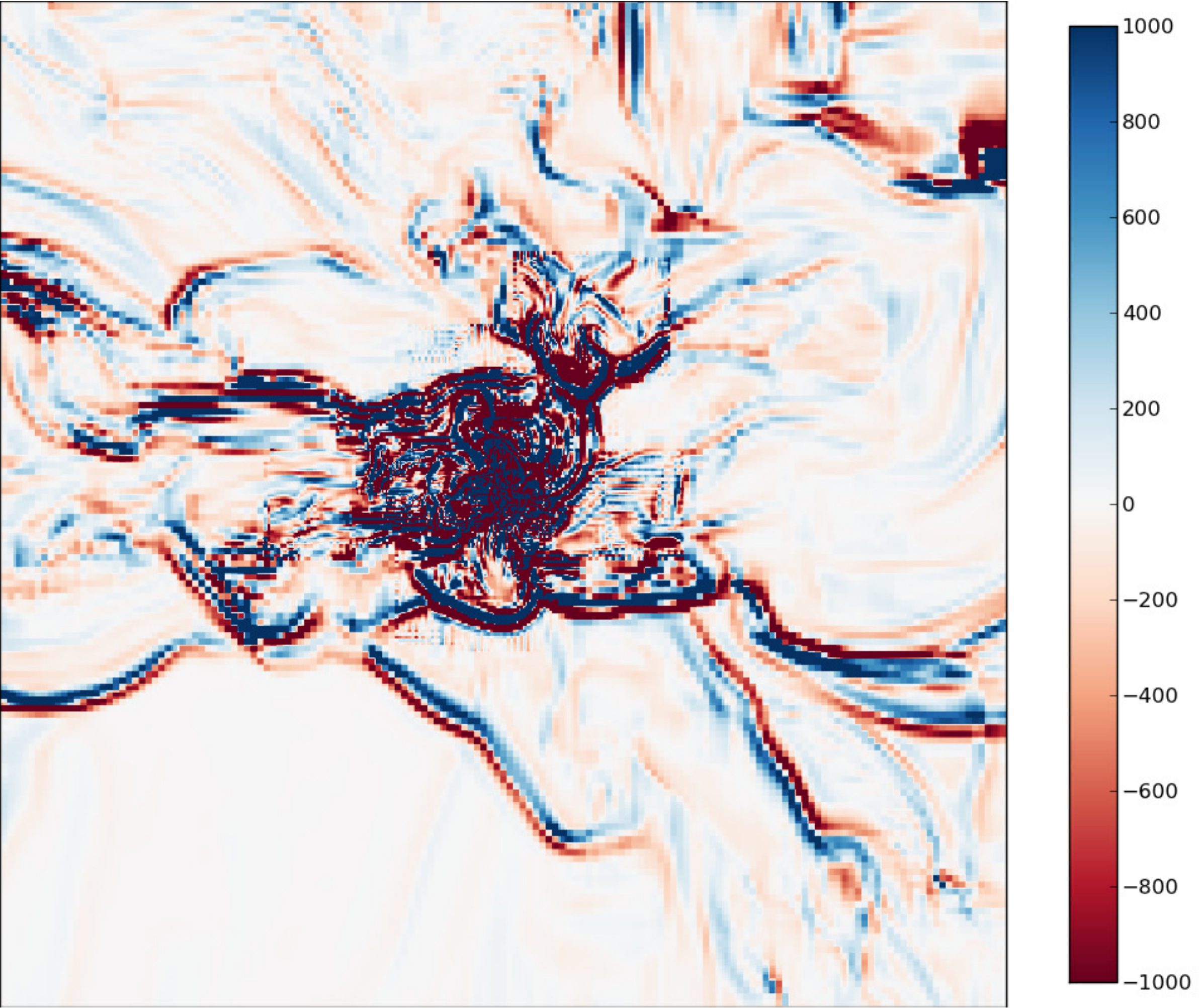}}\qquad
          \subfigure[$P_{\rm m}=B^2/8\pi$]{\includegraphics[width=0.445\linewidth]{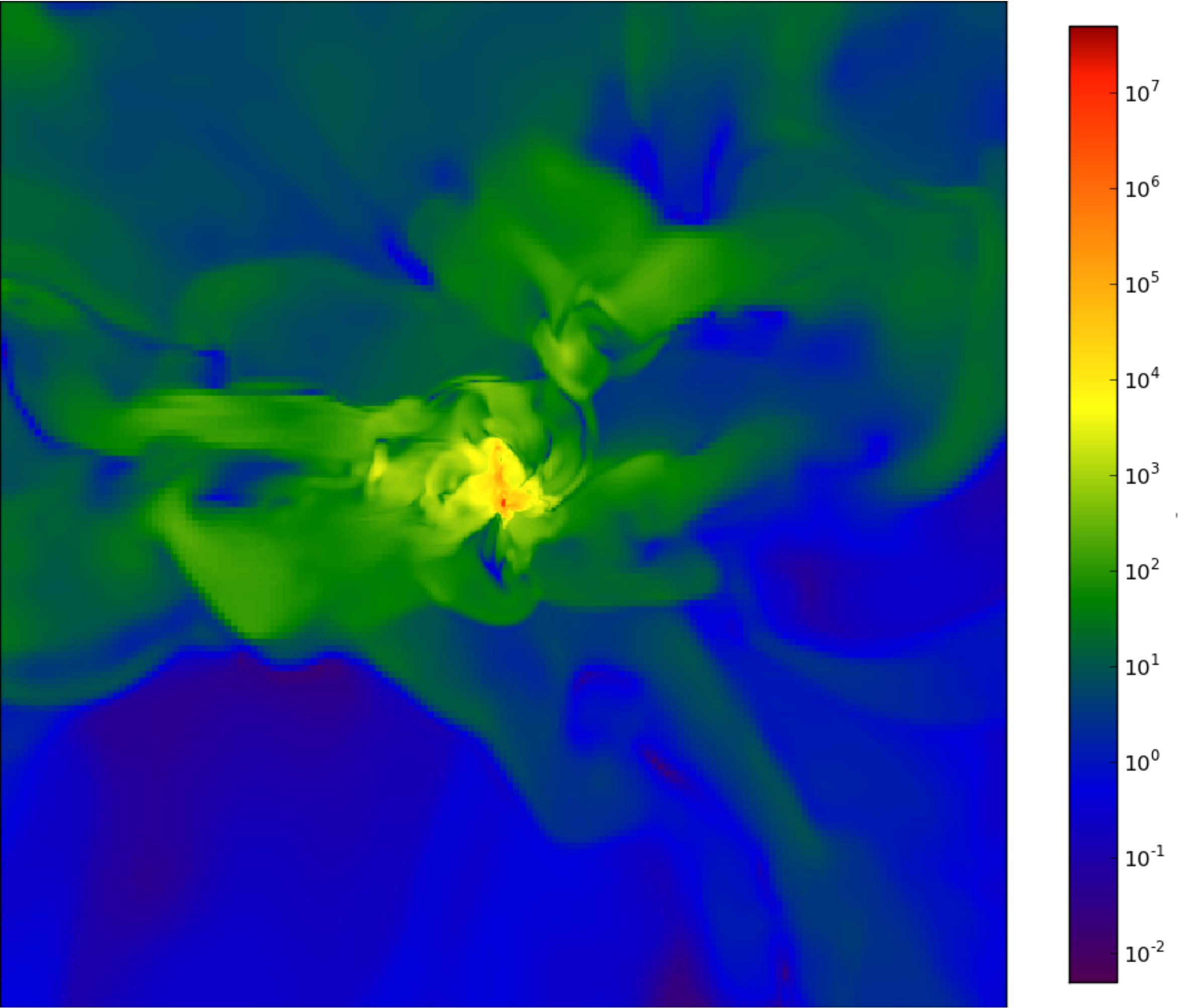}}}
    \caption{Slices of the dimensionless turbulent, thermal, and magnetic support (left column) and associated quantities
    	    (right column) for an MHD turbulence simulation with $\beta_0=20$ at $t=0.5 t_{\rm ff}$. The size
		of the shown region is $0.3$ times the box size. Black lines indicate the boundaries of refinement levels.}
    \label{fg:slices}
  \end{center}
\end{figure*}

As an example, Fig.~\ref{fg:slices} shows slices of the functions~(\ref{eq:support_turb}), (\ref{eq:support_therm}), and 
(\ref{eq:support_magn}) normalized by $4\pi G\rho_0$ for the isothermal MHD simulation
with $\beta_0=20$ \citep[see][]{CollKrit12}. The slices are centered on the density maximum. 
Also shown are the thermal and the magnetic pressures as well as the 
denstrophy $\Omega_{1/2}=\frac{1}{2}|\vecnab\times(\rho^{1/2}\vecv)|^2$ \citep[see][]{KritNor07}. We chose fourth-order finite differences
to compute gradients and Laplacians.\footnote{By using a higher order than the spatial order of the numerical scheme
that is used to integrate the fluid dynamics, we keep the error in the computation of the gradients smaller than
the intrinsic error of the numerical solution.}
Large values of the support function are clearly associated with steep pressure and 
denstrophy gradients. For isothermal turbulence, the thermal pressure is proportional to the density, while the denstrophy 
indicates both density fluctuations and vortices. While negative turbulent support is most prominent, the thermal and 
magnetic support swap their 
signs across shock front. Even in the central, refined region, where the gas density reaches its peak, the support 
is not predominately positive. These qualitative observations are confirmed by the statistical analysis in 
Section~\ref{sc:mhd_stat}.

\begin{table}
\begin{center}
\caption{Simulation parameters.}
\begin{tabular}{l c c c c c c c c c }
\hline
\hline
&Solver & $\mathcal{M}$ & $N$ & $R$ & $L_{\rm{J}}$ & $\chi$ & Continue \\
(1)&(2)&(3)&(4)&(5)&(6)&(7)&(8)\\
\hline
HD  &PPM & 6 & 5 & 4 & 4 & 2/3 & no\\
MHD &PLM & 9 & 4 & 2 & 16 & 0 & yes
\label{table.solver}
\end{tabular}
\end{center}
{\bf Notes:} (1) simulation set; (2) solver used; (3) RMS Mach
number, $\mathcal{M}$;  (4) maximum number of AMR levels, $N$; (5)
refinement factor, $R$; (6) minimum number of zones per Jeans length,
$L_{\rm{J}}$;
(7) ratio of compressive to solenoidal motions, $\chi$;  (8) was
driving continued through the collapse phase.
\end{table}

\section{Simulations}

\label{sc:sim}

The hydrodynamic simulations we analyse here were earlier presented in
\citet{Padoan_ea05} and \citet{KritNor11}, and the MHD simulations were presented
in \citet{CollKrit12}.
Both sets of simulations were run with the Enzo
code\footnote{\tt{http://enzo-project.org}}.
Table \ref{table.solver} summarizes
the differences between the two simulations.
Both simulations began with $512^3$ grids, with uniform density.  Driving was
done as in \citet{MacLow99}, wherein small static velocity perturbations are added
every time step such that kinetic energy input rate is constant.  Once
a steady state was reached, gravity and AMR were initiated. In both simulations the
power in the driving field is distributed approximately isotropically
and uniformly in the interval $k\in[1,2]$.

The hydrodynamic simulation used the third order piecewise parabolic method
\citep{ColWood84}.  It was driven with a mixture of 60\% solenoidal
and 40\% compressive motions, and the energy injection
was such that the RMS Mach number was approximately 6.
The inertial range ratio of compressive to solenoidal motions, and its
dependence on Mach number and magnetic fields, is discussed in
\citet{Kritsuk10b}. At $t=0$, when gravity and AMR are
switched on, the driving is halted, though the kinetic energy decay in the short
duration of the collapse phase is insignificant. Five levels of AMR were
used, with a refinement ratio of 4, and a refinement criterion such that the Jeans
length was resolved by at least 4 zones.

The MHD simulations used the second order piecewise linear method
\citep{Li08}, the divergence-free constrained transport method described in
\citet{Gardiner05}, and the divergence-free interpolation method of
\citet{Balsara01}.  The code is described in detail in \citet{Collins10}.
The driving force was purely solenoidal, and an RMS
Mach number of approximately 9 was reached. Driving continued through the collapse phase of
the simulation.  Four levels of refinement were used, with a refinement ratio of
2, and refinement was such that the Jeans length was resolved by at least 16
zones.
\\

\section{Statistical analysis of hydrodynamic turbulence}
\label{sc:hd_stat}

We use data from the hydrodynamical turbulence simulation performed by \citet{KritNor11} to
calculate statistics of the thermal and turbulent support functions. The virial parameter of this model
is $\alpha=0.25$ and the sonic Mach number is about $6$. Unless otherwise stated, 
we use the final state of the simulation at time $t=0.43 t_{\rm ff}$, where 
$t_{\rm ff}=(3\pi/32G\rho_0)^{1/2}$ is the free-fall time for the mean density $\rho_0$. 

\begin{figure}
\centering
  \includegraphics[width=\linewidth]{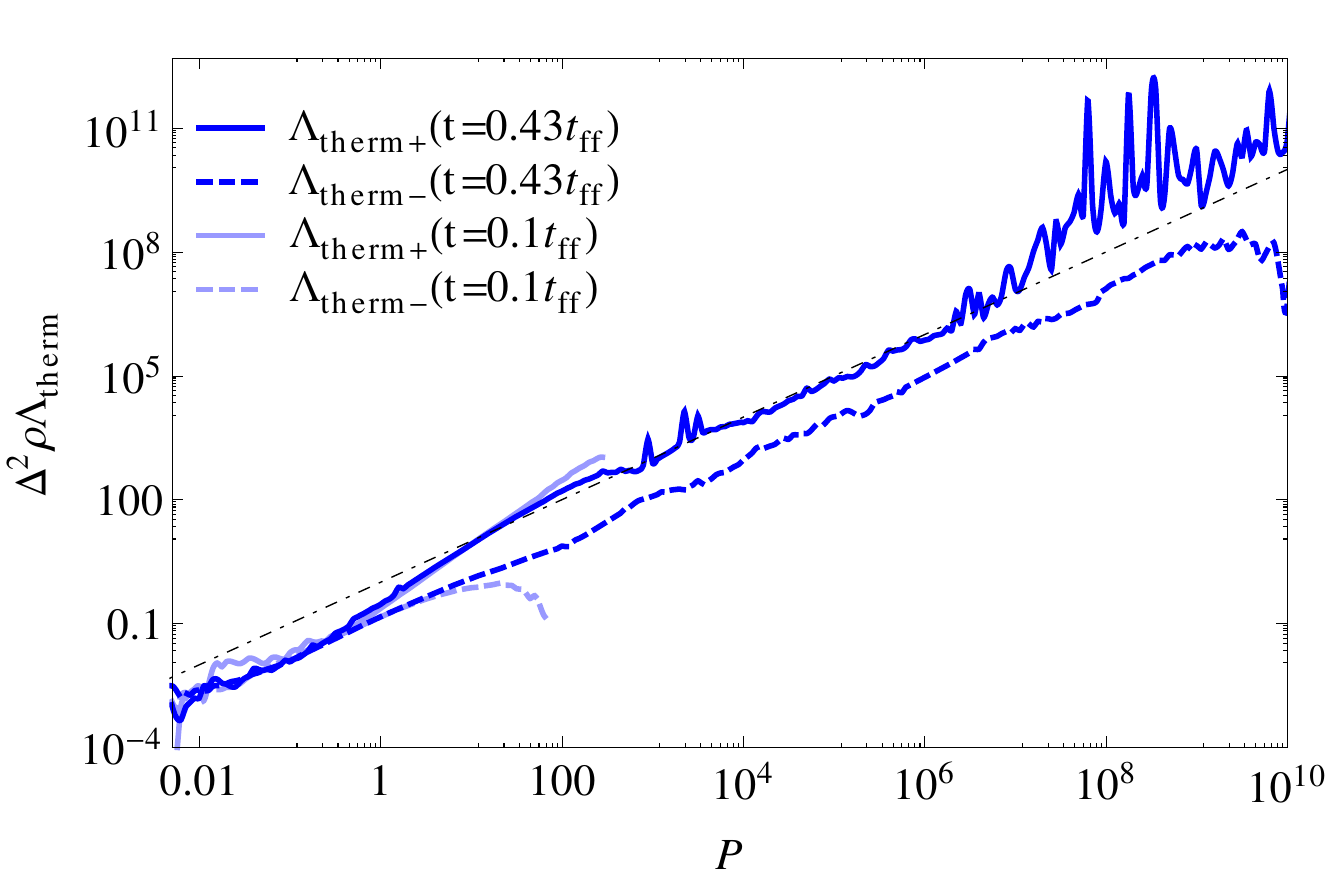}\\
  \includegraphics[width=\linewidth]{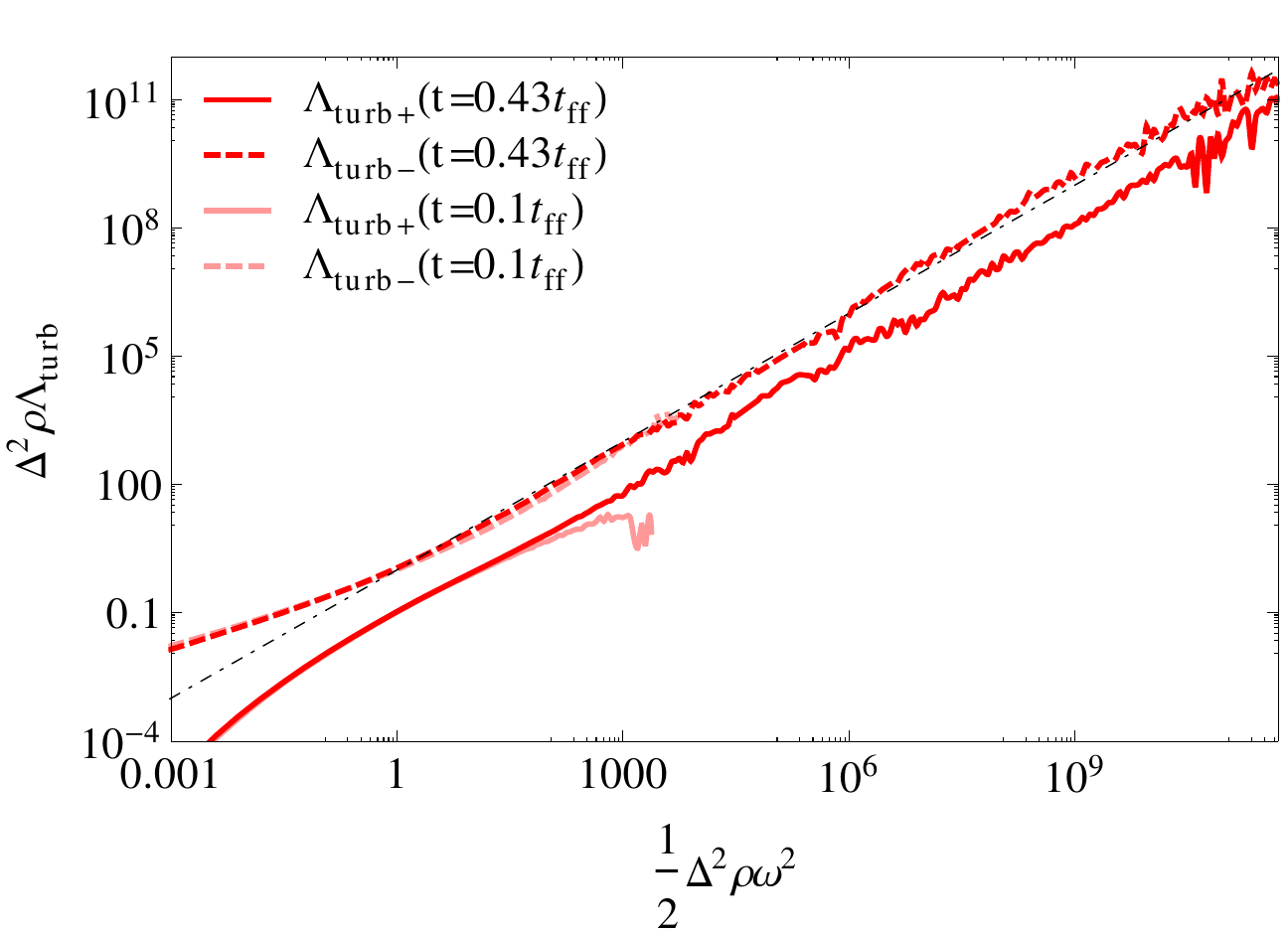}
  \includegraphics[width=\linewidth]{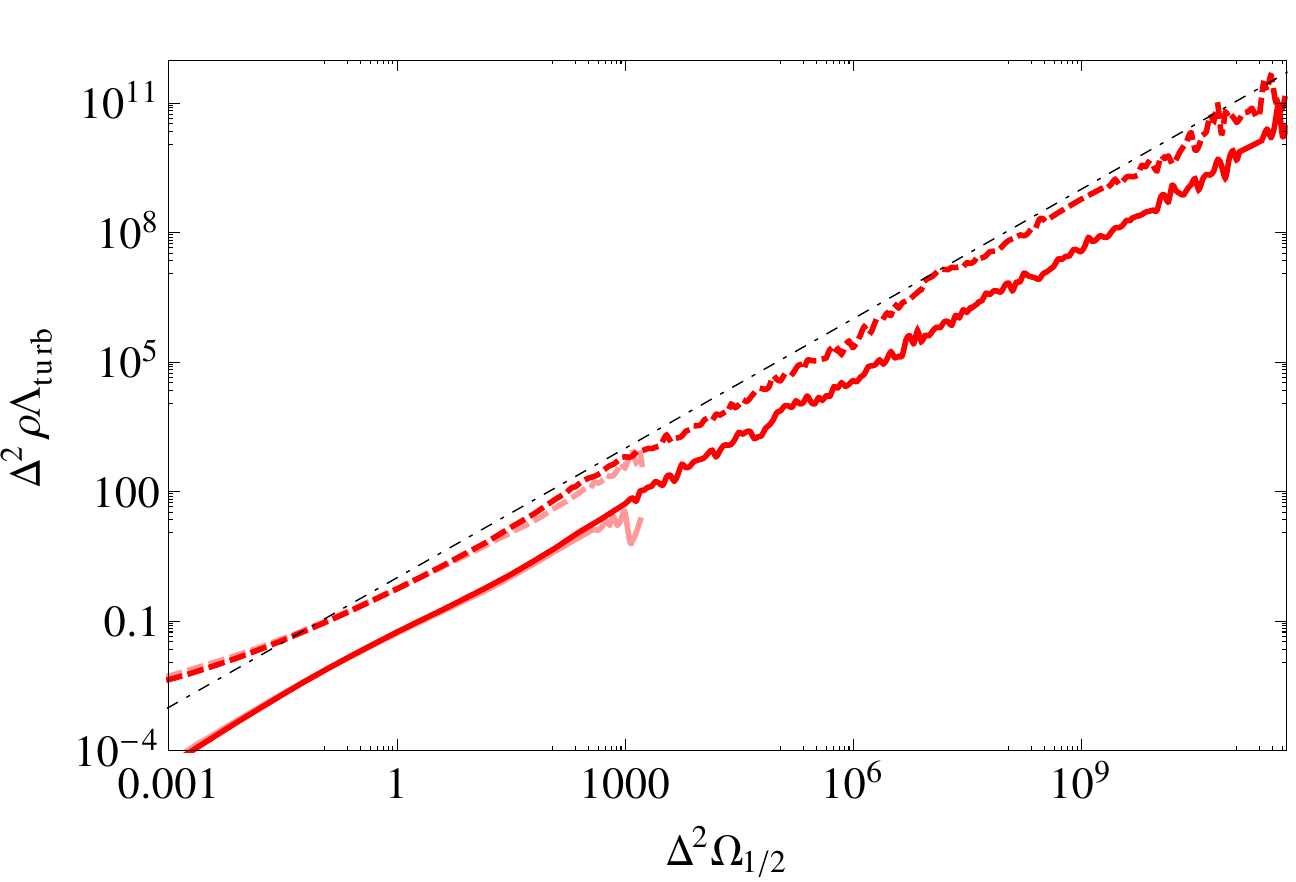}\\
\caption{Averages of the positive and negative thermal support $\Delta^2\rho\Lambda_{\rm therm\,\pm}$
	as functions of the thermal pressure (top) and the averaged turbulent support $\Delta^2\rho\Lambda_{\rm turb\,\pm}$ vs.\
	enstrophy density (middle) and denstrophy (bottom) at time $t=0.43 t_{\rm ff}$. 
	The positive and negative components are respectively shown as solid and dashed lines and the identity functions are indicated by a
	dot-dashed line in each plot. Factors of $\Delta^2$ are applied to 
	obtain quantities with the dimension of pressure. 
	For comparison, the light-coloured lines show the support functions for $t=0.1 t_{\rm ff}$.}
\label{fig:lambda_profiles}
\end{figure}

On the basis of the Jeans criterion, we expect that high-pressure regions are correlated with large, 
positive thermal support $\Lambda_{\rm therm}$ (Eq.~\ref{eq:support_therm}). 
A quantity with the physical dimension of pressure can be obtained by
multiplication of $\Lambda_{\rm therm}$ with the gas density $\rho$
and normalization with the square of the local grid scale $\Delta$. Fig.~\ref{fig:hd_lambda_therm_2d} in
the appendix shows phase plots of the resulting quantity 
$\Delta^2\rho\Lambda_{\rm therm}$ vs.\ $P$, i.~e., two-dimensional histograms of the occupied volume fractions.
Since the support functions are strongly fluctuating quantities, logarithmic scaling is necessary to
investigate the statistics over the full dynamical range of pressure or density. Consequently, we separate the 
support into positive and negative components $\Delta^2\rho\Lambda_{\rm therm\,\pm}$, where generically 
\begin{equation*}
  \label{eq:def_mean}
	\Lambda_+=\left\{\begin{array}{ll}
				\Lambda &\mbox{if}\ \Lambda \ge 0,\\
				0\ &\mbox{otherwise,}\ 
				\end{array}\right.\quad
	\Lambda_-=\left\{\begin{array}{ll}
				-\Lambda &\mbox{if}\ \Lambda \le 0,\\
				0\ &\mbox{otherwise.}\ 
				\end{array}\right.
\end{equation*}
The phase plots of the thermal support show that both positive and negative
contributions are significant, but large values of $\Lambda_{\rm therm\,+}$ appear to be more
frequent in comparison to $\Lambda_{\rm therm\,-}$, particular at high pressures.

\begin{figure}
\centering
  \includegraphics[width=\linewidth]{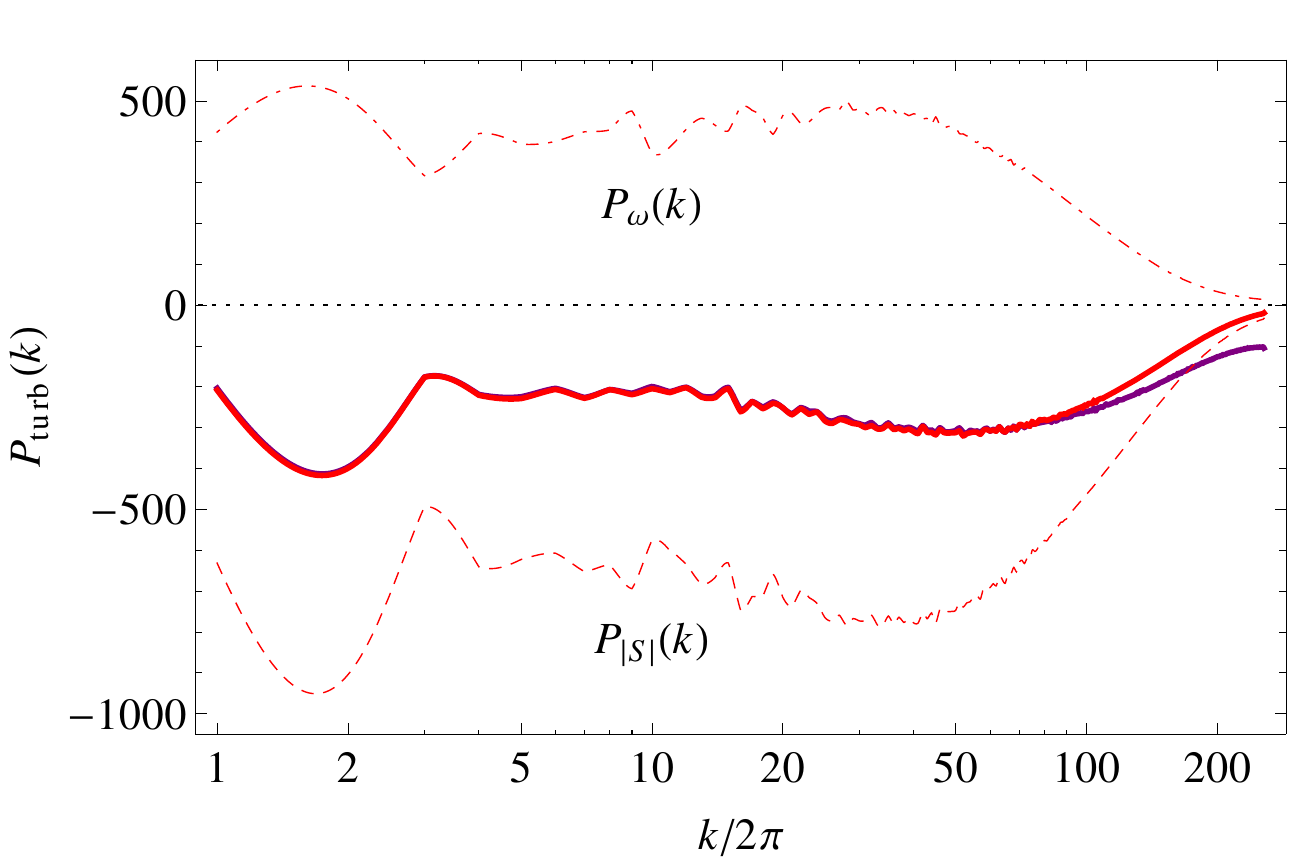}\\
\caption{Power spectrum of the turbulent support, computed with finite differences and pseudo-spectral
	derivatives (solid lines; the finite-difference computation is shown in red in the online version). 
	Also shown are the power spectra of the vorticity $\omega$ and the rate of strain.}
\label{fig:spect_turb}
\end{figure}

For a quantitative comparison that allows us to discern trends, we compute profiles, i.~e., 
mean values for small bins of pressure or other quantities (throughout this article, we use $0.05$ dex as bin width;
see Appendix~\ref{sc:phase} for further details). 
The mean positive and negative components add up to the mean net support, i.~e.,
$\langle f\Lambda\rangle=\langle f\Lambda_+\rangle-\langle f\Lambda_-\rangle$ 
both for volume- and mass-weighted averaging and an arbitrary positive function $f$. 
For $\Delta^2\rho\Lambda$, we calculate volume-weighted averages.
Owing to the factor $f=\Delta^2\rho$, this effectively results in weighing the support by mass.
This allows us to specify the typical magnitude of the support at a given pressure. In 
the top panel of Fig.~\ref{fig:lambda_profiles}, one can see that indeed the positive component of the thermal
support dominates for high pressure. Since gravitational contraction produces much higher densities than the initial 
supersonic turbulence, the range of pressures also increases greatly. This can be clearly seen by comparing to the early 
instant $t=0.1 t_{\rm ff}$, for which the thermal support is also plotted.
For $t=0.43 t_{\rm ff}$, we can approximate the pressure-averaged values by the asymptotic relation
\[
	\langle\Delta^2\rho\Lambda_{\rm therm}\rangle_P\simeq P\quad\mbox{for}\ P\gtrsim 100.
\]
A value of unity corresponds to the thermal pressure at the mean density. In this sense, a higher pressure corresponds 
to an enhanced support against gravity, as expressed by the formula for the Jeans length. But even for very high pressures, 
there is a non-vanishing fraction of negative support. Since $P\simeq c_{0}^2\rho$, the above relation implies the 
typical magnitude $\Lambda_{\rm therm}\sim (c_{0}/\Delta)^2$ at high densities. Naturally, the support functions
depend on the grid resolution $\Delta$. Nevertheless, we can draw meaningful conclusions, as will become clear in the course of
our analysis. For pressures or densities higher than  about $10^7$, there appears to be a systematic deviation of the thermal 
support from the above asymptotic relation. This  becomes even more prominent in the statistics of the thermal support 
relative to gravity, as is shown below.  The strong fluctuations for very high values of the pressure or density are due to
the small samples in each bin. 

Let us now consider the turbulent support. A common interpretation of turbulent support is that turbulent velocity fluctuations 
or, more specifically, eddy-like motions act against gravity. Let us assume
the scaling law\footnote{In code units with appropriate normalization. The actual exponent of the scaling law is not important for our 
reasoning. It can be anything between the Kolmogorov and the Burgers scaling
exponents.} 
$\delta v(\ell)\sim \ell^{1/2}$ for supersonic turbulent velocity fluctuations
on the length scale $\ell$.
Then we have $\omega(\ell)\sim\delta v(\ell)/\ell\sim\ell^{-1/2}$ for the magnitude of the corresponding vorticity. 
This implies that the dominant contribution to the numerically resolved vorticity comes from eddies at the 
(spatially varying) cutoff scale $\Delta$, i.~e., $\omega\propto\Delta^{-1/2}$. Contributions from larger eddies
are suppressed by $\sim(\Delta/\ell)^{1/2}$. The same argumentation applies to the rate of strain $|S|$ and
it follows from Eq.~(\ref{eq:support_turb}) that the local turbulent support $\Lambda_{\rm turb}\sim \Delta^{-1}$. 
Since the turbulent pressure has the dimension of  density times squared velocity fluctuations, 
we consider $\rho\Delta^2\Lambda_{\rm turb}$, where the first term, $\frac{1}{2}\Delta^2\rho\omega^2=\Delta^2\Omega$, 
can be interpreted as the turbulent pressure due to eddies on the grid scale and the second term, $-\frac{1}{2}\Delta^2\rho|S|^2$, 
corresponds to a negative pressure that is produced by the strain. The quantity $\Omega=\frac{1}{2}\rho\omega^2$ is the
enstrophy density. Since the trace of the rate-of-strain tensor is the divergence $d$, large strain is particularly caused by shocks. 
For the gravitational support, the crucial question is whether the positive or the negative component is dominant. 

\begin{figure*}
\centering
  \includegraphics[width=0.48\linewidth]{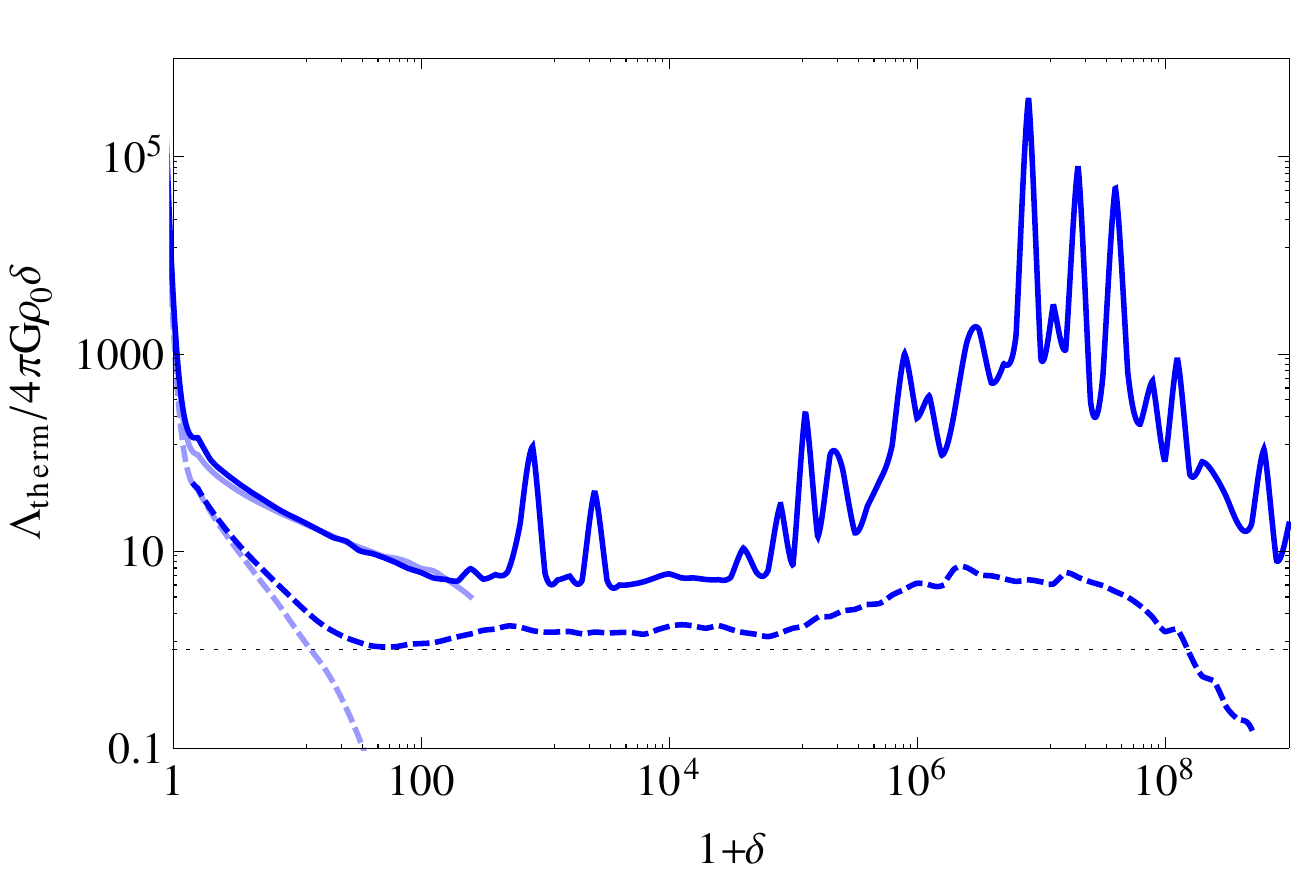}\quad
  \includegraphics[width=0.48\linewidth]{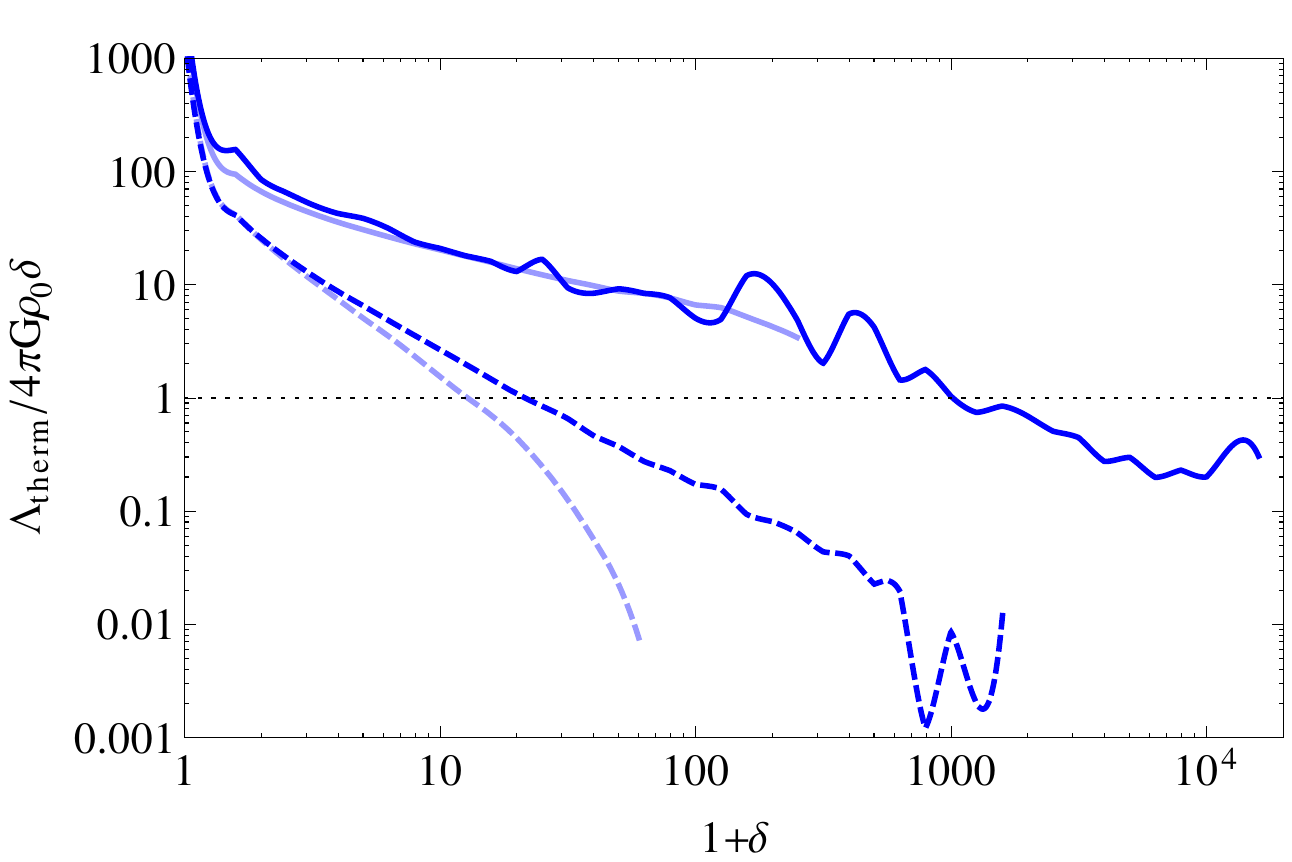}
  \includegraphics[width=0.48\linewidth]{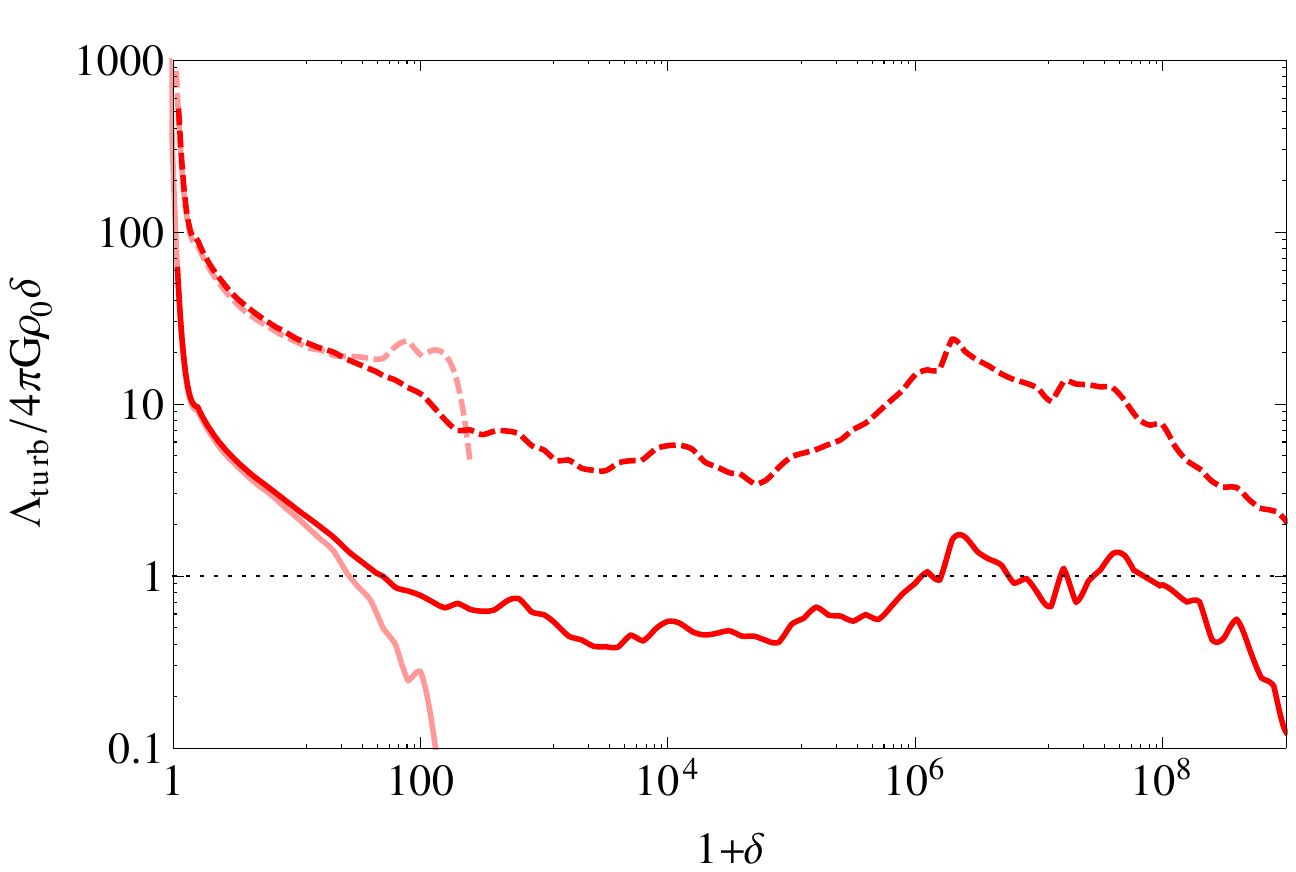}\quad
  \includegraphics[width=0.48\linewidth]{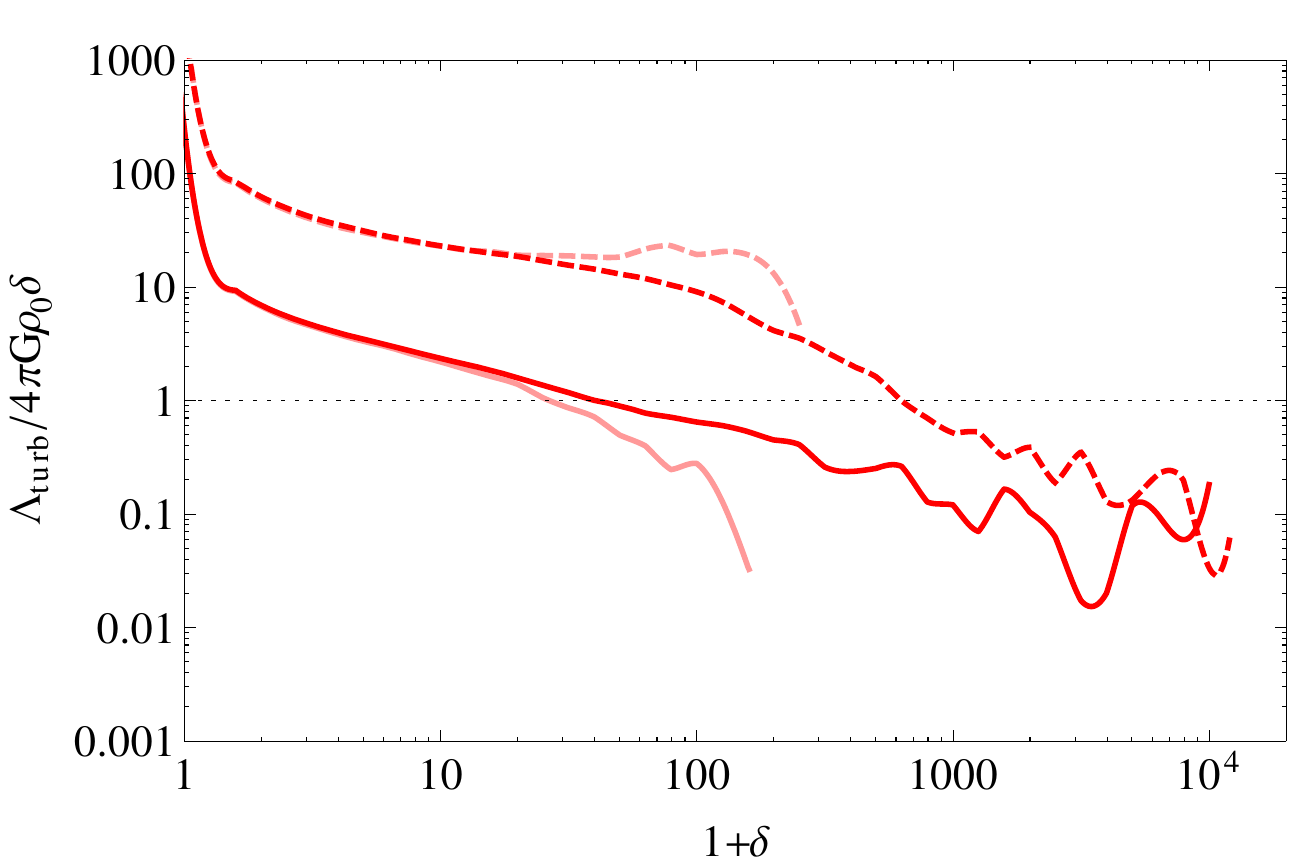}
\caption{Averages of the thermal (top) and turbulent (bottom) support relative to the
	gravitational compression rate in Eq.~(\ref{eq:compr}) as functions of the overdensity ($t=0.43 t_{\rm ff}$). 
	As in Fig.~\ref{fig:lambda_profiles}, positive and negative components are shown as solid and dashed lines and
	the support functions at time $t=0.1 t_{\rm ff}$ are plotted as light-coloured lines.
	The left and right columns of plots show the support functions for the full AMR data and
	the root-grid representations, respectively.}
\label{fig:stability_profiles}
\end{figure*}

Profiles of the turbulent support,
\begin{align*}
	\Delta^2\rho\Lambda_{\rm turb\,+} &=\left\{\begin{array}{ll}
				\Delta^2\rho(\omega^2-|S|^2)/2 &\mbox{if}\ \omega > |S|,\\
				0\ &\mbox{otherwise,}\ 
				\end{array}\right.\\
	\Delta^2\rho\Lambda_{\rm turb\,-} &=\left\{\begin{array}{ll}
				\Delta^2\rho(|S|^2-\omega^2)/2 &\mbox{if}\ \omega < |S|,\\
				0\ &\mbox{otherwise,}\ 
				\end{array}\right.
\end{align*}
are plotted in Fig.~\ref{fig:lambda_profiles}. Regardless of the average enstrophy density, we find that
the negative component $\Delta^2\rho\Lambda_{\rm turb\,-}$ is dominant. For a wide range of enstrophies, both 
components of the turbulent support follow very closely linear relations and the total turbulent support function 
is approximately given by the \emph{negative} turbulent pressure:
\begin{equation}
	\label{eq:turb_asymptot}
	\langle\Delta^2\rho\Lambda_{\rm turb}\rangle_{\Delta^2\Omega}\simeq-\Delta^2\Omega\quad 
	\mbox{if}\ \Delta^2\Omega\gtrsim 1.
\end{equation}
The factor $\Delta^2$ in this relation does not trivially cancel out because values from different refinement 
levels with varying $\Delta$ contribute to the average for a particular value of $\Delta^2\Omega$. 
The above relation also holds for the earlier instant $t=0.1 t_{\rm ff}$, except that the maximum vorticity is much lower. 
The above relation also implies that the typical magnitude of $\rho|S|^2$ is linearly related to
$\Omega$. For dimensional reasons, the turbulent pressure may alternatively be estimated by $\Delta^2\Omega_{1/2}$, where $
\Omega_{1/2}$ is the denstrophy (see Sect.~\ref{sc:support}). As argued by \citet{KritNor07}, the denstrophy combines the 
effects of eddies (through the rotation of $\vecv$) and shocks (through the density gradient). The resulting profile in  
Fig.~\ref{fig:lambda_profiles} (bottom panel) shows a very similar trend and also a nearly linear asymptote. 
The phase plots in Fig.~\ref{fig:hd_lambda_turb_2d} show a substantial overlap between positive and 
negative turbulent support so that locally the support by vortices can be stronger than
the compressive effect of turbulence. For a given vorticity, however, 
$\langle\Delta^2\rho\Lambda_{\rm turb\,+}\rangle_{\Delta^2\Omega}$ could exceed
$\langle\Delta^2\rho\Lambda_{\rm turb\,-}\rangle_{\Delta^2\Omega}$ only if the rate of strain were 
smaller than the vorticity in the majority of cells.

The length scales, which are dominated either by strain or vorticity, can be inferred by decomposing the turbulent support in Fourier space.
The scaling law $\omega(\ell)\sim\ell^{-1/2}$ 
implies a flat power spectrum of $\Lambda_{\rm turb}$.\footnote{The scale-dependence $\Lambda_{\rm turb}\sim \Delta^{-1}$
corresponds to a linear relation with the wavenumber $k$, which cancels with the factor $k^{-1}$ for a spectrum that
measures the power per unit wave number. The turbulent velocity scaling $\delta v(\ell)\sim 
\ell^{1/2}$, on the other hand, corresponds to the spectrum $E(k)\propto k^{-2}$.} 
Since the turbulent support is quadratic in the velocity derivative,
\[
	\Lambda_{\rm turb}=\frac{1}{2}\left(\omega^2-|S|^2\right)=v_{i,j}v_{j,i},
\]
the computation a power spectrum is straight-forward:
\begin{equation}
	\begin{split}
	P_{\rm turb}(k)
	&= -\frac{1}{2}\oint_{|\veck|=k}k_i k_j (\hat{v}_i \hat{v}_j^{\ast} + \hat{v}_i^{\ast}\hat{v}_j)k^2\dd\Omega_{\veck}\\
	&= \frac{1}{2}\oint_{|\veck|=k}(\widehat{v_{i,j}}\,\widehat{v_{j,i}}^{\ast} + \widehat{v_{i,j}}^{\,\ast}
\widehat{v_{j,i}})
		k^2\dd\Omega_{\veck}\\
	&= \oint_{|\veck|=k}
	\left(\frac{1}{2}\hat{\omega}_i\hat{\omega}_i^{\ast} - \widehat{S_{ij}}\widehat{S_{ij}}^{\ast}\right)k^2\dd\Omega_{\veck}
\\
	&= \frac{1}{2}[P_\omega(k)-P_{|S|}(k)].
	\end{split}
\end{equation}
In the above expressions, the Fourier transform of a field $f$ is denoted by $\hat{f}$ and $\hat{f}^{\ast}$ is its complex 
conjugate.
The integrands are symmetrized by complex conjugation and integrated over spherical shells of radius $k$ in Fourier space.
By using the root-grid representations of the data,\footnote{A consistent compuation of the spectra for the higher refinement level
is not possible because of the missing high-wave number modes of turbulence in coarser regions, which are not refined by
the Truelove-like criterion.}
we computed $P_{\rm turb}(k)$ both from the Fourier transforms of the 
finite-difference approximations to $v_{i,j}$ and from the pseudo-spectral derivatives ${\rm i}k_j\hat{v}_i$. The results are 
shown in Fig.~\ref{fig:spect_turb}. The agreement between the two  methods of computation is very good. Only toward the cutoff 
wavenumber, there is a deviation due to Gibbs phenomenon. Remarkably, it turns out that $P_{\rm turb}(k)<0$ for all wave numbers,
i.e., the compression produced by shocks overcompensates the support by turbulent vortices. This can also be seen from the evaluation of the 
power spectra of $\omega$ and $|S|$. We find that $P_\omega(k)$ is smaller than $\frac{1}{2}P_{|S|}(k)$ for all wave numbers. 
Moreover, $P_{\rm turb}(k)$ is nearly flat in between the forcing range at the smallest wavenumbers and the numerical dissipation range, 
as expected from the  aforementioned scaling arguments. Apart from the damping close to the cutoff wavenumber, one can also see 
the typical bottleneck bump \citep[see][]{KritNor07}. 


\begin{figure}
\centering
  \includegraphics[width=\linewidth]{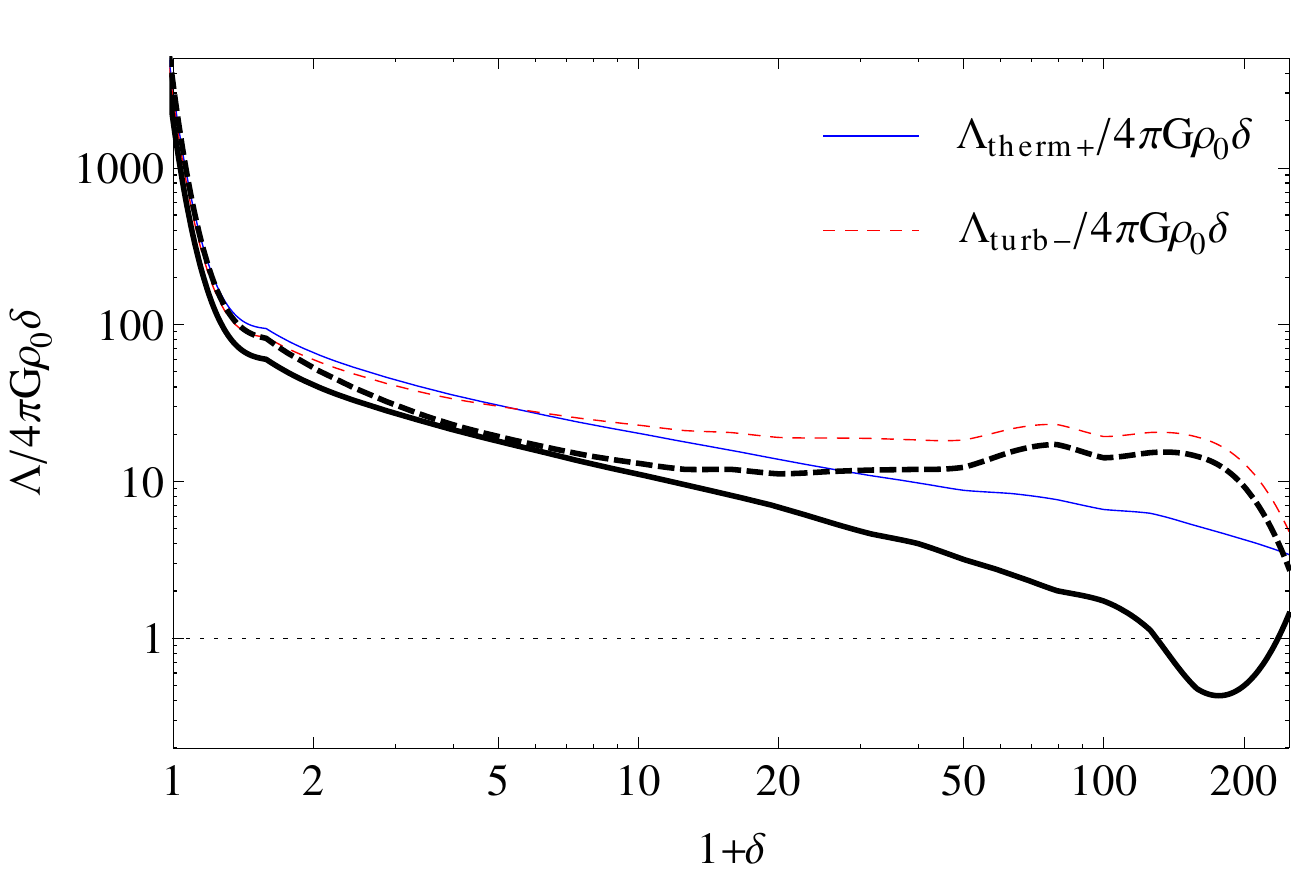}
\caption{Plots of the positive (thick solid) and negative (thick dashed) components of the total support
	at $0.1 t_{\rm ff}$. Also shown are the dominant components
	of the thermal and turbulent support functions (thin lines).}
\label{fig:stability_profiles_tot_early}
\end{figure}

\begin{figure}
\centering
  \includegraphics[width=\linewidth]{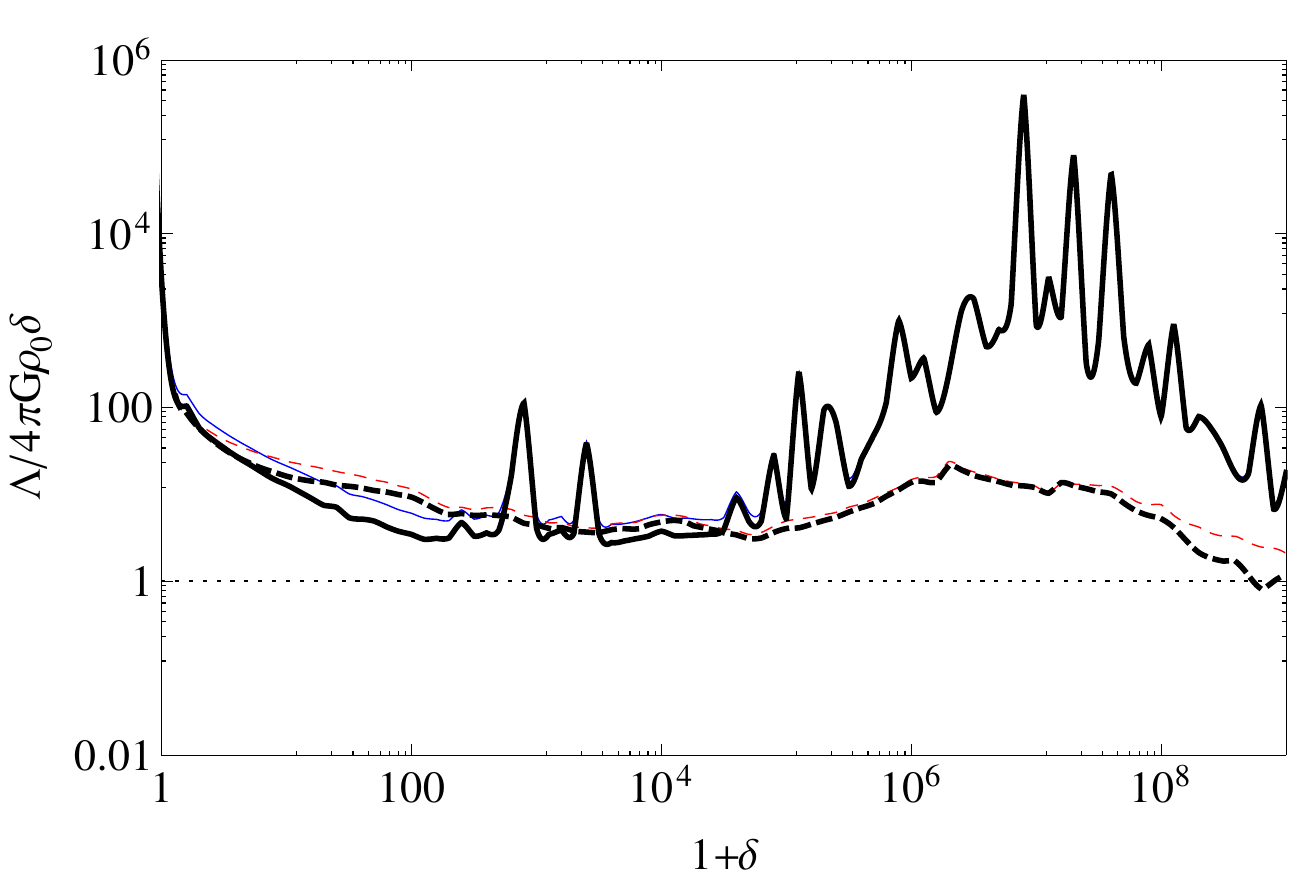}
  \includegraphics[width=\linewidth]{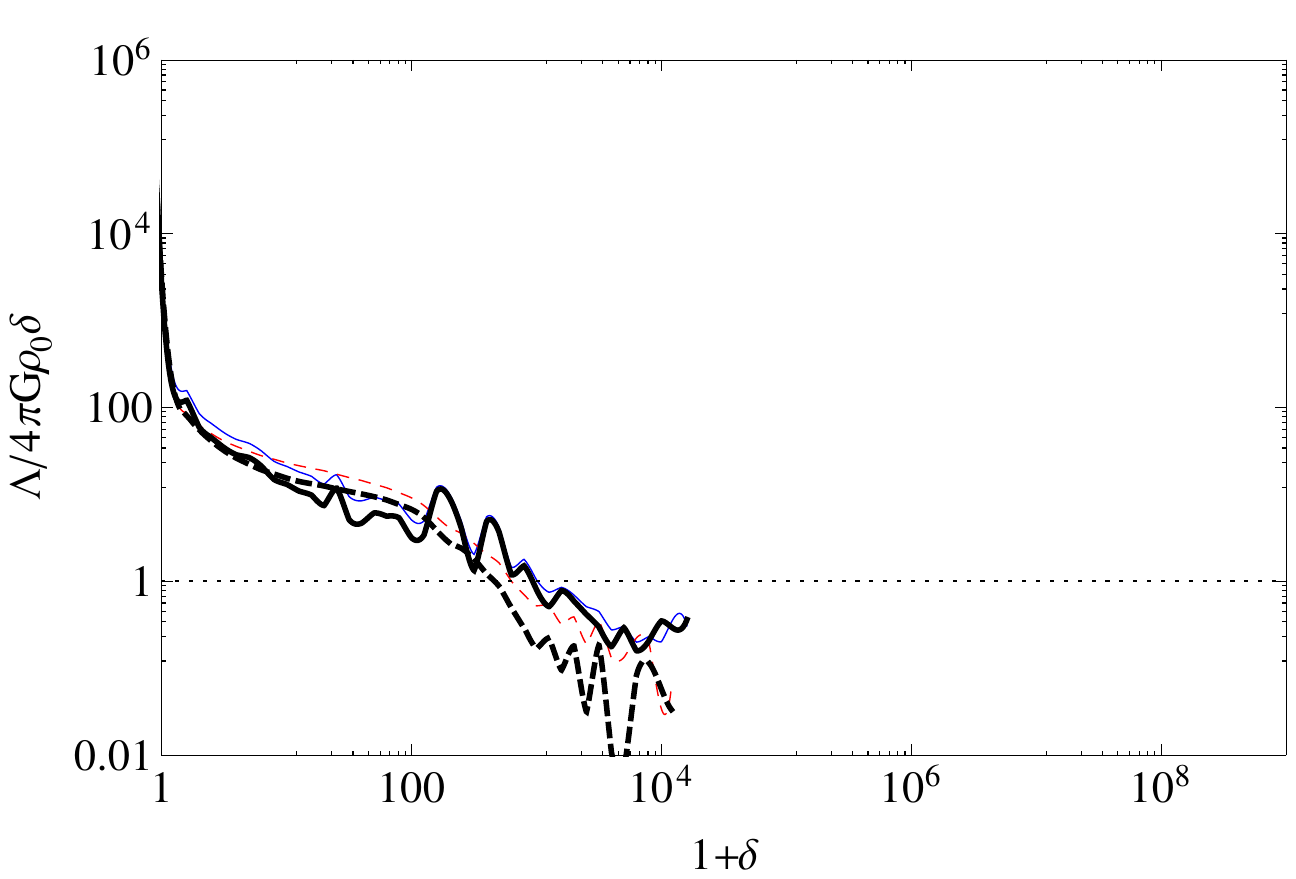}
\caption{Plots of the same functions as in Fig.~\ref{fig:stability_profiles_tot_early} both for the full AMR data (top) 
	and the root-grid data (bottom) at time $t=0.43 t_{\rm ff}$.}
\label{fig:stability_profiles_tot_late}
\end{figure}

\begin{figure}
\centering
  \includegraphics[width=\linewidth]{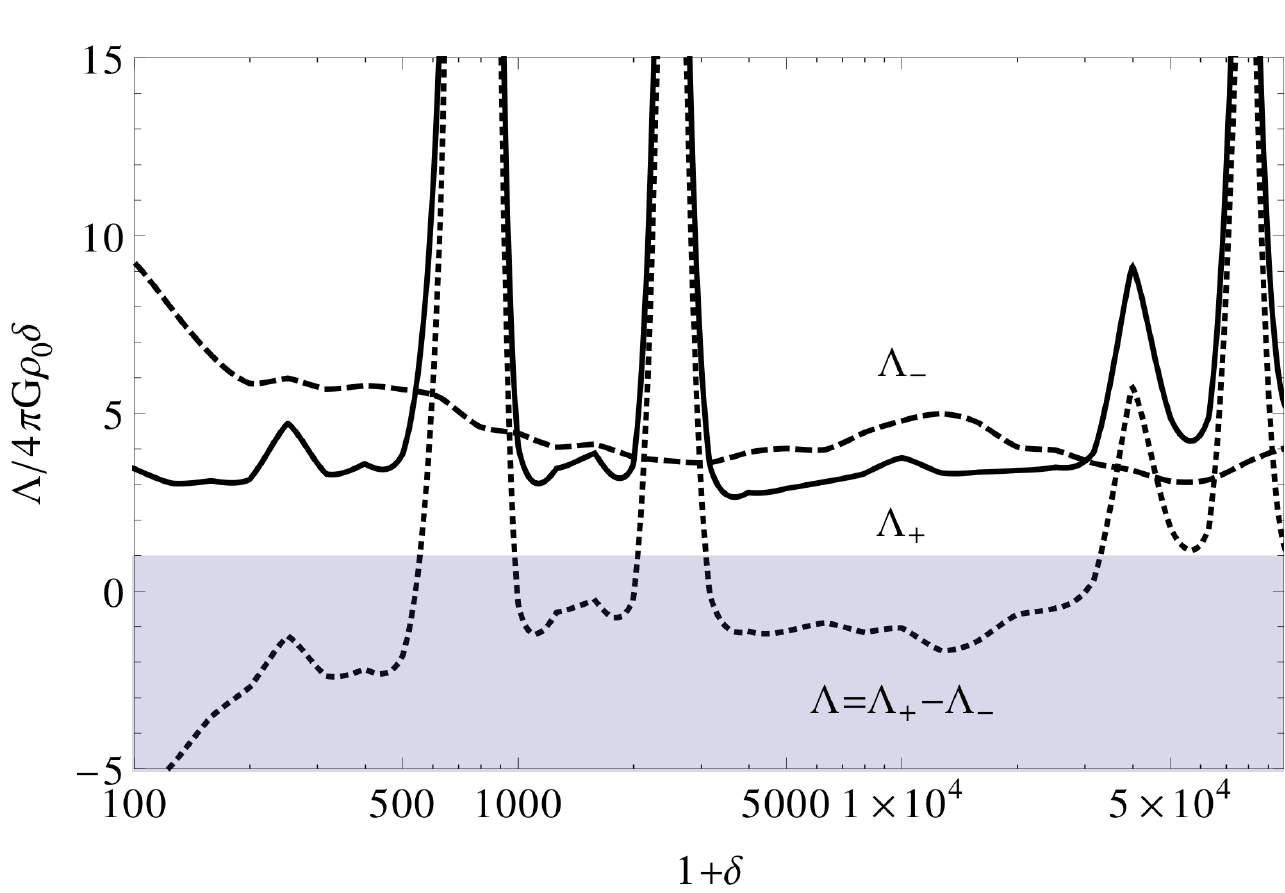}
  \includegraphics[width=\linewidth]{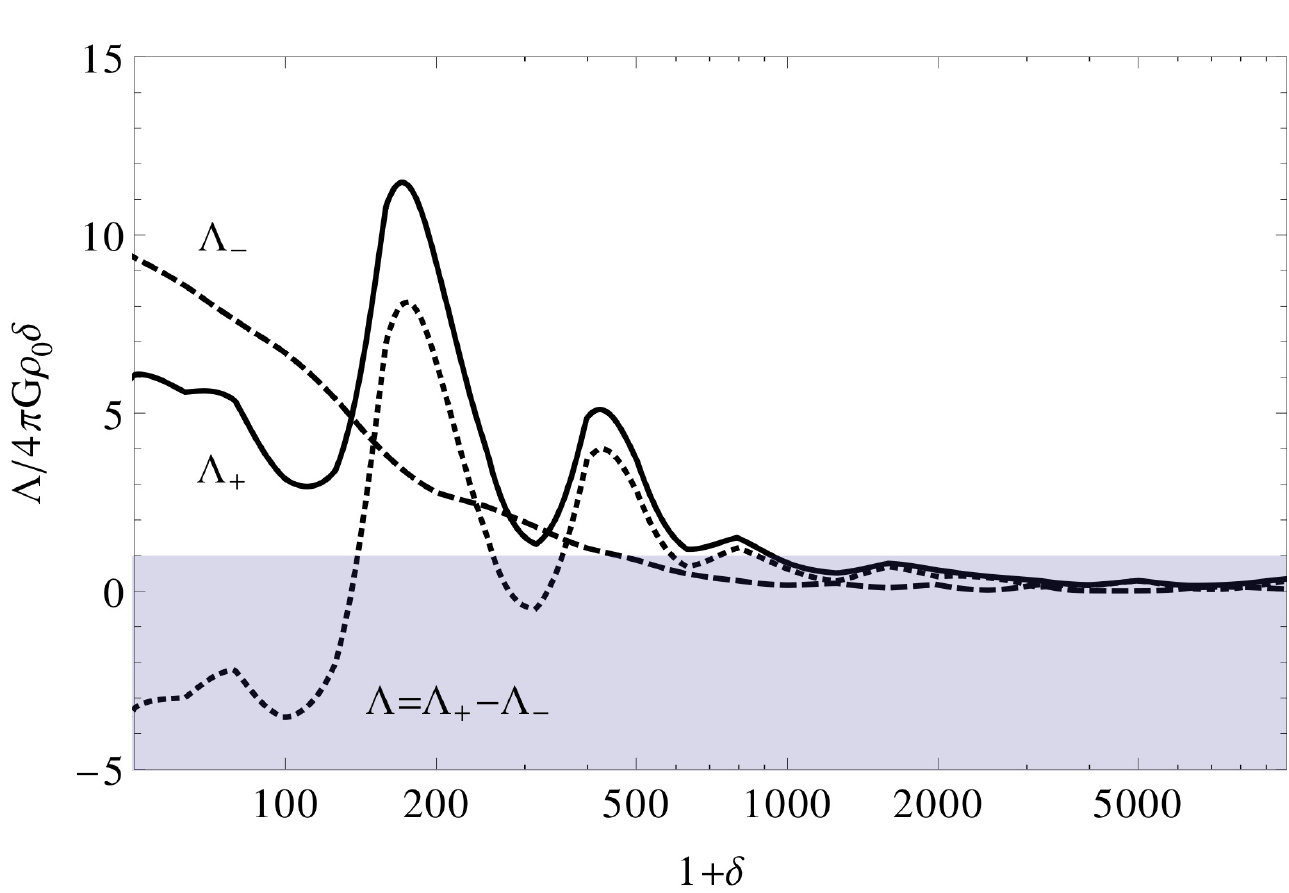}
\caption{Linear plot of the average total support at time $t=0.43 t_{\rm ff}$, showing the positive and negative
	components as well as the net support. The grey shaded region indicates the region for which
	$\Lambda<4\pi G\rho_0\delta$, i.~e., negative compression rate.
	The full AMR data are plotted on the left, the root-grid representation on the right.}
\label{fig:stability_profiles_lin}
\end{figure}

To investigate the magnitude of the local support function $\Lambda$ relative to the gravitational compression rate $4\pi G\rho_0\delta$ in Eq.~(\ref{eq:compr}), 
we compute averages of the ratio $\Lambda_{\pm}/4\pi G\rho_0\delta$ for logarithmic bins of the overdensity $1+\delta=\rho/\rho_0$.
For this ratio, we use the mass per cell as weighing function.
First, we consider the thermal and turbulent support functions separately (see Fig.~\ref{fig:stability_profiles}). In the early phase ($t=0.1 t_{\rm ff}$), the the support functions simply decrease with density. 
The total support is plotted together with the dominant thermal and turbulent components in Fig.~\ref{fig:stability_profiles_tot_early}. 
The critical value of unity is approached for the highest densities ($\delta\sim 100$), which indicates that the gas is self-gravitating. Since $\Lambda_{\rm turb\,-}/4\pi G\rho_0\delta$ is large, however, the gas is mainly compressed by shocks. For $t=0.43 t_{\rm ff}$, the statistics plotted in Figs.~\ref{fig:stability_profiles} and~\ref{fig:stability_profiles_tot_late} clearly show that the positive component of the thermal support and the negative component of the turbulent support continue to dominate for all $\delta>0$. Compared to $t=0.1 t_{\rm ff}$, the support remains nearly unaltered for the lower range of densities, but an intermediate range $100\lesssim\delta \lesssim 10^5$ has formed, in which the ratio of support to gravity is nearly constant and $\Lambda_{\rm therm\,+}$ is roughly balanced by $\Lambda_{\rm turb\,-}$ 
(see top panel in Fig.~\ref{fig:stability_profiles_tot_late}). 
For this range of densities, the simulations resolve the physics of self-gravitating turbulence
very well. In the phase plots of $\Lambda_{\rm therm}/4\pi G\rho_0\delta$ and $\Lambda_{\rm turb}/4\pi G\rho_0\delta$ (see Figs.~\ref{fig:hd_stab_therm_2d} and~\ref{fig:hd_stab_turb_2d}), one can see that the strong thermal support at high densities is associated with rare strong fluctuations, while the negative turbulent support extends toward higher values than the positive component. At all densities, there is a significant volume fraction for which the support relative to the compression rate is small. This is simply a consequence of the support function fluctuating between positive and negative values. The gas is collapsing only if the mean support
is sufficiently small.

In Fig.~\ref{fig:stability_profiles_lin}, the positive and negative components of the total support are plotted in linear scaling together with the profile of the net support $\Lambda/4\pi G\rho_0\delta$, where $\Lambda=\Lambda_+-\Lambda_-$.
Except for the strong spikes, which probably correspond to collapsing filaments that start to feel strong pressure support, the net support is negative and of the order of the gravity term. For the floor, we have $|\Lambda|\sim 4\pi G\rho_0\delta$ and a median value
$\Lambda/4\pi G\rho_0\delta\approx -0.67$. This indicates an accelerated contraction of the gas, as supersonic gas compression
triggers gravitational collapse. The absence of a dominant positive thermal pressure component in this range of densities 
also supports the explanation for the deviation of the power-law index of the density PDF tail from $-1.5$ given in \citet{KritNor11}. 
There, the pressureless collapse solution of \cite{Penst69} is used to explain a steeper, $\approx -1.7$, slope of the tail observed in the simulation. At densities around $10^7$, however, the thermal pressure of 
the gas tends to overcompensate gravity. This does not necessarily mean that the gas is expanding ($d>0$), but that the contraction 
slows down ($\DD d/\DD t>0$) toward higher densities. \citet{KritNor11} show that an overdensity of $10^7$ marks the transition to 
rotationally supported cores, which entails a decreasing importance of thermal support. This transition can be seen as a flattening 
of the power-law slope of the density  PDF (see Fig.~1 in their paper). For $\delta \gtrsim 10^7$, both the thermal and the 
turbulent support decrease relative to gravity. 
However, the gas mass per cell ceases to be a small fraction of the local Jeans mass for extremely dense gas (the Jeans 
length is comparable to the cells size at the highest refinement level for $\delta\sim\times 10^{10}$) and, consequently, the
gas dynamics is not well resolved.\footnote{
	Regardless of the significance of the local Jeans length for gravitational instability, this is suggested by
	simple time-scale arguments.
} 

A different picture emerges if we compute the support functions from the root-grid representation of the data. This amounts to a smoothing or filtering operation that averages the data at the higher refinement levels down to the much coarser cells at the root grid. The
averaging over finer cells conserves mass and momentum.
Compared to the highest refinement level, the ratio of the linear grid scales is $4^5=1024$. 
Consequently, the smoothing of the densest structures in the regions with maximum refinement is substantial. This is
reflected in the much smaller density range of the support functions plotted on the right-hand side of Fig.~\ref{fig:stability_profiles}, 
where we computed the derivatives in Eqs.~(\ref{eq:support_therm}) and~(\ref{eq:support_turb}) from the smoothed 
density, velocity, etc.\ fields. Most importantly, we find that the $\Lambda_\pm/4\pi G\rho_0\delta$ gradually decreases to values below unity for both the thermal and the turbulent support, thus indicating that the densest gas is strongly self-gravitating. The transition occurs
at densities around $\delta\sim 10^3$. Moreover, the net support is positive, as one can see
in the plot with linear scaling shown in Fig.~\ref{fig:stability_profiles_lin} (bottom plot). Consequently, the properties of the
smoothed structures are markedly different from the collapsing structures with negative net support that appear at higher refinement levels. Apart from that, the positive turbulent support shown in Fig.~\ref{fig:stability_profiles}
increases relative to the negative component for $\delta\gtrsim 1000$, which suggests that
the smoothed-out effect of vorticity in the dense, self-gravitating gas becomes more important. This does not contradict the power spectra
in Fig.~\ref{fig:spect_turb}, where the vorticity and the rate of strain are
averaged for given wavenumbers. The volume weighted power spectra are not
sensitive to the high density, low volume fraction self-gravitating structures.  Here, we find a relatively
large vorticity compared to the rate of strain for very high densities.  
A composite plot of the support functions for the root-grid data is shown in the bottom panel of 
Fig.~\ref{fig:stability_profiles_tot_late}.
The large values of $\Lambda_\pm/4\pi G\rho_0\delta$ for $\delta\gtrsim 10^4$, which can be seen for the full AMR data in the top panel,
are absent. The very large peaks of the thermal support at the higher refinement levels are likely due to strongly localized and to some degree
under-resolved structures that tend to be over-pressurized. In contrast, gravity in refined regions is more dominant for a given density 
at the root grid because the mass elements are larger, whereas local fluctuations tend to be averaged out. Thus, we observe the simple trend of gravity becoming increasingly dominant for increasing density.

\begin{table}
\begin{center}
\caption{Ratio of mean thermal to magnetic pressures for the MHD simulations.}\label{tb:beta}
\begin{tabular}{lll}
\hline
$t/t_{\rm ff}$ & $\beta_0$ & $\beta(t)$ \\
\hline
\hline
0.1 & 20 & 0.2037 \\
0.5 & 20 & 0.2056 \\
\hline
0.1 & 0.2 & 0.0336 \\
0.5 & 0.2 & 0.0339 \\
\hline
\end{tabular}
\end{center}
\end{table}

\section{Statistical analysis of magnetohydrodynamical turbulence}
\label{sc:mhd_stat}

To analyze the support in magneto-turbulent fluids, we use data from the simulations by \citet{CollKrit12} for
a weak magnetic field with a mean thermal-to-magnetic pressure ratio
$\beta_0=20$ and a moderately strong field with $
\beta_0=0.2$. The case of hydrodynamical turbulence considered in the previous section corresponds to the limit $\beta_0\rightarrow\infty$.
Since tha parameter $\beta_0$ specifies the strength of the initial, uniform field, we calculated the ratio of the mean thermal and
magnetic pressures, which are summarized in Table~\ref{tb:beta}. As one can see, the magnetic pressure of the
turbulent gas is much higher than the initial value. For brevity, we use the parameter $\beta_0$
to tag the two MHD models. The virial parameter of both models is $1$ and the sonic Mach number $\mathcal{M}$ is about $9$. Since 
$\mathcal{M}^2=v_{\rm rms}^2/c_{\rm s}^2$ and $\beta=8\pi\rho_0/c_{\rm
s}^2\langle B^2\rangle$, it follows that mean magnetic energy is comparable to
the mean kinetic energy for $\beta_0=0.2$ (corresponding to $\beta\approx0.034$), while it is significantly smaller for 
$\beta_0=20$ (corresponding to $\beta\approx 0.2$). Compared to the hydrodynamical model, gravity is weaker in the MHD simulations, but the
turbulence energy is higher. Moreover, in contrast to the hydrodynamical simulation, we have
a significantly lower limit of $\delta\approx 6.35\times 10^3$ for resolved overdensities
(see Sect.~\ref{sc:sim}).

\begin{figure*}
\centering
  \mbox{\subfigure[$\beta_0=20$]{\includegraphics[width=0.46\linewidth]{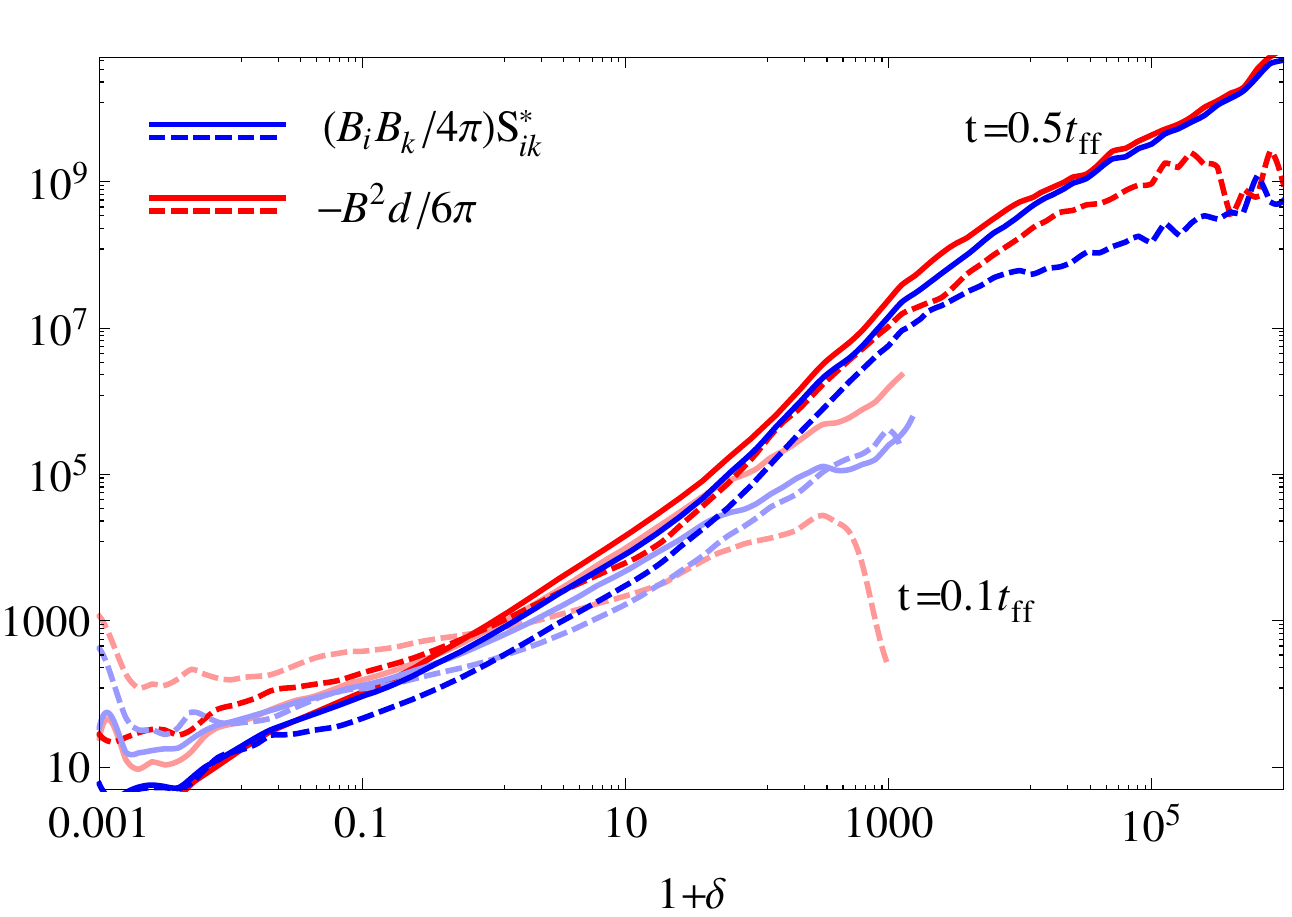}}\quad
  	\subfigure[$\beta_0=0.2$]{\includegraphics[width=0.46\linewidth]{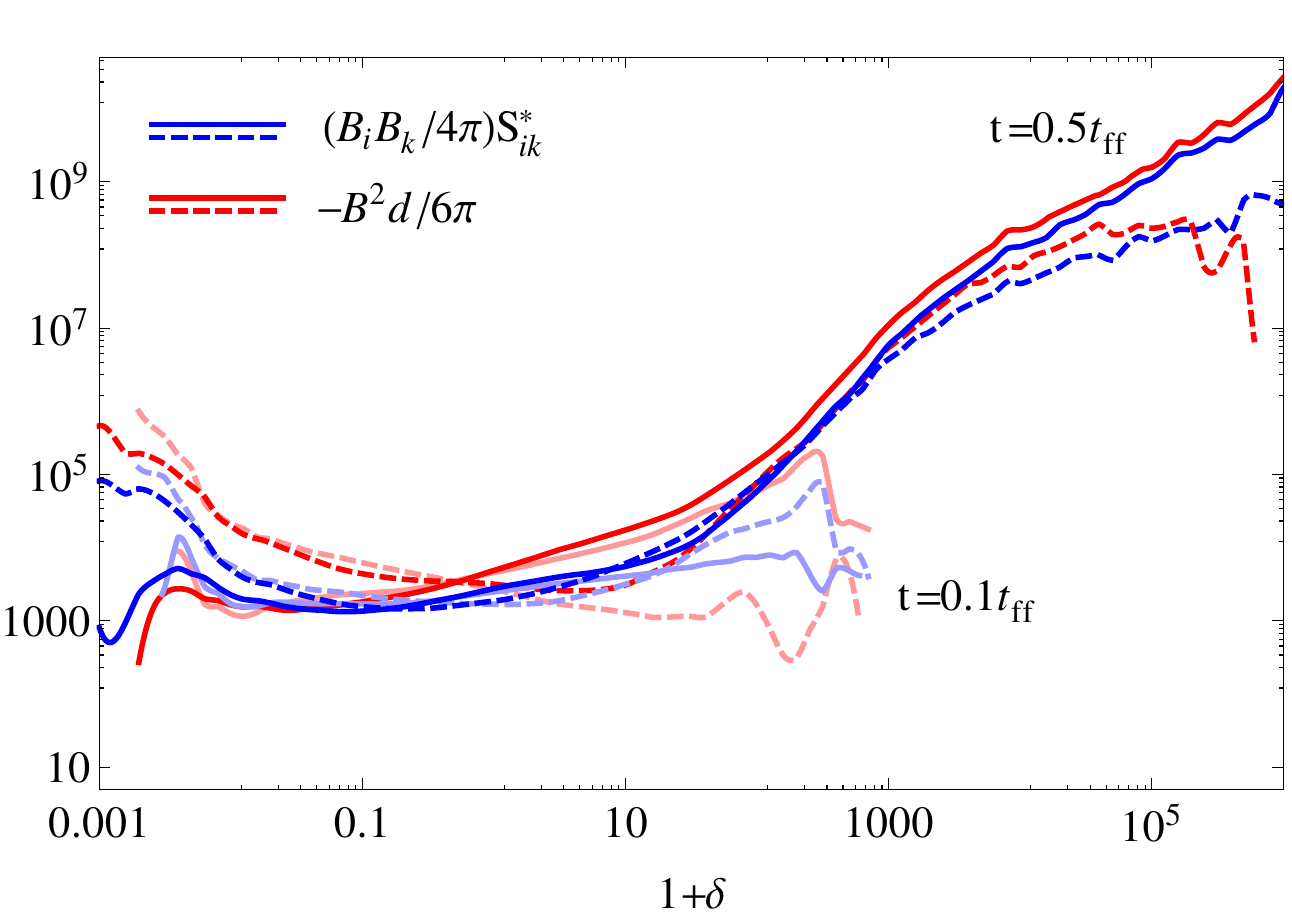}}}
\caption{Plots of the positive (solid lines) and negative (dashed lines) components of the
	shear-induced and compressive magnetic field amplification (see Eq.~\ref{eq:dynamo}) as functions of overdensity at time 
	$t=0.1t_{\rm ff}$ (light colour) and $t=0.5t_{\rm ff}$ (full color).}
\label{fig:dynamo_dens}
\end{figure*}

Since the impact of magnetic fields depends on the amplification by shear and compressions, we first 
consider the two corresponding source terms in Eq.~(\ref{eq:dynamo}) for the time evolution of the magnetic pressure. 
The first term arises from pure (trace-free) shear, while the second term is due 
to gravity and shocks (both are associated with highly negative divergence, which results in positive $-B^2 d$).
The amplification of the magnetic field induced by the shear of the turbulent flow can be understood as small-scale dynamo action.
The mean positive and negative components of these terms are plotted as functions of the overdensity in Fig.~\ref{fig:dynamo_dens}. 
There is clearly a net amplification of the magnetic field in the high-density gas for both $\beta_0=0.2$ and $20$. 
However, in the initial phase ($t=0.1 t_{\rm ff}$), the total amplification is small because the amplification at high densities 
(solid light lines) is roughly counterbalanced by the weakening of the field at low densities (dashed light lines). 
This indicates that the magnetic field is close to saturation,
which makes sense because we start with statistically stationary turbulence at time $t=0$. 
But the graphs for $t=0.5t_{\rm ff}$ show that gravitational contraction causes further amplification in overdense gas ($\delta>1$), 
which is nearly matched by shear-induced amplification. In comparison to $\beta_0 = 0.2$, the field amplification  
in the \emph{weak}-field case ($\beta_0=20$) is noticeably \emph{larger} for overdensities in the range $1\lesssim\delta\lesssim 10^3$.
Below, we show that a similar trend is found for the magnetic support. 

\begin{figure*}
\centering
  \mbox{\subfigure[$\beta_0=20$]{\includegraphics[width=0.46\linewidth]{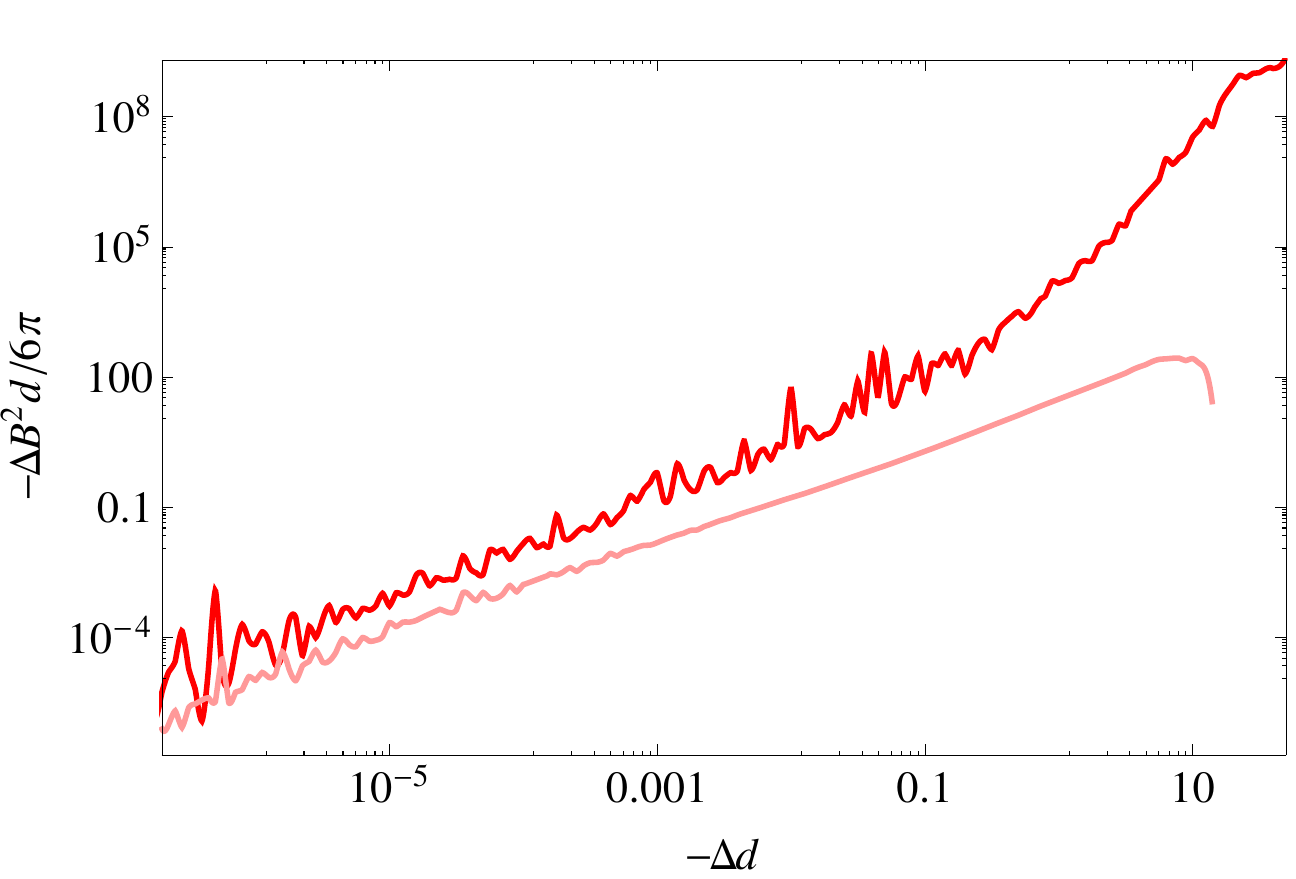}}\quad
  	\subfigure[$\beta_0=0.2$]{\includegraphics[width=0.46\linewidth]{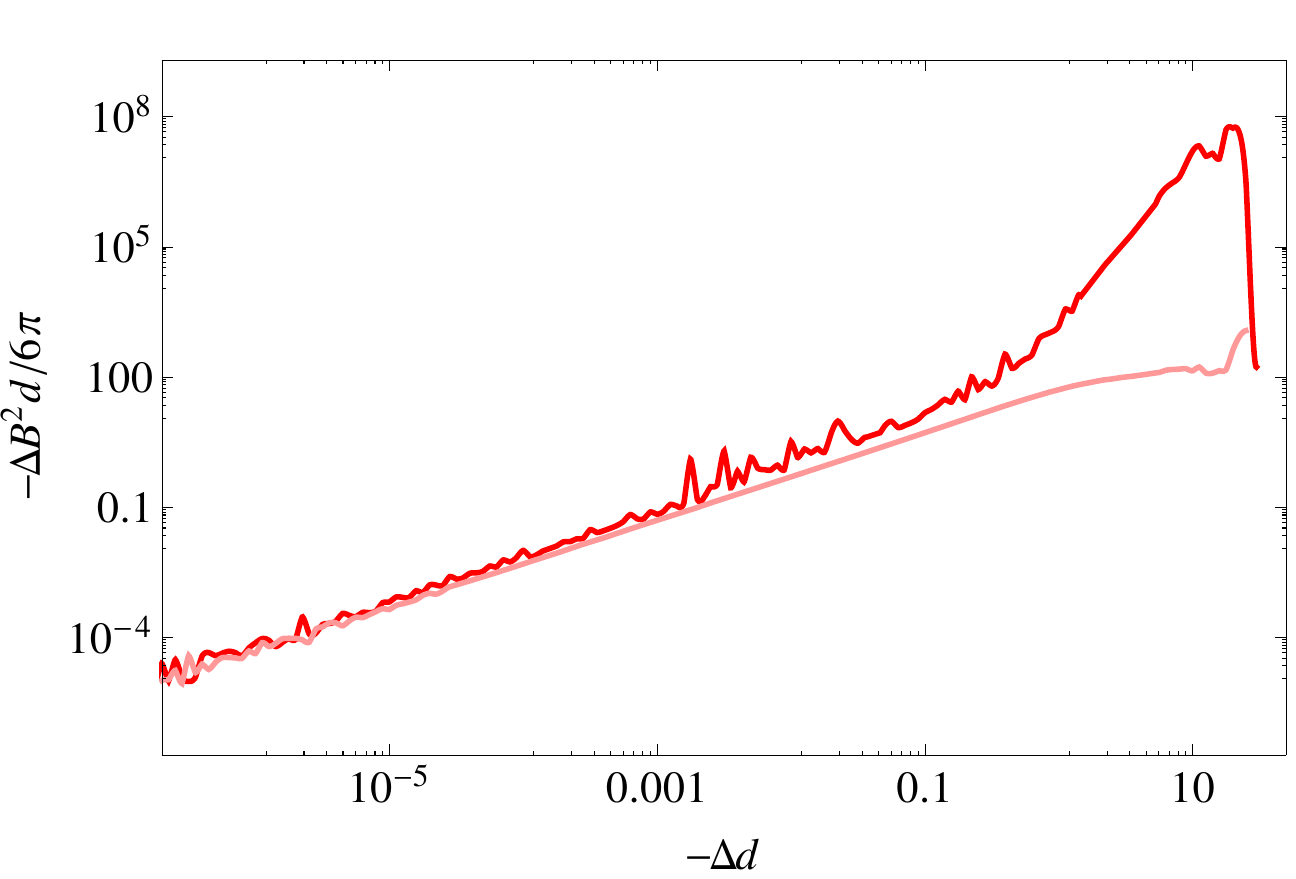}}}
\caption{Plots of the magnetic field amplification by contraction as functions of negative divergence at time 
	$t=0.1t_{\rm ff}$ (light colour) and $t=0.5t_{\rm ff}$ (full color), as in Fig.~\ref{fig:dynamo_dens}.}
\label{fig:ampl_div}
\end{figure*}

\begin{figure*}
\centering
  \mbox{\subfigure[$\beta_0=20$]{\includegraphics[width=0.46\linewidth]{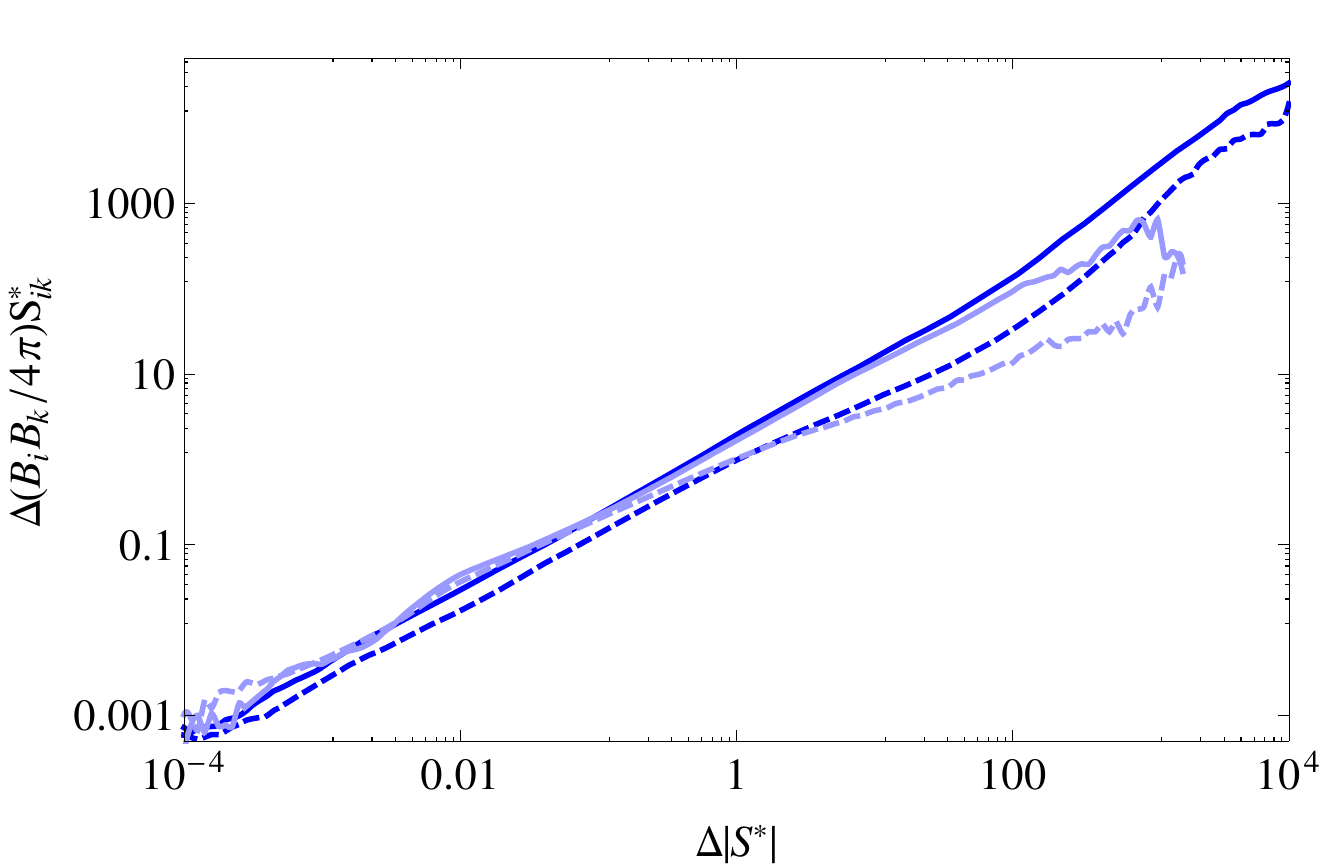}}\quad
  	\subfigure[$\beta_0=0.2$]{\includegraphics[width=0.46\linewidth]{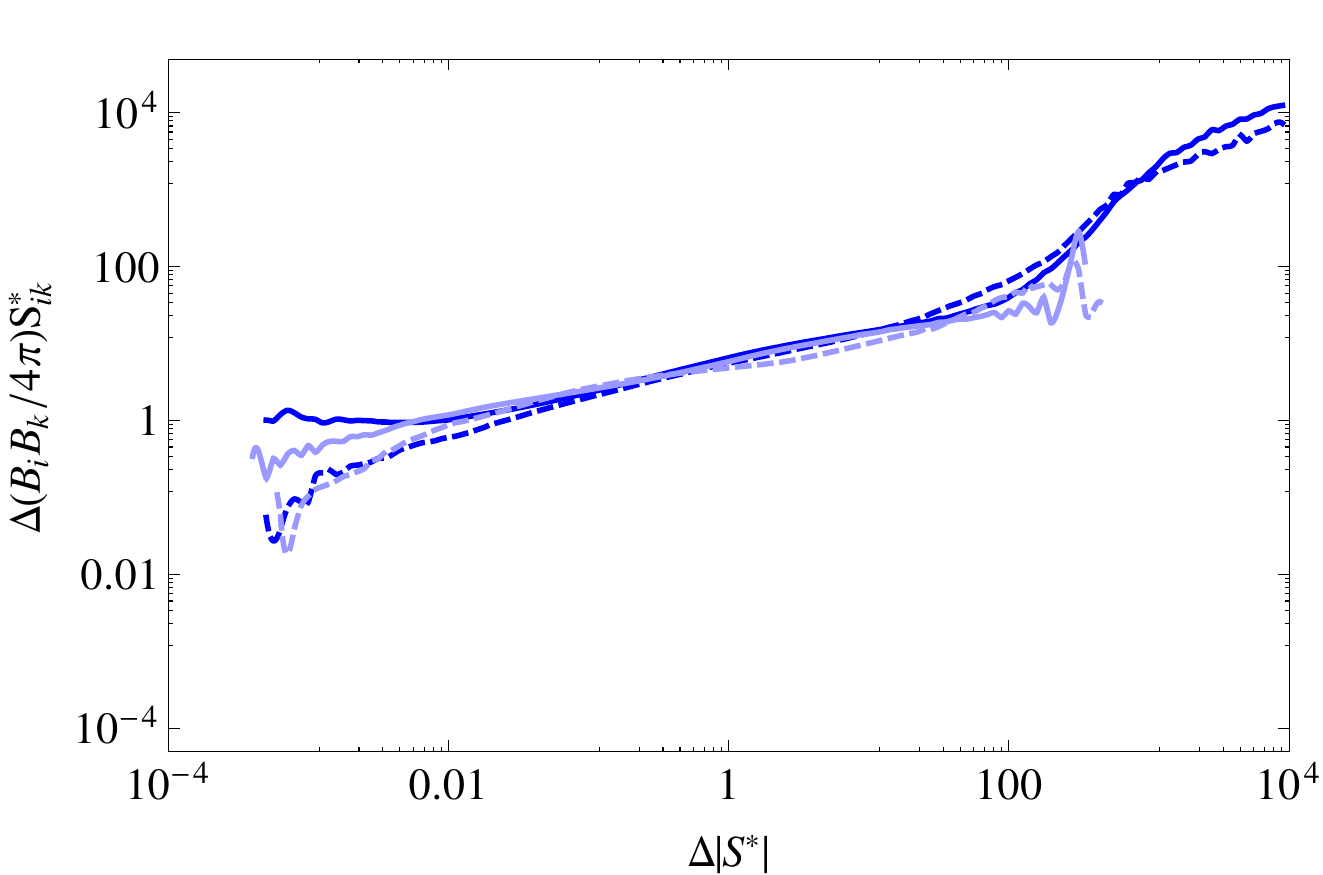}}}
\caption{Plots of the positive (solid lines) and negative (dashed lines) components of dynamo term in Eq.~(\ref{eq:dynamo}) 
	as functions of trace-free rate-of-strain scalar at time $t=0.1t_{\rm ff}$ (light colour) and $t=0.5t_{\rm ff}$ 
	(full color), as in Fig.~\ref{fig:dynamo_dens}.}
\label{fig:dynamo_strain}
\end{figure*}

\begin{figure}
\centering
  \includegraphics[width=\linewidth]{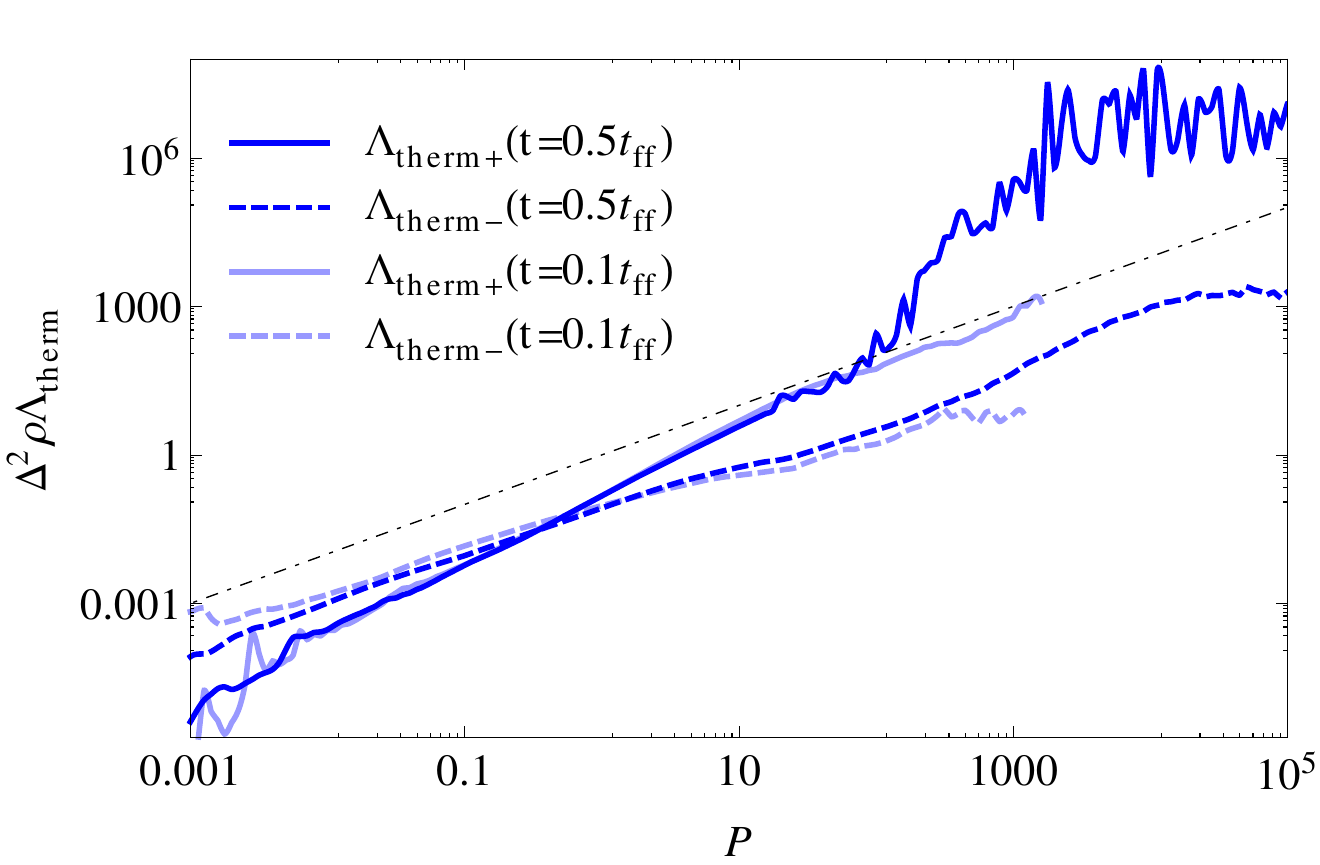}\\
  \includegraphics[width=\linewidth]{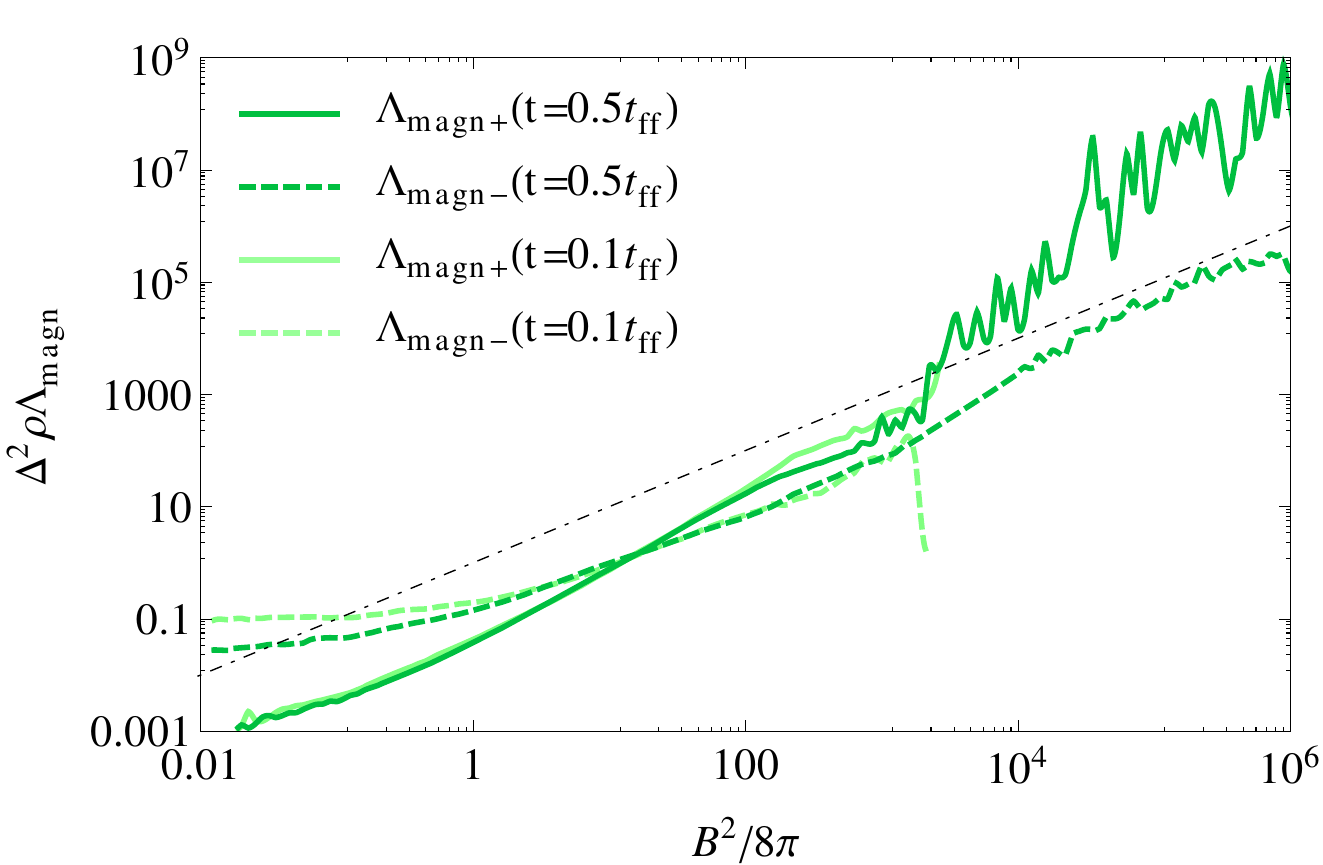}\\
  \includegraphics[width=\linewidth]{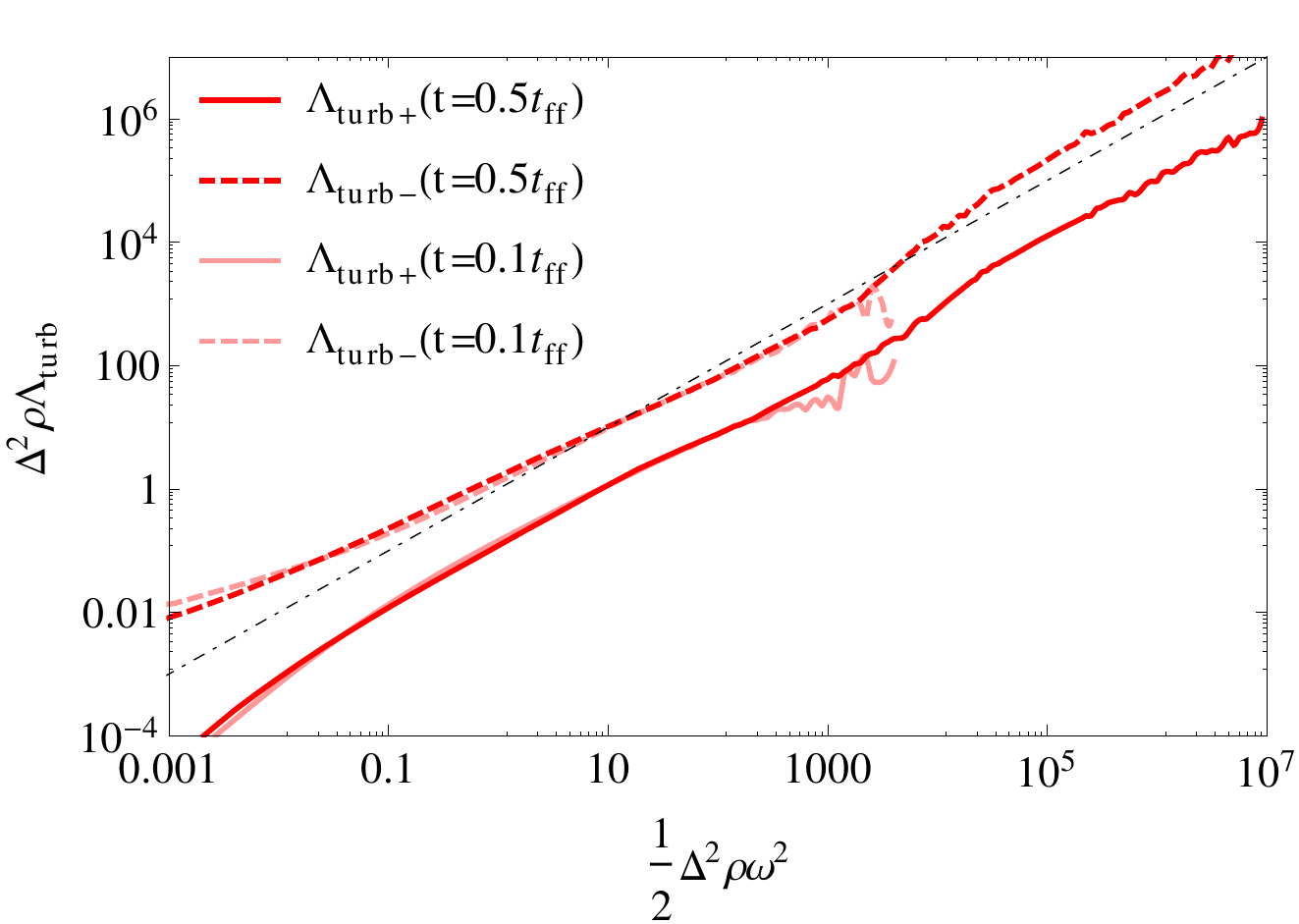}
\caption{Averages of $\Delta^2\rho\Lambda_{\rm therm\,\pm}$
	as functions of the thermal pressure (top), $\Delta^2\rho\Lambda_{\rm magn\,\pm}$ 
	as functions of the magnetic pressure (middle), and $\Delta^2\rho\Lambda_{\rm turb\,\pm}$ as functions of the
	enstrophy densities (bottom) for a weak magnetic field ($\beta_0=20$) at $t=0.5 t_{\rm ff}$.
	As in Fig.~\ref{fig:lambda_profiles}, positive and negative components are shown as solid and dashed lines, respectively,
	the light colored lines are obtained form $t=0.1 t_{\rm ff}$, and identity functions are indicated by dot-dashed lines.}
\label{fig:lambda_profiles_weak}
\end{figure}

\begin{figure}
\centering
  \includegraphics[width=\linewidth]{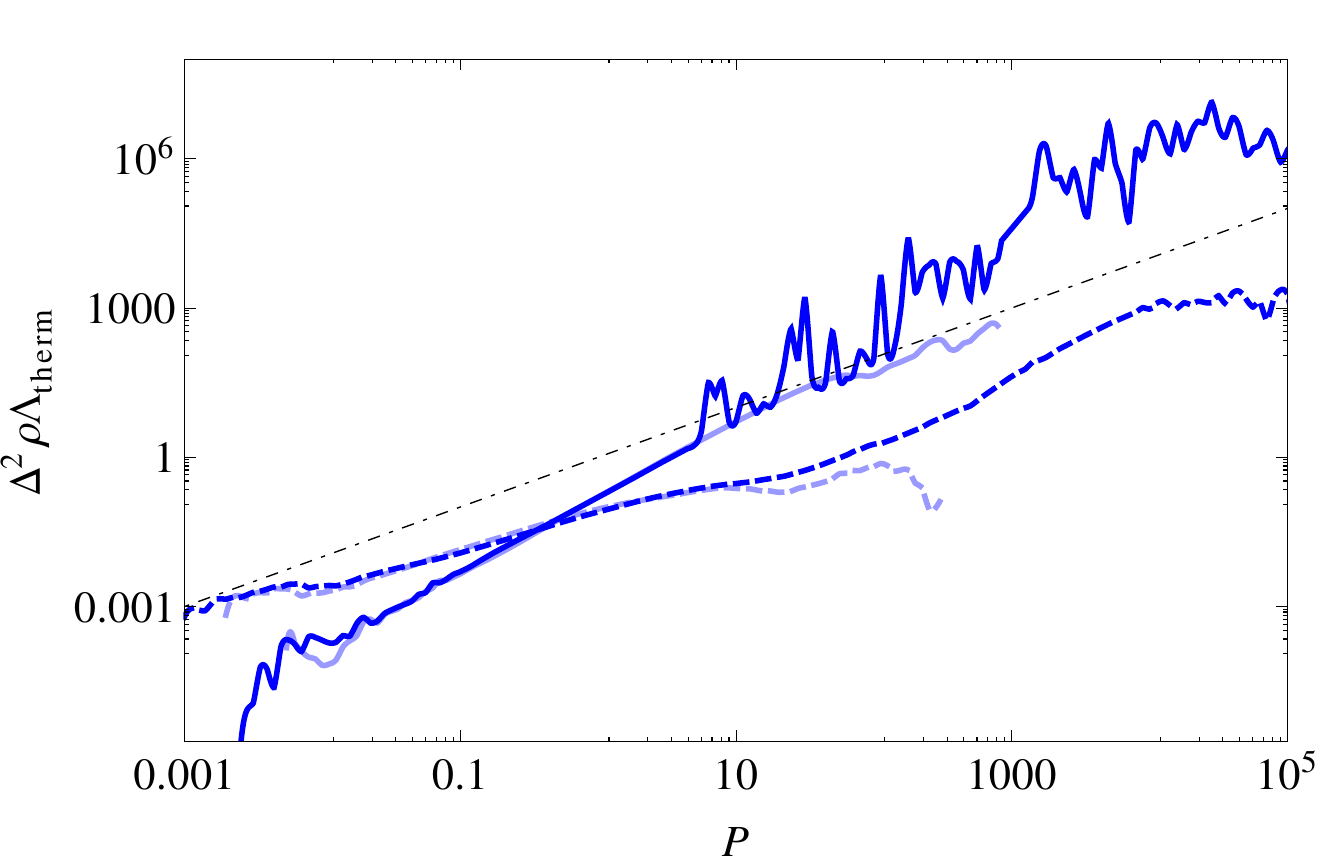}\\
  \includegraphics[width=\linewidth]{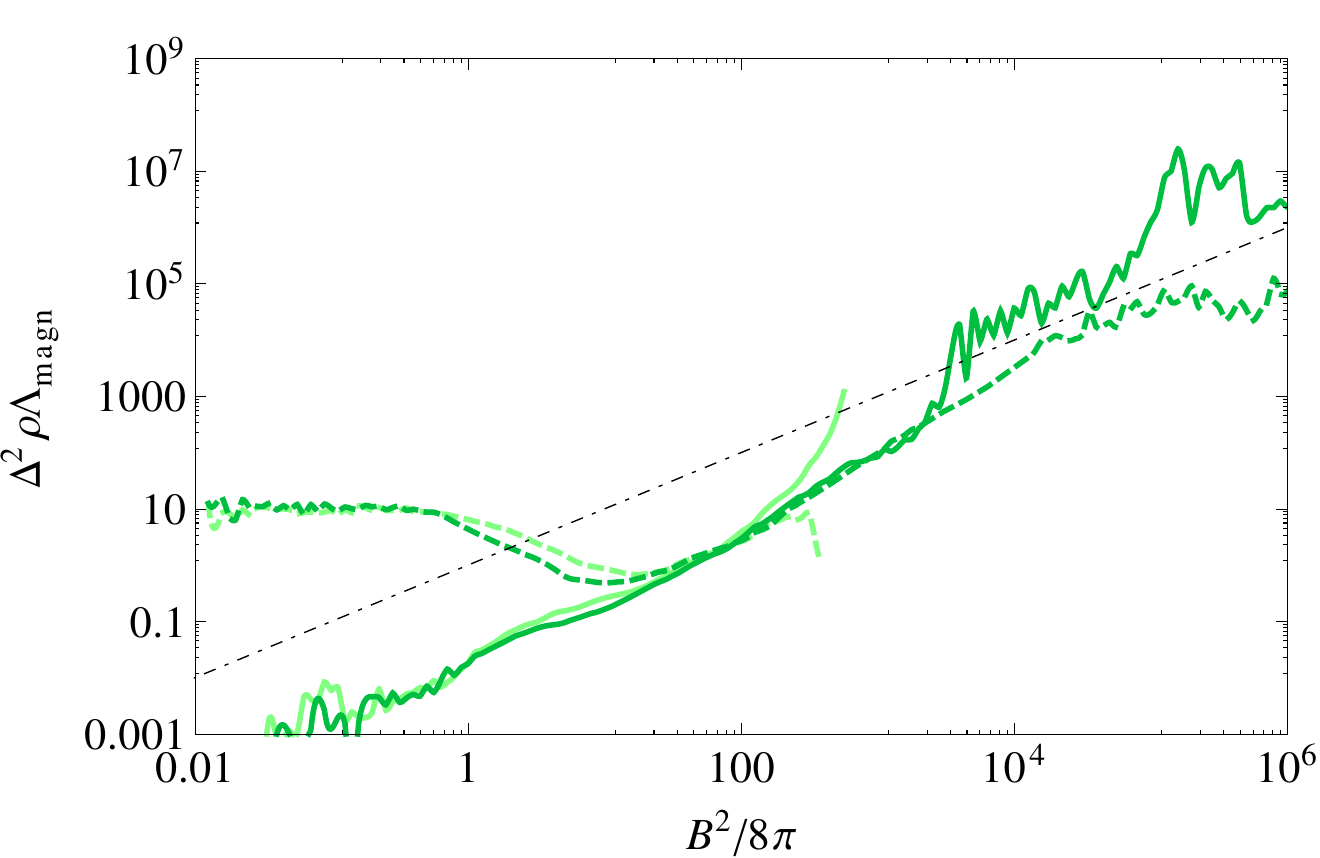}\\
  \includegraphics[width=\linewidth]{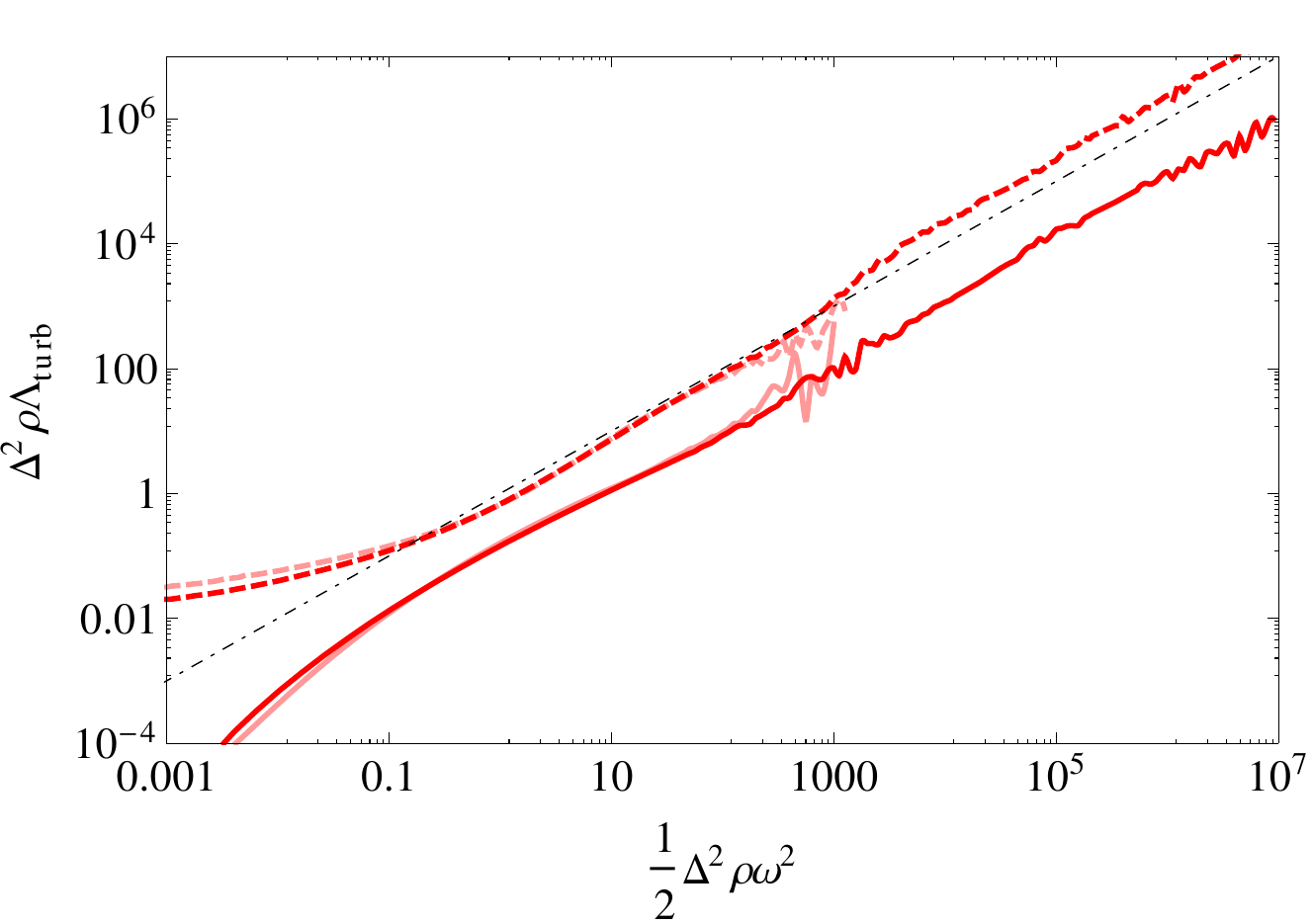}
\caption{Same functions as in Fig.~\ref{fig:lambda_profiles_weak} for a strong magnetic field ($\beta_0=0.2$).}
\label{fig:lambda_profiles_strong}
\end{figure}

Figs.~\ref{fig:ampl_div} and~\ref{fig:dynamo_strain} show the dependence of the magnetic field amplification on the 
divergence and the shear, where velocity derivatives are multiplied by the grid scale $\Delta$ to level out the different 
refinement levels. Let us first consider the weak-field case ($\beta_0=20$). In the case of compressive amplification, 
only the contribution from negative divergence is plotted. Except for strongly negative values of $d$ at time $t=0.5t_{\rm ff}$, 
where the effect of gravitational compression becomes manifest, $\langle-B^2 d\rangle_d$ is
nearly proportional to $d$. This implies that regions of higher divergence are not associated with systematically higher 
magnetic fields. In a similar way, $\langle-B_i B_j S_{ik}^{\ast}\rangle_{|S^{\ast}|}$ varies roughly linearly with the rate-
of-strain scalar $|S^\ast|$. Thus, the magnetic field cannot remain attached to turbulent structures that efficiently amplify 
the field. In other words, there is no strong backreaction of the field on these structures. This is not a odds to the 
frozen-in-motion in ideal MHD because neither shear nor divergence are properties carried by fluid elements. On the 
other hand, the field tends to be captured in the collapsing gas and this is why we see the super-linear increase for $-d\Delta
\gtrsim 1$ after 0.5 free-fall time scales. For the strong field ($\beta_0=0.2$), the behaviour of the compressive 
amplification is qualitatively similar, but the dynamo action is markedly different. The graph of $\langle-B_i B_j S_{ik}
^{\ast}\rangle_{|S^{\ast}|}$ is significantly flatter than for $\beta_0=20$, but the amplification rises steeply for strong 
shear. In contrast to $\beta_0=0.2$, this indicates that back-reaction effects cause a strongly non-linear interaction between the 
turbulent gas flow and the magnetic field. There is only little net amplification though, as the positive and negative components 
nearly balance each other.

As motivated in Sect.~\ref{sc:hd_stat}, the mean thermal, magnetic, and turbulent support functions are plotted as 
functions of $P$, $P_{\rm m}$ and $\frac{1}{2}\Delta^2\rho\omega^2$, respectively. The results for $\beta_0=20$ are
shown in Fig.~\ref{fig:lambda_profiles_weak}. At $t=0.5t_{\rm ff}$, the thermal support (top panel) is qualitatively similar 
to the case of hydrodynamic turbulence (see Fig.~\ref{fig:lambda_profiles}), but the range of pressures, 
for which the mean $\Delta^2\rho\Lambda_{\rm therm\,+}$ approximately matches $P$, is very narrow. 
This is probably due to the significantly lower resolution limit. 
The magnetic support function (middle panel) shows interesting properties. The overall behaviour is quite similar to the thermal support, 
which suggests that the magnetic support is basically pressure-like, althogh the non-diagonal Maxwell 
stresses are included in the support function defined by Eq.~(\ref{eq:support_magn}).
This becomes also apparent in the phase plots 
of the thermal and magnetic support functions (Figs.~\ref{fig:b20_lambda_therm_2d} and~\ref{fig:b20_lambda_magn_2d}).
Only the negative magnetic support (compression due Maxwell stresses) is relatively large at lower magnetic pressure.
For high magnetic pressures, $P_{\rm m}\gtrsim 1000$, the magnetic field produces strong support,
which becomes comparable to the thermal support.
Thus, magnetic fields have a significant influence on the
stabilization of the gas against gravity even if the field is relatively weak. Negative turbulent support (bottom panel)
again dominates, although the linear relation found for hydrodynamic turbulence is
not as closely matched in the magnetic case. For the strong field ($\beta_0=0.2$), Fig.~\ref{fig:lambda_profiles_strong} shows
qualitatively similar results as for the weak field, except for significantly less magnetic support. This result seems
at first counter-intuitive, but it can be understood by considering the probability density functions of the magnetic field (see Fig.~13 in 
\citealt{CollKrit12}).
Although the mean field is stronger for $\beta_0=0.2$, it turns out that the magnetic field fluctuations have a much wider tail 
toward high field intensities for $\beta_0=20$. This is why the local support function, which is mainly caused by strong
fluctuations, tends to be stronger in the latter case. 

\begin{figure*}
\centering
  \mbox{\subfigure[$\beta_0=20$, $t=0.1 t_{\rm ff}$]{\includegraphics[width=0.46\linewidth]{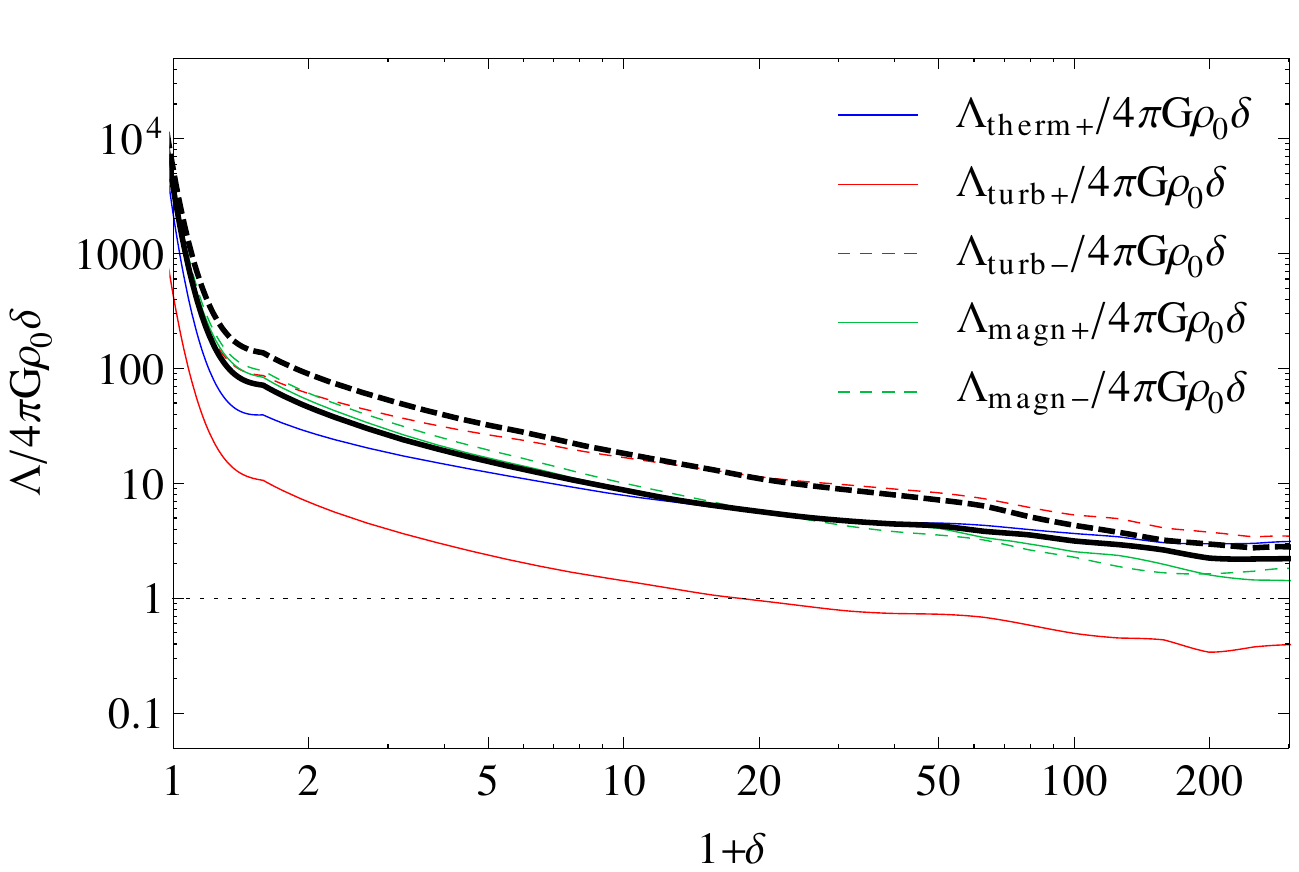}}\quad
  	\subfigure[$\beta_0=0.2$, $t=0.1 t_{\rm ff}$]{\includegraphics[width=0.46\linewidth]{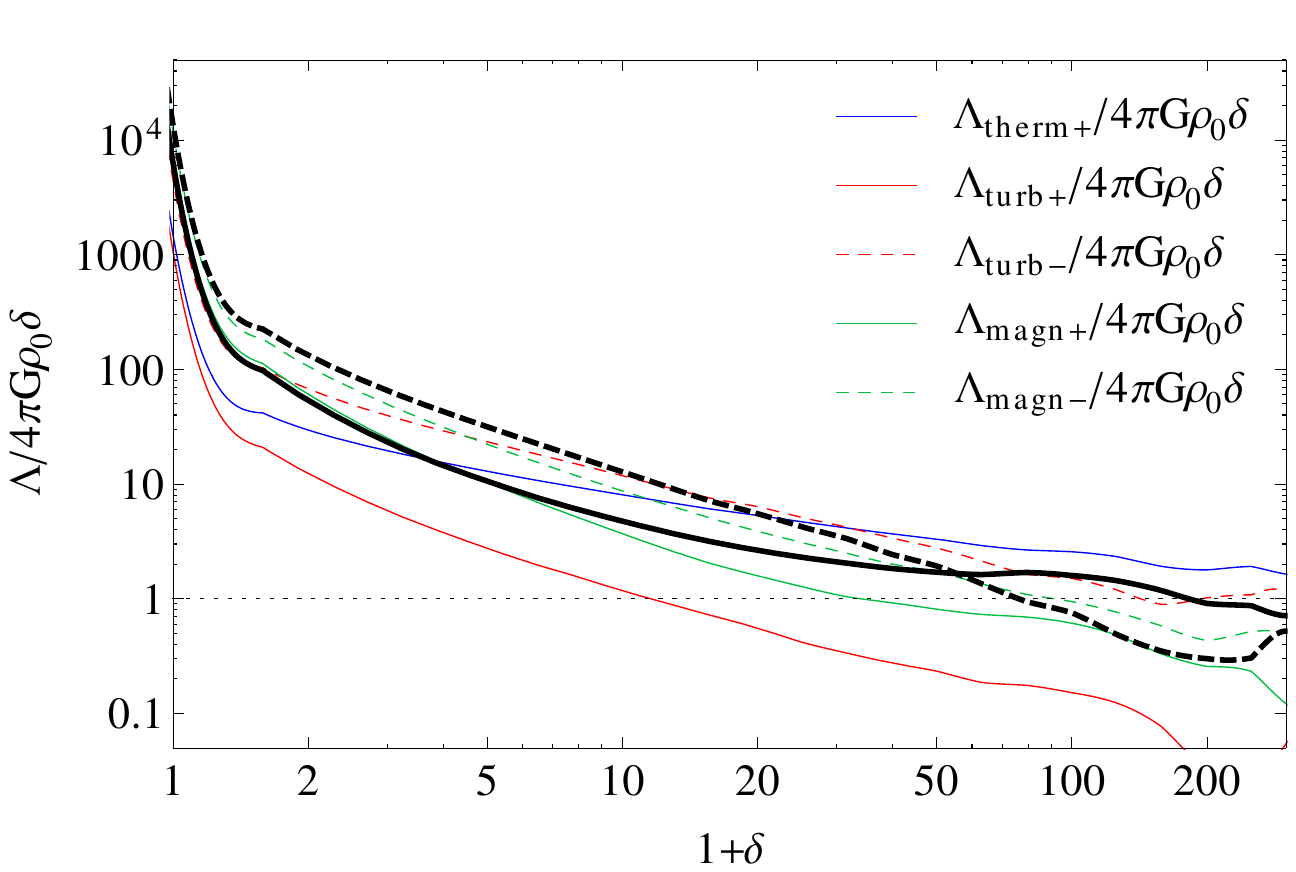}}}
  \mbox{\subfigure[$\beta_0=20$, $t=0.5 t_{\rm ff}$ (all levels)]{\includegraphics[width=0.46\linewidth]{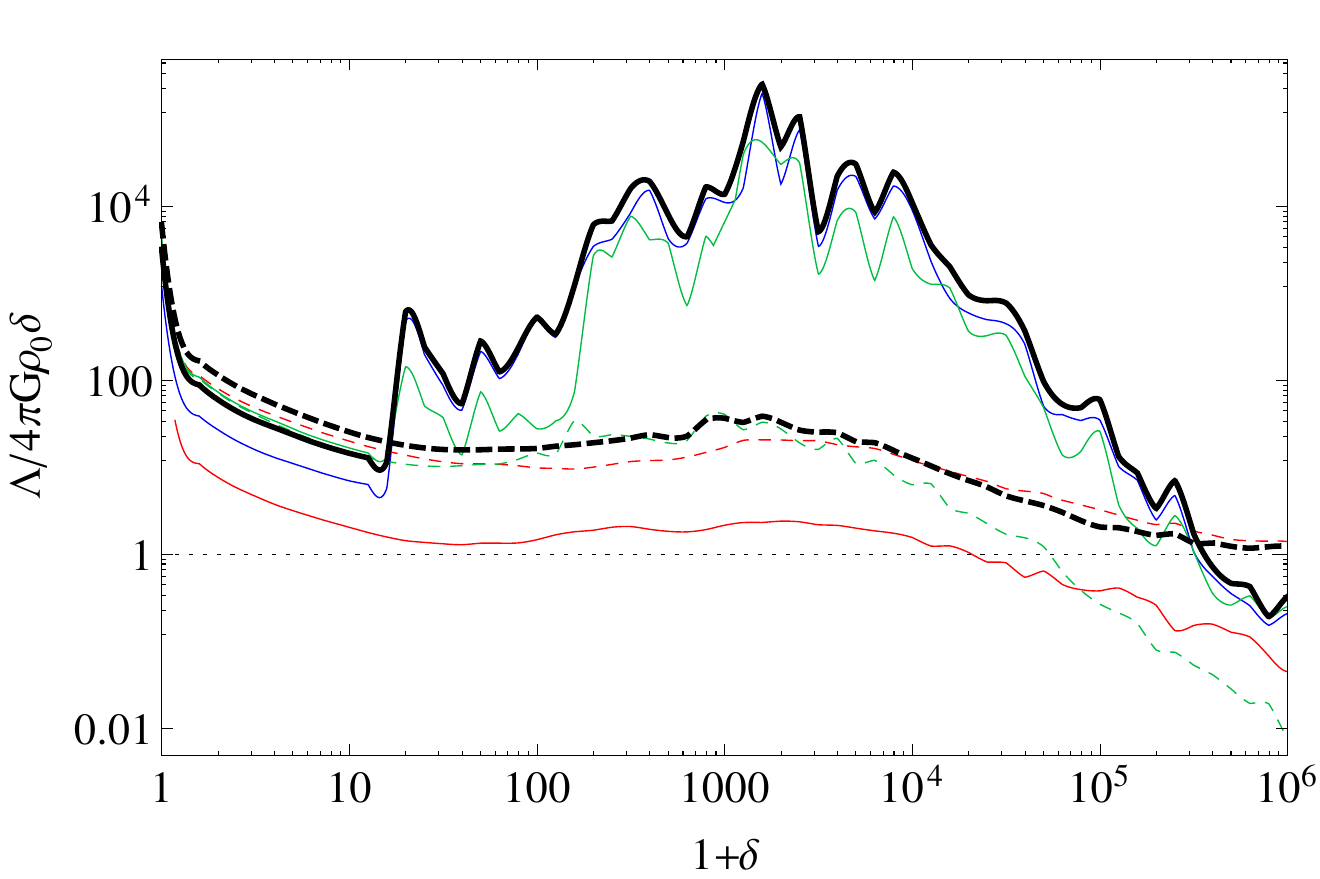}}\quad
	\subfigure[$\beta_0=0.2$, $t=0.5 t_{\rm ff}$ (all levels)]{\includegraphics[width=0.46\linewidth]{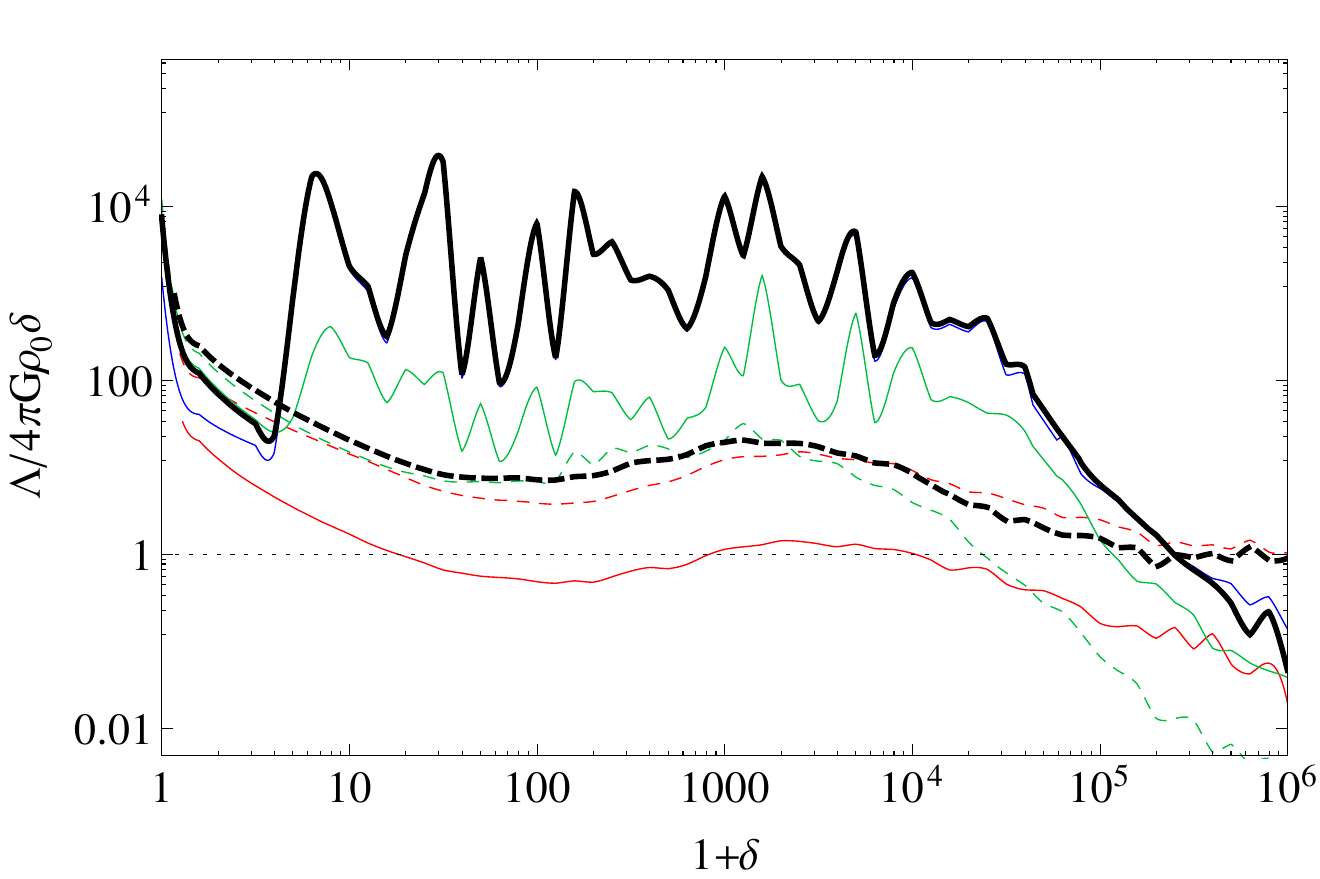}}}
  \mbox{\subfigure[$\beta_0=20$, $t=0.5 t_{\rm ff}$ (root level)]{\includegraphics[width=0.46\linewidth]{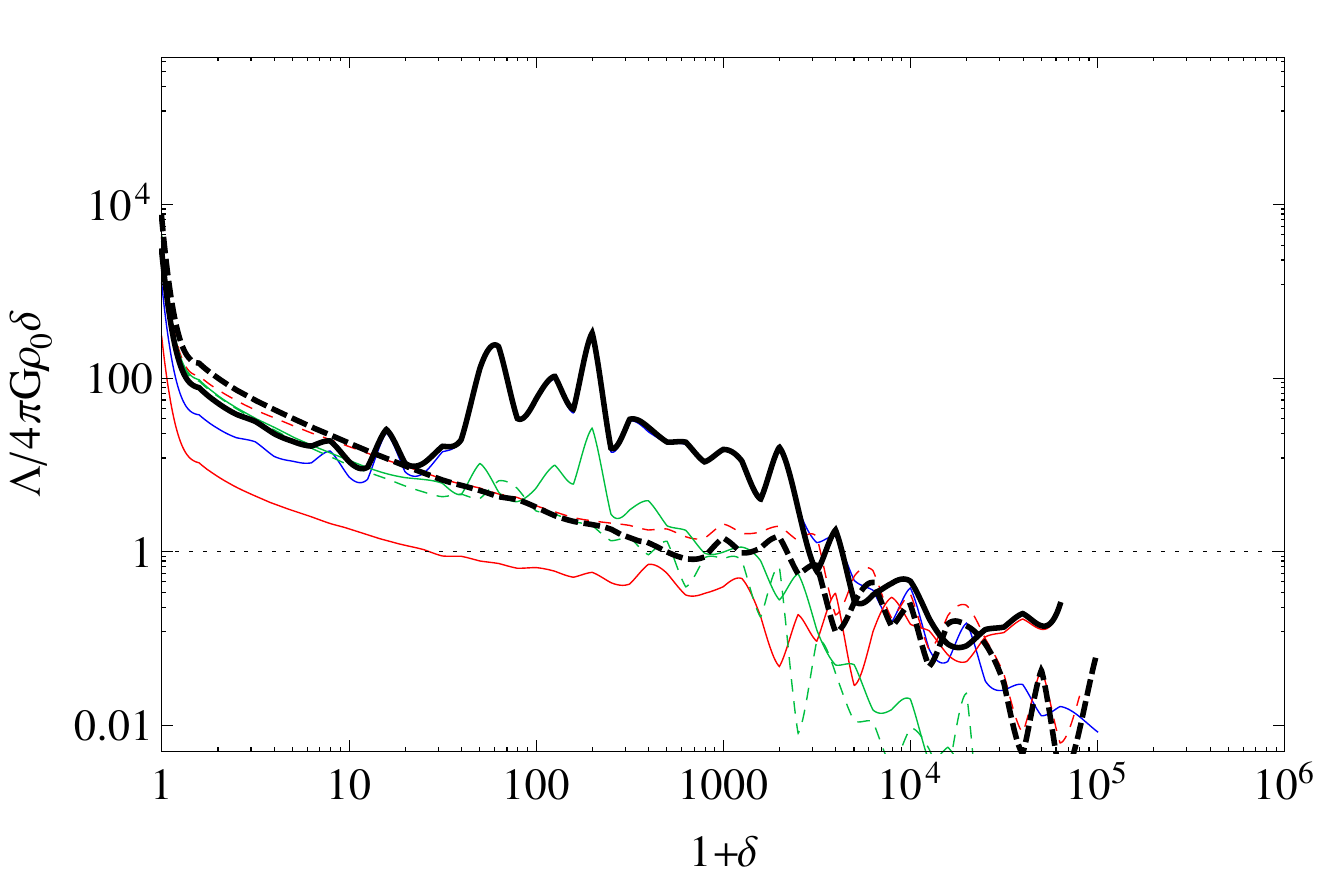}}\quad
	\subfigure[$\beta_0=0.2$, $t=0.5 t_{\rm ff}$ (root level)]{\includegraphics[width=0.46\linewidth]{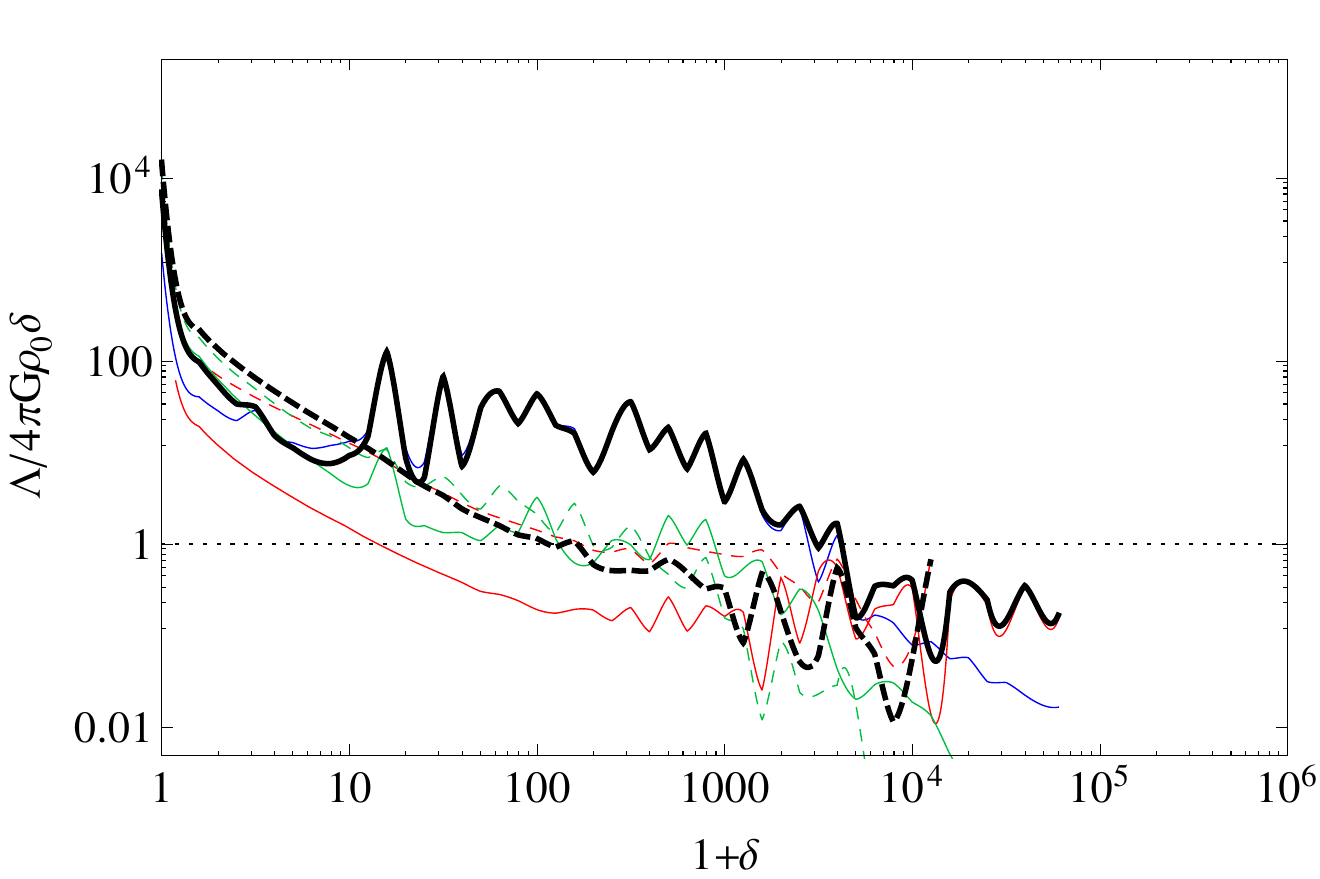}}}
\caption{Plots of the positive (thick solid) and negative (thick dashed) components of the total support
	relative to the gravitational compression rate in Eq.~(\ref{eq:compr}) as functions of overdensity. The left and right column of plots show the
	cases of weak and strong magnetic fields, respectively. The plots on the top are obtained for $t=0.1 t_{\rm ff}$.
	For $t=0.5 t_{\rm ff}$, the results from both the full AMR data (middle) and the root-grid data (bottom) are plotted.
	In each plot, the dominant components of the thermal and turbulent support functions and both components of the magnetic support are indicated by
	thin lines.}
\label{fig:stability_profiles_tot_mhd}
\end{figure*}

Profiles of $\Lambda/4\pi G\rho_0\delta$ against the overdensity $1+\delta$ are plotted in Fig.~\ref{fig:stability_profiles_tot_mhd}.
At the early instant $t=0.1 t_{\rm ff}$, the support gradually decreases with the
density and approaches unity for $\delta$ above 200. Clearly, supersonic compression is stronger than the magnetic support of the gas, 
although the magnetic compression rate $\Lambda_{\rm magn\,-}$ exceeds $\Lambda_{\rm turb\,-}$ at moderate overdensities in the
case $\beta_0=0.2$. 
In the advanced stage of gravitational collapse ($t=0.5 t_{\rm ff}$), there is
large positive support for overdensities below $10^5$. $\Lambda/4\pi G\rho_0\delta$ becomes smaller than unity only for $\delta\sim 10^6$.
The main support is thermal, particularly in the strong-field case ($\beta_0=0.2$). This corresponds well to the
trends for the pressure-like quantities in Figs.~\ref{fig:lambda_profiles_weak} and~\ref{fig:lambda_profiles_strong}. 
For $\beta_0=20$, however, the magnetic support is almost as strong as the support by thermal pressure
In both cases, the positive component of the turbulent support is much smaller. 
The negative turbulent component tends to be comparable to the negative magnetic
component, although the former dominates for high densities ($\delta\gtrsim 10^5$) at
time $t=0.5 t_{\rm ff}$. The large impact of magnetic fields on the support of the gas for $\beta_0=20$
is further demonstrated by the phase plots shown in Figs.~\ref{fig:b20_stab_therm_2d} to~\ref{fig:b20_stab_magn_2d}. The
trend of $\Lambda_{\rm therm\,+}/4\pi G\rho_0\delta$ to become smaller than unity is particularly pronounced in this case. 
Apart from that, the distributions of the  thermal and turbulent support functions vs.\ density and the turbulent support spectra 
(not included in this paper) are qualitatively similar to the hydrodynamical case. 

The profiles computed from the root-grid representation of the data are plotted in the bottom panels of Fig.~\ref{fig:stability_profiles_tot_mhd}. 
In this case, the support drops below unity for 
densities above the critical density, which is set by the resolution criterion based on the local Jeans length \citep[see][]{CollKrit12}. 
As discussed in Sect.~\ref{sc:hd_stat}, this indicates that the high-density gas is dominated by gravity when smoothed out over
the length scale of the root grid. In contrast, the pressure of smaller structures at higher refinement levels can locally 
oppose gravity up to very high densities, but these structures are not sufficiently resolved relative to the Jeans length in these
MHD models. A further important result is that magnetic support is relatively weak in the range of densities, 
where $\Lambda/4\pi G\rho_0\delta \lesssim 1$.
Consequently, the relatively strong magnetic support that can be seen for the full AMR data in the case $\beta_0=20$ stems from local
field compression and folding, which largely average out over the root grid cells. On the other hand, the positive contribution from 
turbulence becomes comparable to the thermal support at the highest densities and exceeds the support by magnetic fields. Thus, 
the statistics of $\Lambda_{\rm turb}/4\pi G\rho_0$ plotted in Fig.~\ref{fig:stability_profiles_tot_mhd} show that vortices on the root-grid can play an important role in providing support against gravity in the most dense gas, while shock compression generally dominates on the smaller scales of the higher refinement levels.

\begin{figure}
\centering
  \includegraphics[width=\linewidth]{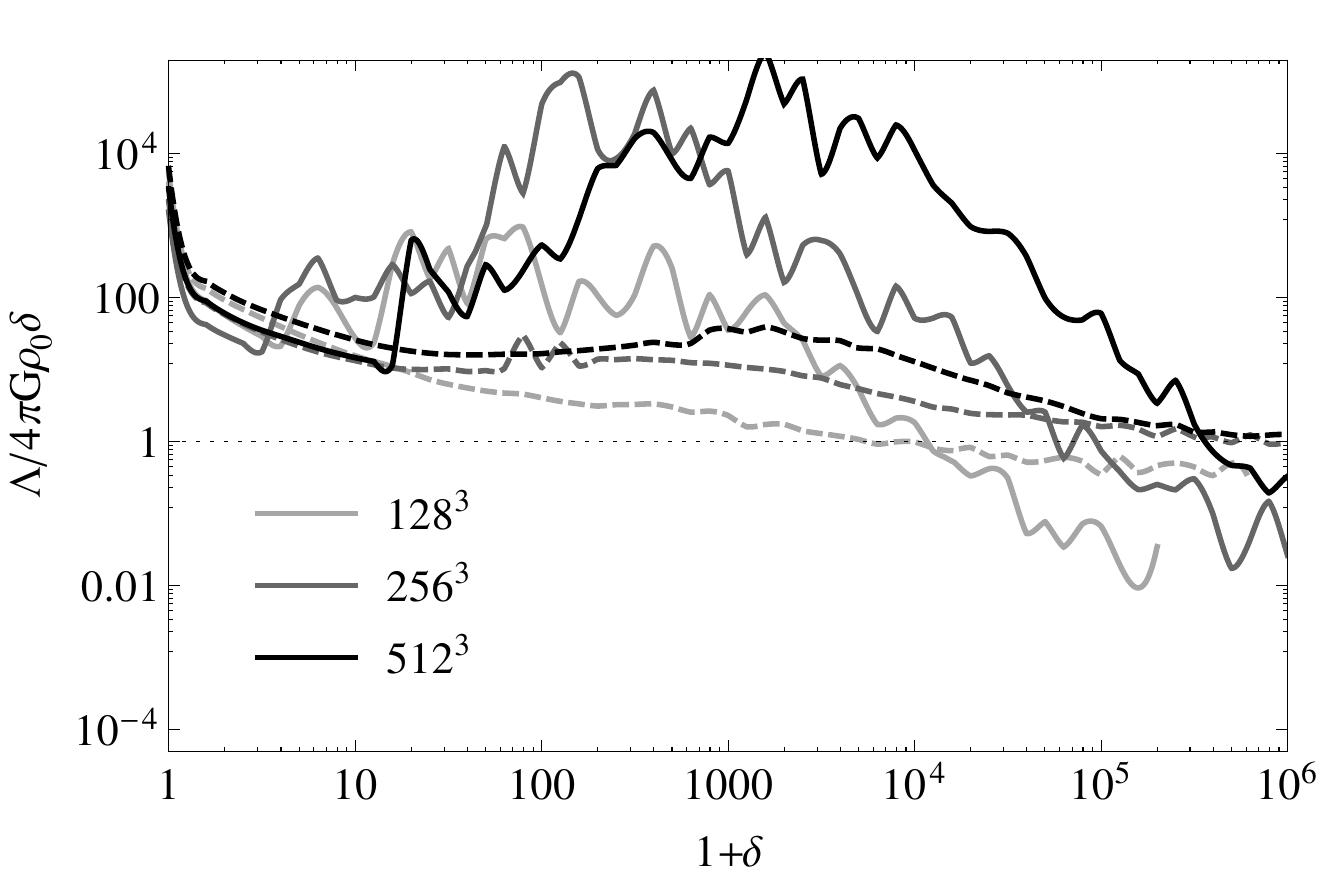}
\ \includegraphics[width=\linewidth]{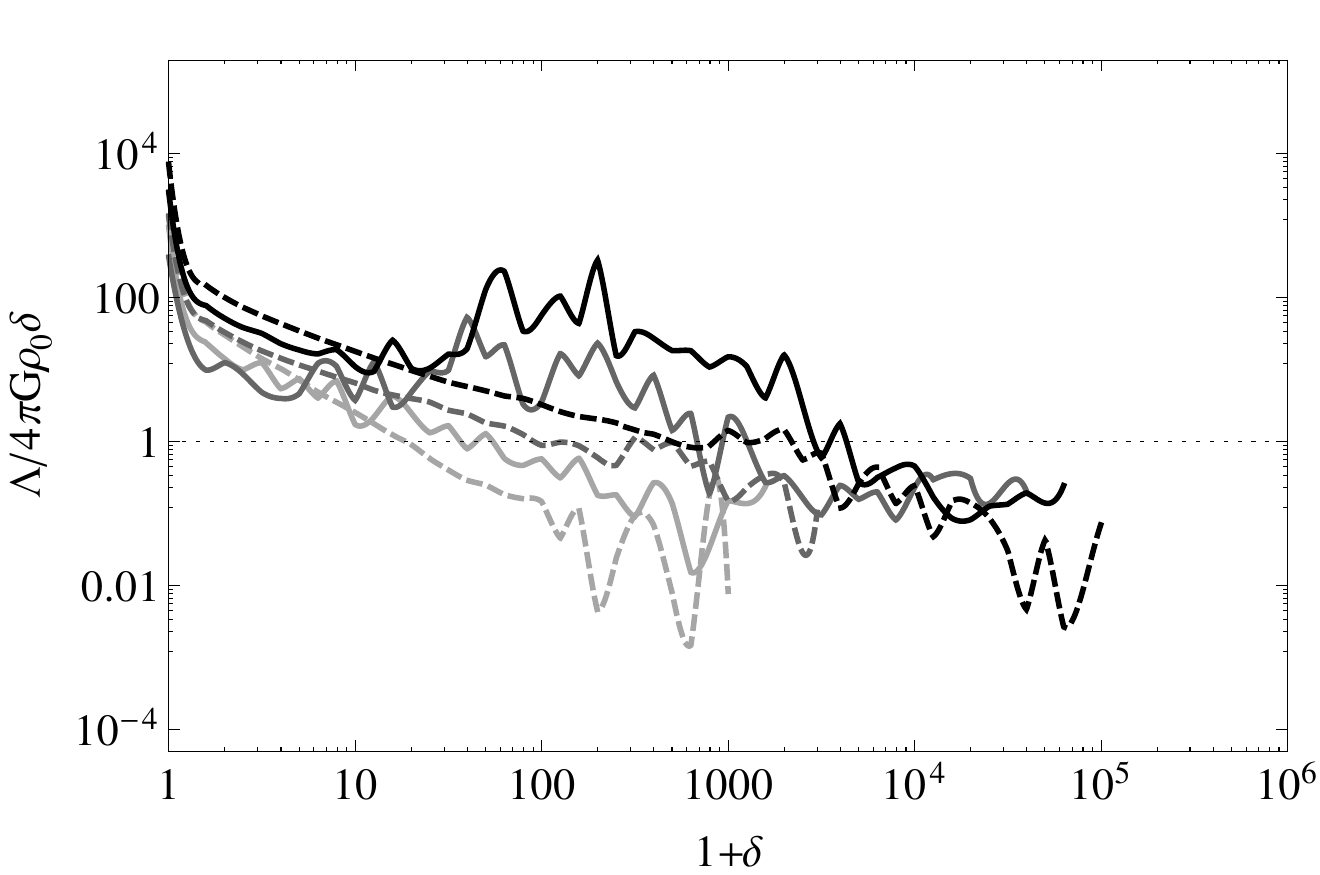}
\caption{Plots of the total support as in Fig.~\ref{fig:stability_profiles_tot_mhd} for different resolutions of the root-grid
	in the case of weak field ($\beta_0=20$) and $t=0.5 t_{\rm ff}$. The statistics is plotted both for the full AMR data (top) 
	and the root-grid representation of the data (bottom).}
\label{fig:stability_profiles_res_weak}
\end{figure}

\begin{figure}
\centering
  \includegraphics[width=\linewidth]{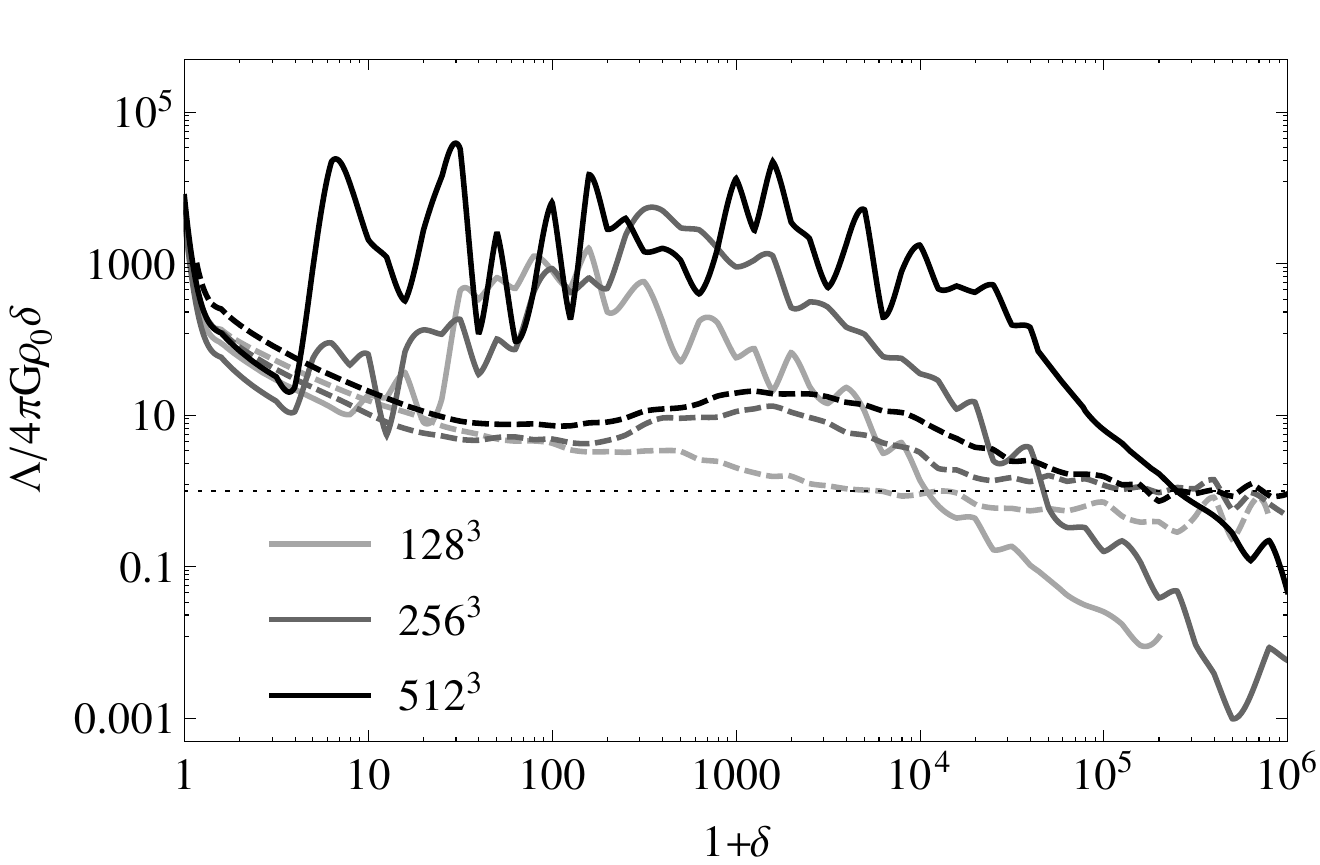}
  \includegraphics[width=\linewidth]{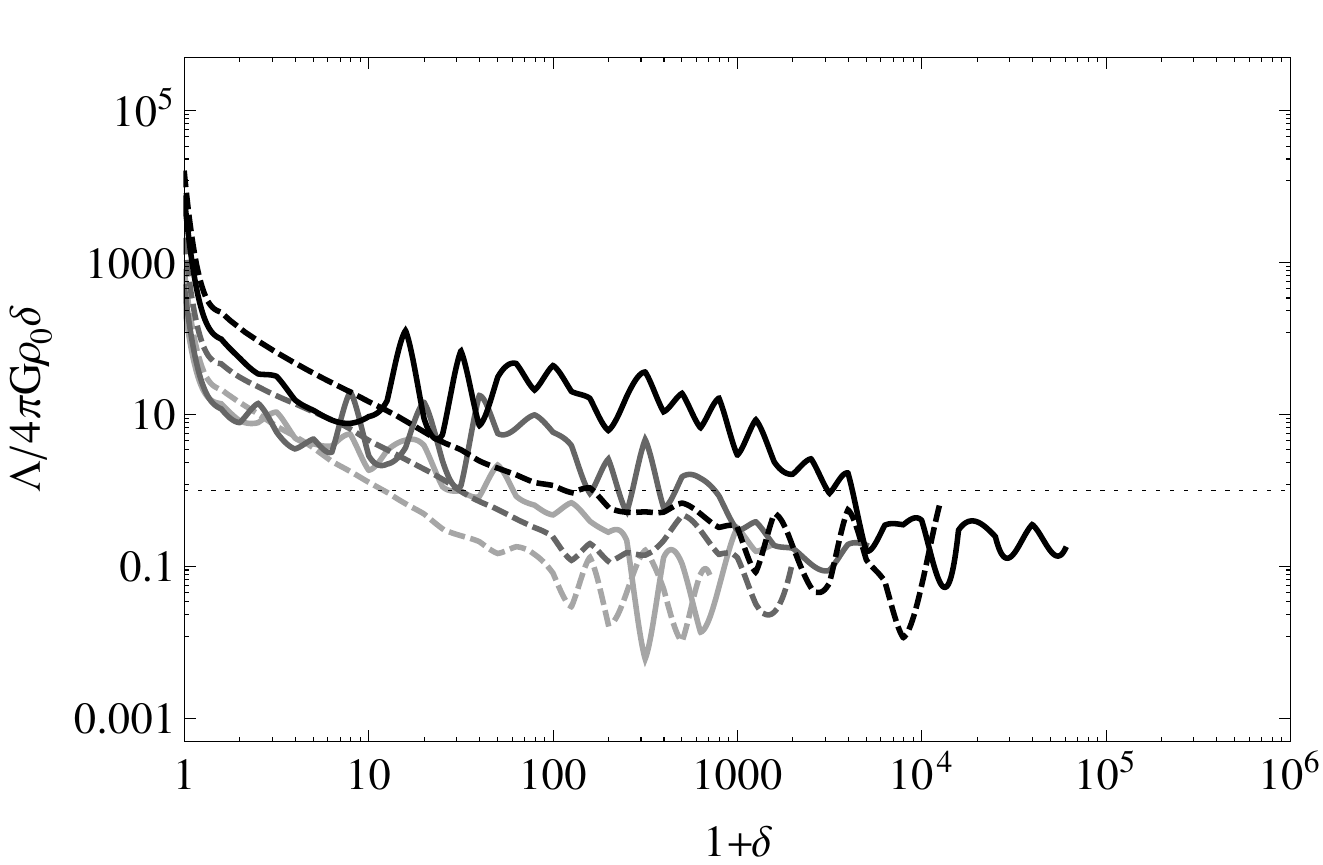}
\caption{The same plots as in Fig.~\ref{fig:stability_profiles_res_weak} for $\beta_0=0.2$.}
\label{fig:stability_profiles_res_strong}
\end{figure}

\begin{figure}
\centering
  \includegraphics[width=\linewidth]{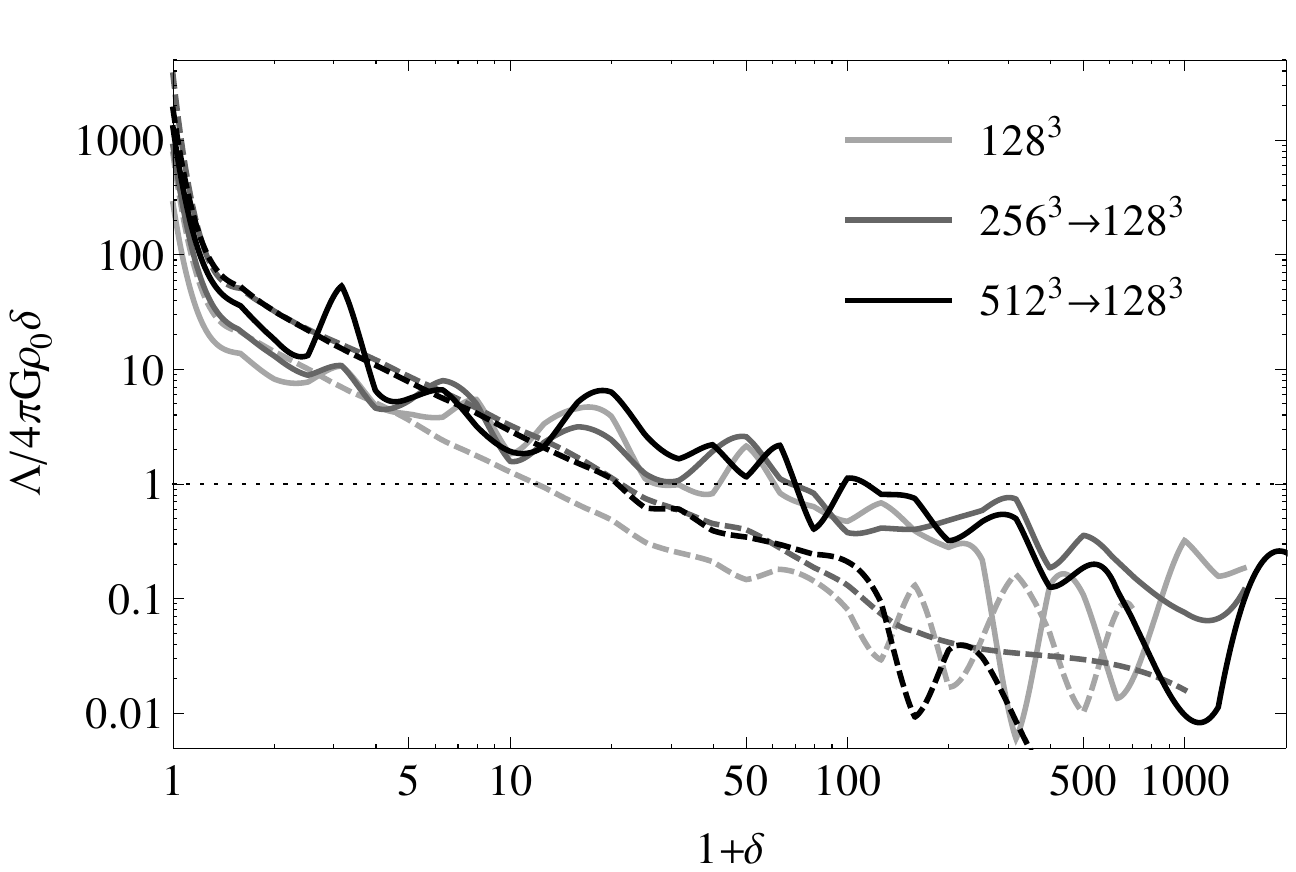}
\caption{Plots of the root-grid representation of the total support down-sampled to $128^3$ for $\beta_0=0.2$ and $t=0.5 t_{\rm ff}$.}
\label{fig:stability_profiles_downsample}
\end{figure}

Since gravity breaks the scale invariance of turbulence, the properties of collapsing turbulent structures in a numerical simulation
are inevitably resolution-dependent. If the grid scale decreases, self-gravitating overdense gas will collapse to higher densities, unless the physical length scale on which the collapse ends by whatever mechanism is resolved. In the context of star formation in the interstellar
medium, the end of the collapse would be reached when the gas becomes optically thick. Since this is beyond the achievable maximum resolution for a box size of the order of parsecs, the collapse could be stopped numerically by dumping mass into sink particles beyond a critical density that is set by the resolution limit. For the reasons outlined in Sect.~\ref{sc:intro}, however, no sink particles were applied in our simulations. 
Having said that, gradients of the pressure, the velocity, and the magnetic field become steeper as smaller length scales are resolved, which give rise to stronger fluctuations and thus enhance the magnitude of the local support. It is difficult to see \emph{a priori} how this affects the ratio $\Lambda/4\pi G\rho_0\delta$. So the question arises how robust the trends we observe are if the resolution is varied. We investigate this question by changing the resolution of the root grid, while keeping the number of refinement levels constant. Then
the size of the densest structures that can be resolved at the highest refinement level decrease proportional to the root-grid resolution.
Apart from that, the overall turbulent velocity field becomes coarser, corresponding to a higher numerical viscosity. 
We plot the results for $\Lambda_{\pm}/4\pi G\rho_0\delta$ for the three different root-grid resolutions $128^3$, $256^3$, and $512^3$ in 
Figs.~\ref{fig:stability_profiles_res_weak} and~\ref{fig:stability_profiles_res_strong}. The Truelove number is $16$ for all simulations.
Clearly, there are systematic differences in the positive support at high densities. Structures in refined regions resist gravity up to higher densities for increasing resolution. 
This is expected, according to the discussion above. Owing to the randomization and the strong intermittency of local turbulent structures, the spikes in the support functions differ substantially. Nevertheless, the trend with density is qualitatively similar for the three simulations. In this regard, time-averaging would be advantageous, but this is not feasible because self-gravitating turbulence is not statistically stationary. If we consider the support computed from the smoothed fields on the root grid, we also see a systematic drift of the density range with $\Lambda/4\pi G\rho_0\delta\lesssim 1$. This implies that the structures captured by the refined grids at higher
resolution indeed have a noticeable influence even when they are smoothed out over the root-grid scale. The increase of the support with resolution at a given overdensity in the range $10\lesssim\delta\lesssim 10^3$ can be attributed to the steepening of the gradients in Eq.~(\ref{eq:support}), as the effective numerical viscosity becomes less. Naturally, for the reasons outlined above, convergence cannot be achieved as long as the gas is not prevented from collapse to ever higher densities. However, by conservative down-sampling of the support functions for the higher resolutions to $128^3$ (see Fig.~\ref{fig:stability_profiles_downsample}), we also see that the main difference can be attributed to the averaging from larger to smaller scales. Thus, the general trends of the support we outlined above are not substantially altered by the numerical resolution.

\section{Conclusions}

In the highly non-linear, turbulent regime, we propose to infer the local contributions of turbulence, thermal pressure, and 
magnetic fields to the support of the gas against gravity from the dynamical equation for the rate of compression,
$-\DD d/\DD t$ \citep{Schmidt09}. We performed a statistical analysis of the source terms, i.~e., the local support $\Lambda$ 
and the gravitational compression rate $4\pi G\rho_0\delta$, where $\rho_0$ is the mean density and $\delta$ the relative 
density fluctuation, for three AMR simulations of self-gravitating turbulence. One simulation is purely hydrodynamical 
\citep{KritNor11}, 
the other two are MHD simulations with different mean field strengths \citep{CollKrit12}. In the following, we summarize the 
main results of our study:

\begin{enumerate}
\item The statistics of the local support function~(\ref{eq:support_turb}) does not provide evidence
	of positive turbulent support. On the contrary, the net turbulent support is negative at all densities and 
	for arbitrarily high vorticity. This result is obtained both for hydrodynamic turbulence with mixed
	forcing and MHD turbulence with purely solenoidal forcing at high Mach numbers. 
	The spectra of this function also show a dominant negative component for all wave 
	numbers. The negative component can be interpreted as negative pressure produced by the strain of the turbulent velocity
	field. For hydrodynamical turbulence, we find a simple linear relation (Eq.~\ref{eq:turb_asymptot}) between the vorticity 
	and the negative pressure. Magnetic fields cause deviations from this linear relation.
\item The largest contribution to the support of the gas comes from fluctuations of the thermal pressure, also in the case of MHD. 
	For isothermal gas, this means that compression rate is mainly given by the density structure of the gas because both the
	thermal support function~(\ref{eq:support_therm}) and the gravitational compression rate $4\pi G(\rho-\rho_0)$ are solely determined by 
	$\rho$. 
\item Magnetic fields contribute significantly to the support. For an initial field with 
	$\beta_0=20$ (corresponding to $\beta\approx 0.2$ for statistically stationary turbulence), 
	the positive component of the local magnetic support function~(\ref{eq:support_magn})
	is comparable to the thermal support. However, the support does not further increase for a stronger field with $\beta_0=0.2$
	($\beta\approx 0.03$).
	The reason for this is the wider tail of the PDF of $B$ for larger $\beta_0$. While turbulence causes the initial
	compression of the gas, the turbulent support function in overdense gas is small relative to the magnetic support function after 
	$0.5$ free-fall time scales (except for very high, underresolved densities). 	
\item Gravitational collapse locally enhances the magnetic field strength. In addition to the compression by gravity and shocks,
	a roughly equal amplification is caused by shear (small-scale turbulent dynamo). 
	Consistent with our findings for the magnetic support, the amplification
	in high-density gas tends to be stronger for a relatively weak mean field ($\beta_0=20$).
\item The local support function, as defined by Eq.~(\ref{eq:support}), varies between values greater or
	smaller than the gravitational compression rate. For moderate overdensities, we find a negative
	net support, i.~e., gravity and supersonic turbulence cause the gas to contract. At higher densities,
	however, positive pressure support becomes strong and the average ratio of the support function to the gravitational 
	compression rate is much greater than unity, with strong fluctuations due to intermittent structures.
	Gravitational compression is dominant only for very high densities at the resolution limit.
\item We compared our results to smoothed data, which can be obtained from the root-grid representation. The main
	effects of the smoothing are that the total support becomes small compared to gravitational compression for
	much lower densities, marginal turbulent support is found at the highest densities, 
	and the support by magnetic fields is reduced. Thus, it appears that the gas is self-gravitating
	when averaged over large volume elements, while smaller structures that are only resolved at higher
	refinement levels are supported by strong fluctuations of the pressure and the magnetic field. 
\end{enumerate}

The behaviour that emerges from our statistical analysis is unexpected in several aspects, most noticeably, the lack of a positive
turbulent support. Global support by turbulent pressure is clearly an implication of the generalized virial theorem for an isolated 
object. For a smaller subregion inside a turbulent cloud, however, the boundary terms will significantly modulate the volume 
integrals of the thermal, turbulent, and magnetic energy densities \citep[see, for instance,][]{Lequeux}, 
depending on conditions that vary in space and time. 
The local support functions we use in this article are source terms of the partial differential equation~(\ref{eq:div}) for the divergence
of the velocity, which can be directly related to the mass density source of the gravitational potential in the Poisson equation. 
The positive and negative variations of the local support function imply that a particular fluid element frequently experiences
contraction phases with $\Lambda/4\pi G\rho_0\delta\lesssim 1$ and then crosses regions with
strong positive support, where $\Lambda/4\pi G\rho_0\delta\gtrsim 1$. Only if it accumulates enough density so that 
gravitational compression is dominant over a sufficiently long period of time, it will be torn into irreversible collapse. 
To improve our understanding of the mechanism of core formation, Lagrangian statistics might turn out to be valuable. 
The role of supersonic turbulence is ambiguous in this regard. Shocks can compress the gas to sufficiently high density, but also 
disrupt dense cores. On the average, the compression effect is dominant. The analysis of simulations of self-gravitating
turbulence by \citet{FederKless12} also suggest that the main role
of turbulence is to enhance gravitational collapse. It is possible though that non-turbulent, 
rotation-free flow produced by the pull of gravity biases the turbulent support toward compression 
because the negative divergence of such flows contributes to the rate-of-strain scalar. 
Currently, it is unclear how to separate this from the genuinely turbulent velocity fluctuations.
But the contribution of gravity to the turbulent support function appears to be small compared to the impact of shocks
for several reasons. Firstly, we see the dominance of negative turbulent support 
already at early stages, where the gas only begins to contract under its self-gravity. Secondly, if the negative component of the
turbulent support were strongly influenced or even dominated by gravity, it should decrease relative to the positive
component towards low densities, as gravity becomes increasingly weak. This is also not the case. Moreover, one can clearly
see in Fig.~\ref{fg:slices} that strongly negative turbulent support is associated with shocks. Thirdly,
the velocity power spectra are not significantly affected by gravity \citep{CollKrit12,FederKless13}.

In the case of hydrodynamical self-gravitating turbulence, the extremely high dynamical range of the
simulation allowed us to identify a nearly self-similar regime with a \emph{negative net} support of magnitude
$|\Lambda|\sim 4\pi G\rho_0\delta$. 
This regime could be roughly described by the free-fall collapse solution of \citet{Penst69}. However, we find a subdominant, yet
non-negligible positive component of the thermal support. The idea of a hierarchy of quasi-virialized structures proposed by
\citet{BigDia88}, on the other hand, does not appear to be consistent with our simulations, because this would imply a
positive net support of the order of the gravitational compression rate. 
 
We attempted to bridge the gap between local and large-scale support by computing the support functions from the smoothed 
data on the root grid of our AMR simulations. Since the refinement follows the gas density, this is roughly equivalent 
to integrations over larger volumes of overdense gas and, indeed, the positive component of the turbulent support 
tends to become larger. On the other hand, the statistics computed from the fine grid data reflects the dynamics of the collapsing
cores, at least up to densities, for which the physics of self-gravitating turbulence is sufficiently resolved. 
It is an interesting question whether the different behavior of the support we see on the coarse root grid and on the refined
grids could be carried over to observations of molecular clouds. Indeed, the root-grid resolution, which is about $0.01$ pc roughly corresponds the pixel size of current CO maps \citep[e.~g.,][]{GoldNara08}. Some observed prestellar molecular cores have 
Bonnor-Ebert-like profiles with central densities up to several $10^5\,\mathrm{cm^{-3}}$ \cite[for a recent review, see][]{BergTaf07}.
For the corresponding overdensities $\delta\sim 10^3$ at the root grid of our hydrodynamical simulation, we find a positive net support
$\Lambda$ that is comparable to the gravitational compression rate $4\pi G\rho_0\delta$. This could mimic Bonnor-Ebert-like cores in quasi-hydrostatic
equilibrium. In stark contrast, however, the small-scale structure that is resolved by AMR reveals collapsing cores resulting from
supersonic compression, while the thermal pressure support is subdominant. These considerations are supported by calculations of the instantaneous virial balance of clumps and cores, including the surface terms, by \citet{DibKim07}. They show that clumps and cores 
are non-equilibrium structures  because surface terms are generally of the same order as the volume terms. 
In particular, they find structures that are undergoing supersonic compression, but not all of them are gravitationally bound. 
As a future extension of our study, clump finders or, alternatively, the dendrogram analysis technique developed by 
\citet{RosPin08} could be applied to constrain the statistics of the support functions to clump-like objects 
on different length scales. This might also help to interpret observational data in the light of high-resolution numerical data 
and to clarify which biases are introduced by relatively coarse observational resolutions. The dendrogram analysis of
numerical and observational data presented in \citet{RosPin08}, for example, reveals scale-dependent discrepancies 
in the distribution of self-gravitating objects that are characterized by the virial parameter.

Magnetic fields locally have a significant stabilizing effect \citep[see also][]{FederKless12}, 
but we find that magnetic support does not dominate
for large mean magnetic pressure ($\beta\ll 1$). The stronger support in the case of a weak field ($\beta_0=20$) 
can be explained by the turbulent amplification in the initial turbulence simulation and the resulting
wider tails of the magnetic field strength in comparison to $\beta_0=0.2$ \citep{CollKrit12}. In conclusion, 
the mean magnetic pressure is not a suitable quantity to infer the importance of magnetic fields in stabilizing self-gravitating gas.
By computing the magnetic support function for the smoothed root-grid data, we find a weaker effect in comparison to both the thermal 
and turbulent support functions. This suggests that the magnetic field is significantly amplified in local, collapsing structures.
Indeed, we find a large amplification effect by gravitational and supersonic compression at high densities, which is roughly
balanced by shear-induced amplification.
This is different from the results of \citet{SurSchl10} and \citet{TurkOishi12}, who run simulations of 
collapsing halos, where subsonic turbulence is solely produced by the gravitational collapse. 
Since turbulence is vortex-dominated on length scales below the Jeans length, the amplification by shear becomes strong
compared to compressive amplification in this case if the numerical resolution is high enough. 

The importance of magnetic fields in molecular clouds has been notoriously
difficult to pin down \citep[see][]{Crutch12}.  There are observations that show fields are possibly
weak, such as the column density power spectra study of \citet{PadJim04}, as
well as observations that they are dynamically important \citep{LiDow09}. 
\citet{KritUsty11} argue that the super-Alfv\'{e}nic nature of turbulence in molecular clouds 
is a natural outcome of the cloud formation process. Our simulations show that the relative importance 
of magnetic fields is dependent on density scale as well as averaging scale.  
Thus it is important to carefully
consider both effects when interpreting observations using numerical data.

In our resolution study, we demonstrate that the effect of numerical viscosity has a potential impact on the statistics
of the local support function. However, a direct estimate of the effect of numerical viscosity, for example, as proposed by \citet{PanKrit09},
is infeasible because the mean numerical viscosity can only be estimated from ensemble averages. These can be approximated by
global averages for uniform-grid simulations, but not for the different refinement levels in AMR simulations. An improvement could be made, 
however, by computing the effect of unresolved turbulence with a subgrid scale model. Unfortunately, a subgrid scale model for
highly compressible MHD turbulence is not available yet.
For hydrodynamical turbulence, on the other hand, the subgrid scale model by \citet{SchmFeder10} can be applied.
By adding the subgrid scale stresses to the equation for the compression rate, not only the
effect of turbulent viscosity, but also the  contribution form the turbulent pressure due to velocity fluctuations below the grid scale 
can be calculated. \citet{IapiSchm11} show that this contribution is
small, yet non-negligible in numerical simulations of turbulence in the intergalactic medium. For supersonic turbulence in 
the interstellar medium, however, the unresolved fraction of the turbulent pressure can become comparable to the 
thermal pressure. So far, it is not clear whether this effect would enhance or further decrease the support of the gas.
A further implication of our study is that the analysis of turbulent pressure support by \citet{ZhuFeng10} and \citet{IapiSchm11} 
falls short of the possibly dominating negative component, at least for the highly compressible turbulence in the intergalactic medium. It is 
clearly necessary to consider both the positive and the negative contributions to the local support. 

Our findings also bear consequences on theoretical and numerical approaches to the clump or core mass function in star-forming clouds.
For example, the analytic theory by \citet{HenneChab08} assumes that turbulence has a scale-dependent stabilizing effect,
which follows from the incorporation of a power law for the turbulent pressure into the Jeans criterion. As 
shown by \citet{SchmKern10}, this has a significant impact on the mass spectrum of gravitationally unstable clumps in 
hydrodynamical turbulence simulations (without explicit treatment of self-gravity). In a similar way, an effective equation
of state is used in the statistical excursion-set model for fragmentation in self-gravitating turbulent media, which
was recently put forward by \citet{Hop11}. From our discussion, however, it follows that an
effective local Jeans mass for the clumpy substructure of supersonic turbulence is difficult to consolidate with the properties of 
the turbulent support function. This does not mean that the \citet{Chandra51} relation for the effective speed of sound is invalid per se.
Such a relation can be rigorously derived for hydrodynamical turbulence by means of scale decomposition \citep[see][]{SchmFeder10} and, 
on length scales larger than the forcing scale, also for self-gravitating turbulence \citep{BonaPer92}. But the fully non-linear dynamics
of gas compression, which is determined by the local balance between thermal, turbulent, and magnetic support versus gravity, depends on 
the Laplacians and gradients of the pressure components. As opposed to the thermal and magnetic pressures,
our statistical analysis implies that large turbulent pressure does not give rise to a strong mean support against gravity for highly resolved self-gravitating turbulence. As a consequence, the Lagrange identity form of the virial theorem,
$2(E_{\rm kin}+E_{\rm int})+E_{\rm magn}-E_{\rm grav}=0$, is not applicable to the substructure in molecular clouds.

\section*{Acknowledgments}

We thank Jens Niemeyer and Dominik Schleicher for many useful comments. We also acknowledge the yt toolkit by \citet{TurkSmith11} that
was used for our analysis of numerical data. A.~K.\ was supported in part by NSF grants AST-0808184, AST-0908740,
AST-1109570. D.C. is supported by Advanced Simulation and Computing Program (ASC) and LANL which is
operated by LANS, LLC for the NNSA of the U.S. DOE under Contract No. DE-AC52- 06NA25396. We utilized computing resources
provided by SDSC and NICS through the XRAC allocations MCA07S014 and TG-AST090110
This work used the Extreme Science and Engineering Discovery Environment (XSEDE),
which is supported by the NSF grant OCI-1053575.

\bibliography{grav_support}

\appendix

\section{Phase plots}
\label{sc:phase}

In the following, two-dimensional distributions are shown, where the volume or mass fractions occupied by each bin
are indicated by colour scales. 
In each case, the positive and negative contributions are plotted separately so that we can use logarithmic scaling
in the plots (we do not show, however, distributions of logarithmic variables).
The distributions are computed from the full AMR data sets at time $t=0.43 t_{\rm ff}$ for the hydrodynamical simulation
and $t=0.5 t_{\rm ff}$ for the MHD simulations. The curves show averages of the quantity on the $y$-axis vs.\ the quantity
on the $x$-axis normalized by the integrated volume or mass of those cells in which 
the quantity on the $y$-axis
is non-zero. This normalization avoids a bias of the averages relative to the two-dimensional distributions. Even so,
cuts through the distributions in $y$-direction can have very wide, asymmetric and irregular tails, which lead to
pronounced fluctuations in the mean values. In mathematical terms, the distribution $\dd V(X,Y)$ is a measure of the differential volume 
occupied by particular values of the random variables $X$ and $Y$. In the phase plots, we show the piecewise constant functions 
$\widetilde{\dd V}(X_i,Y_{k\,\pm})$, which approximate $\dd V(X,Y_\pm)$. The mean value of $Y_{\pm}$, given a particular value $X=X_i$, 
is
\[
      \overline{Y_\pm}_i = \sum_k Y_{k\,\pm}\,\widetilde{\dd V}(X_i,Y_{k\,\pm})
      \left(\sum_k \widetilde{\dd V}(X_i,Y_{k\,\pm})\right)^{-1},
\]
where it is understood that $\widetilde{\dd V}(X_i,Y_{k\,\pm})>0$ only if $Y_{k\,\pm}>0$ because, for example, all bins for which 
$Y_k < 0$ are unoccupied by $Y_{k\,+}$.
  
In Figs.~\ref{fig:lambda_profiles}--\ref{fig:stability_profiles_downsample}, on the other hand, we normalize by the unconstrained volume
or mass of each bin so that $\langle\Lambda\rangle=\langle\Lambda_+\rangle-\langle\Lambda_-\rangle$ is satisfied (see Sect.~\ref{sc:hd_stat})
and the profiles of negative and positive components can be directly compared (the bin size is also smaller by a factor of $2$ in these plots and, for as smoother representation of the data, curves are computed with Mathematica's higher-order interpolation function
rather than the linear interpolation between data points that is used by yt).
This means that the decomposition of the mean value of $Y$ for $X=X_i$ is given by
\begin{align*}
    \langle Y\rangle_i &= \sum_k Y_k\,\widetilde{\dd V}(X_i,Y_k)\left(\sum_k \widetilde{\dd V}(X_i,Y_k)\right)^{-1} =\\
	& = \sum_k (Y_{k\,+}-Y_{k\,-})\,\widetilde{\dd V}(X_i,Y_k)\left(\sum_k \widetilde{\dd V}(X_i,Y_k)\right)^{-1} =\\
	& = \langle Y_+\rangle_i - \langle Y_-\rangle_i\;.
\end{align*}
Note that $\langle Y_{\pm}\rangle_i \ne \overline{Y_\pm}_i$ because the sums in the denominators are
generally different. Analogous expressions hold for
mass-weighted averages with $\dd m(X,Y)$ in place of $\dd V(X,Y)$.

\begin{figure*}
\centering
  \includegraphics[width=0.48\linewidth]{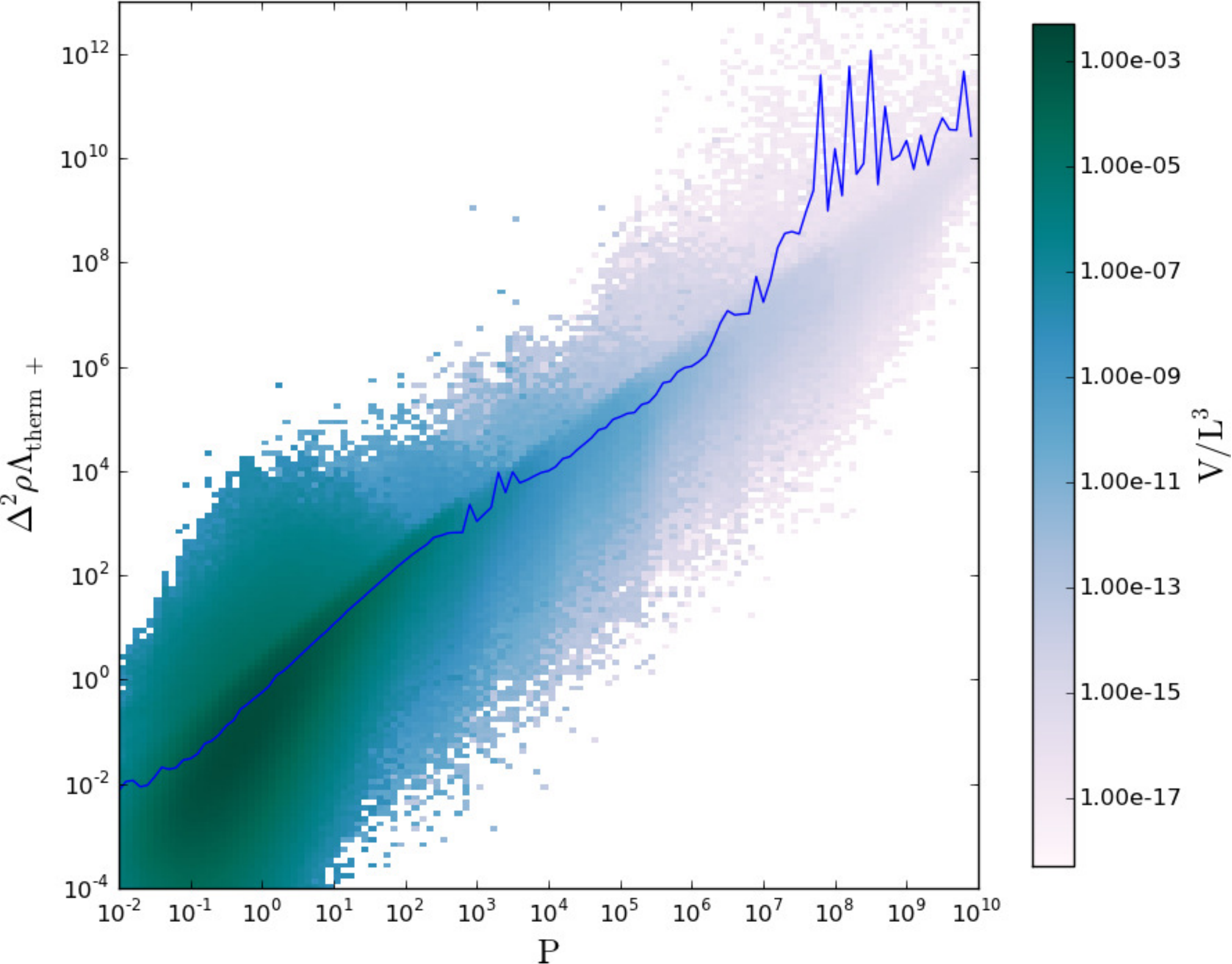}\quad
  \includegraphics[width=0.48\linewidth]{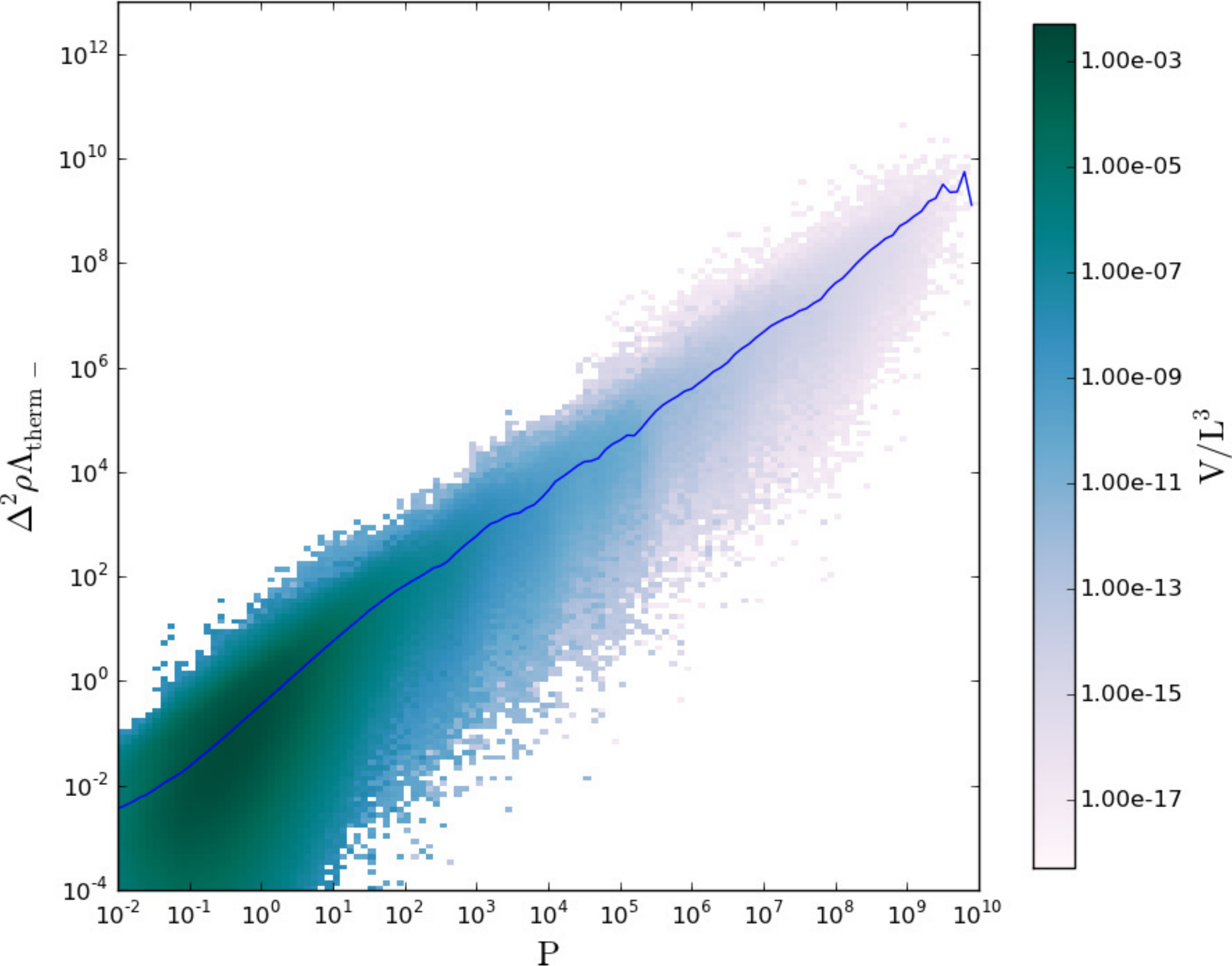}
\caption{Phase plots of of the
	normalized thermal support against the thermal pressure (no magnetic field).}
	\label{fig:hd_lambda_therm_2d}
\end{figure*}

\begin{figure*}
\centering
  \includegraphics[width=0.48\linewidth]{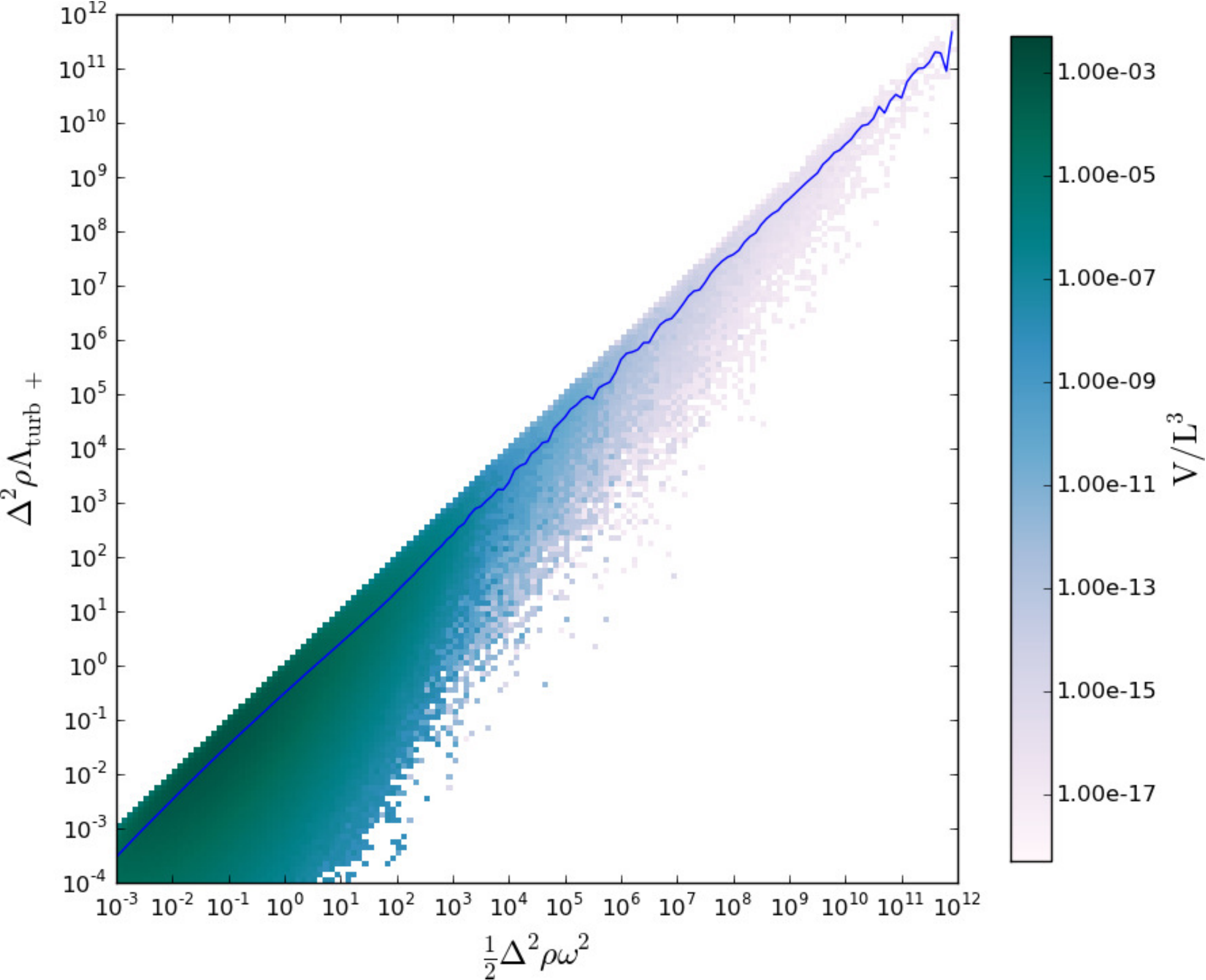}\quad
  \includegraphics[width=0.48\linewidth]{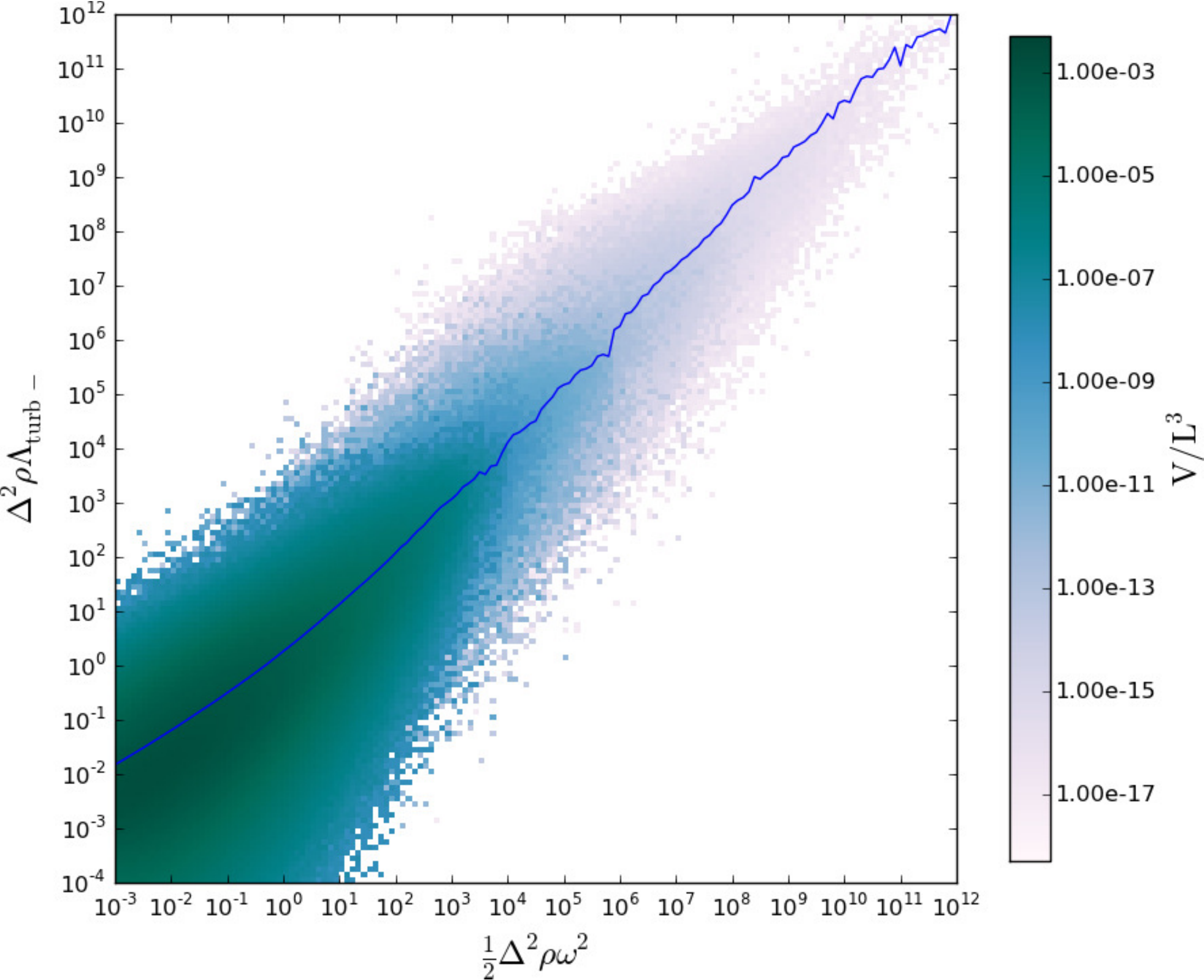}
\caption{Phase plots of normalized turbulent 
	support against the enstrophy density (no magnetic field).}
\label{fig:hd_lambda_turb_2d}
\end{figure*}

\begin{figure*}
\centering
  \includegraphics[width=0.47\linewidth]{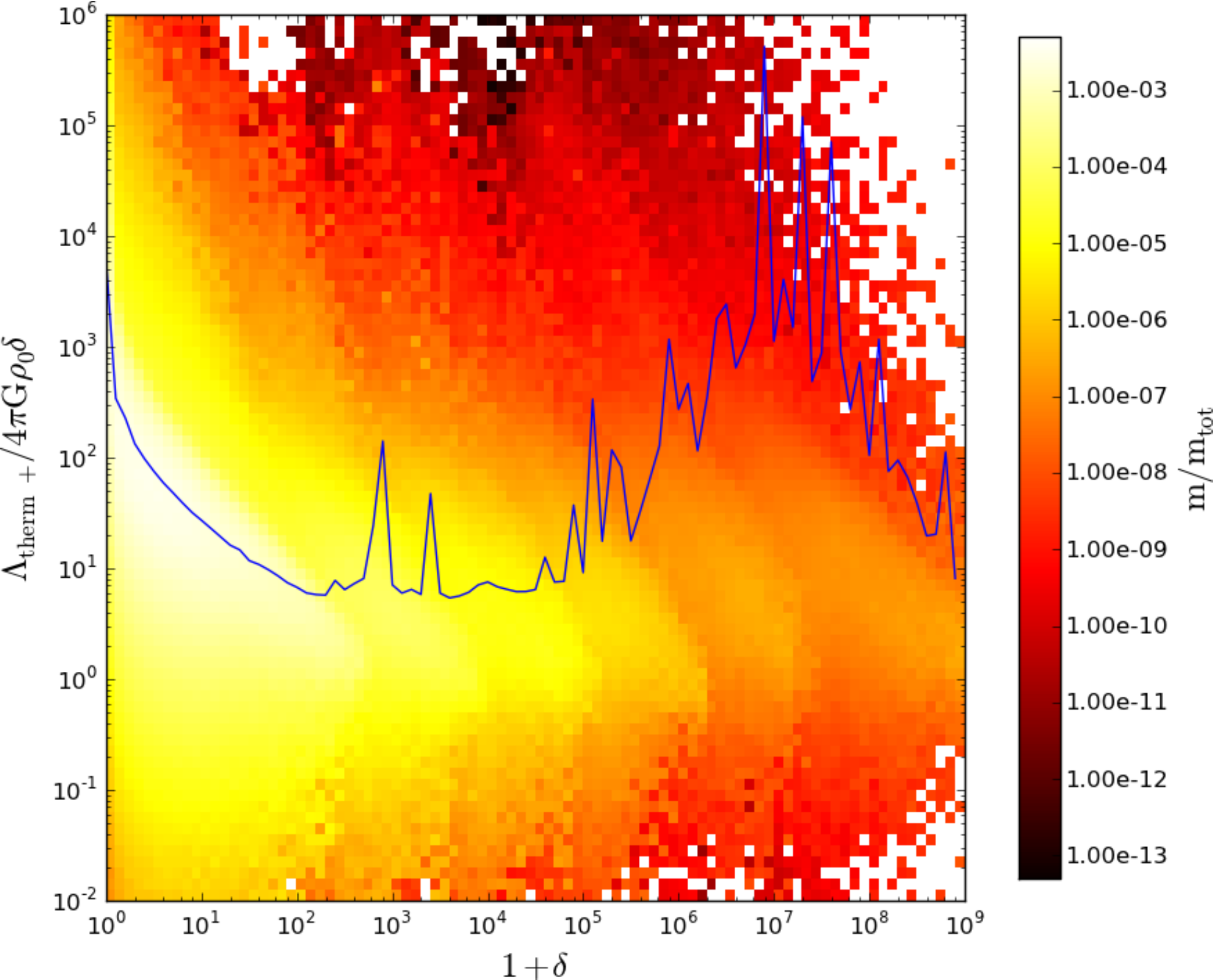}\quad
  \includegraphics[width=0.47\linewidth]{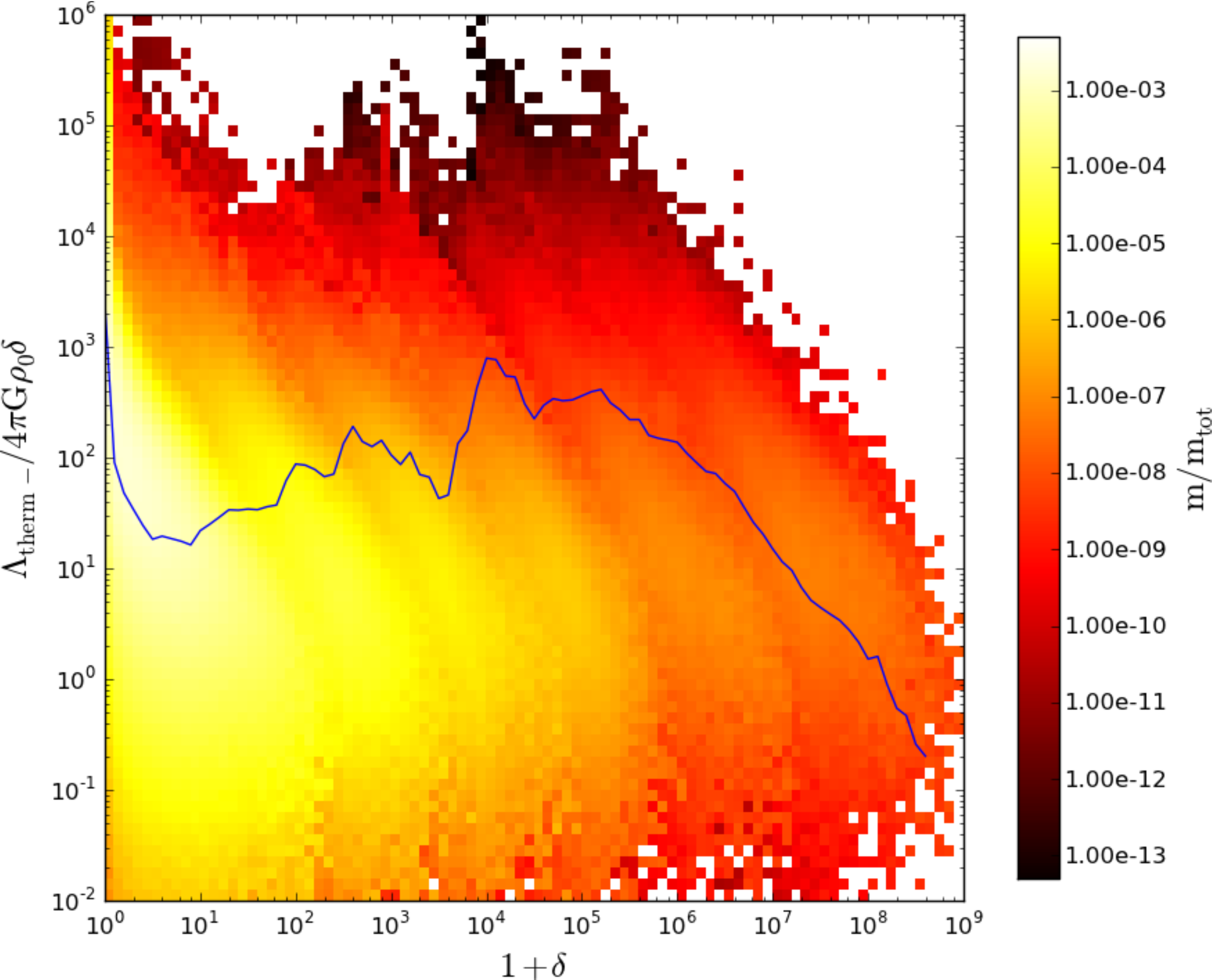}
\caption{Phase plots of the
	thermal support relative to the gravitational compression rate (no magnetic field).}
	\label{fig:hd_stab_therm_2d}
\end{figure*}

\begin{figure*}
\centering
  \includegraphics[width=0.47\linewidth]{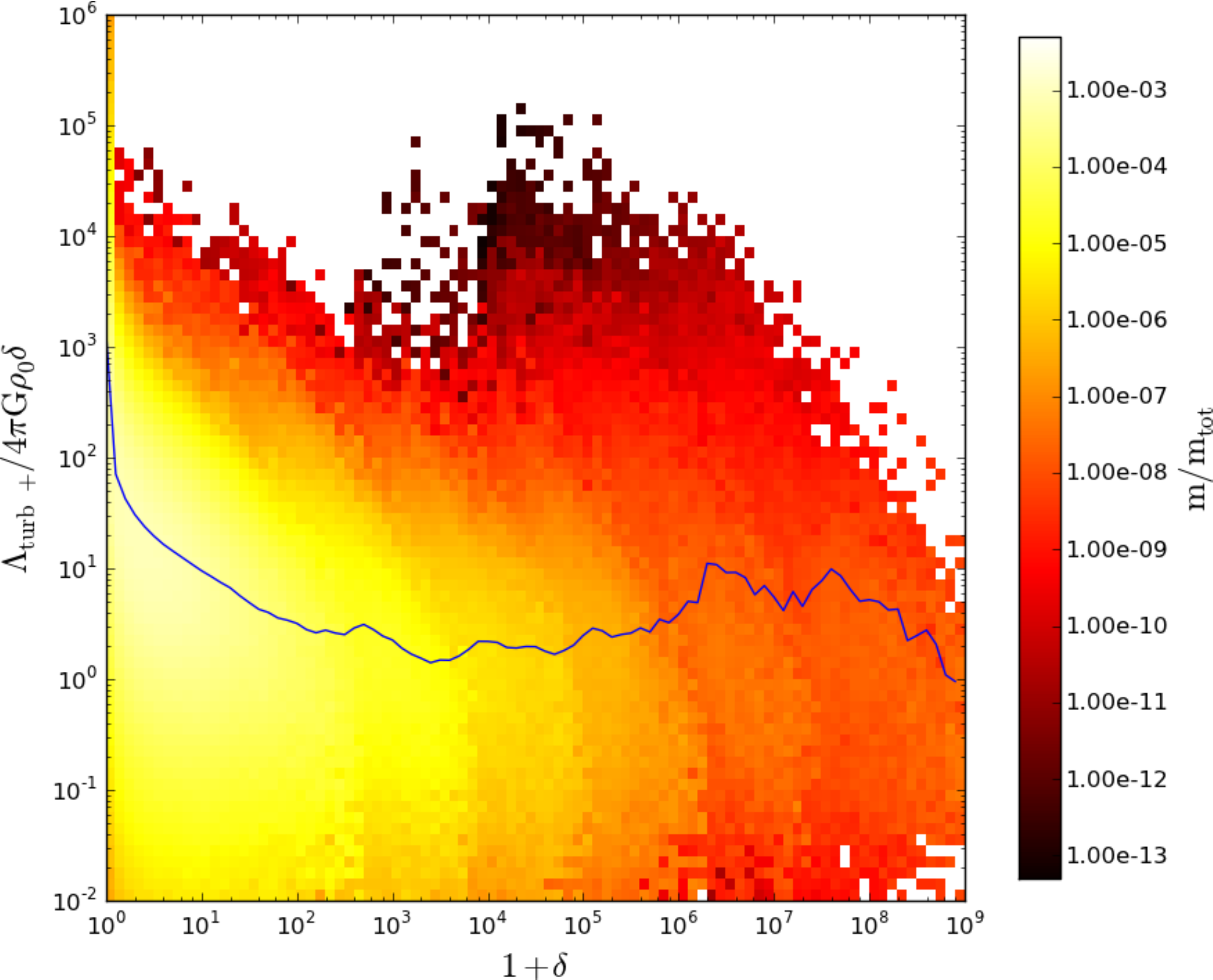}\quad
  \includegraphics[width=0.47\linewidth]{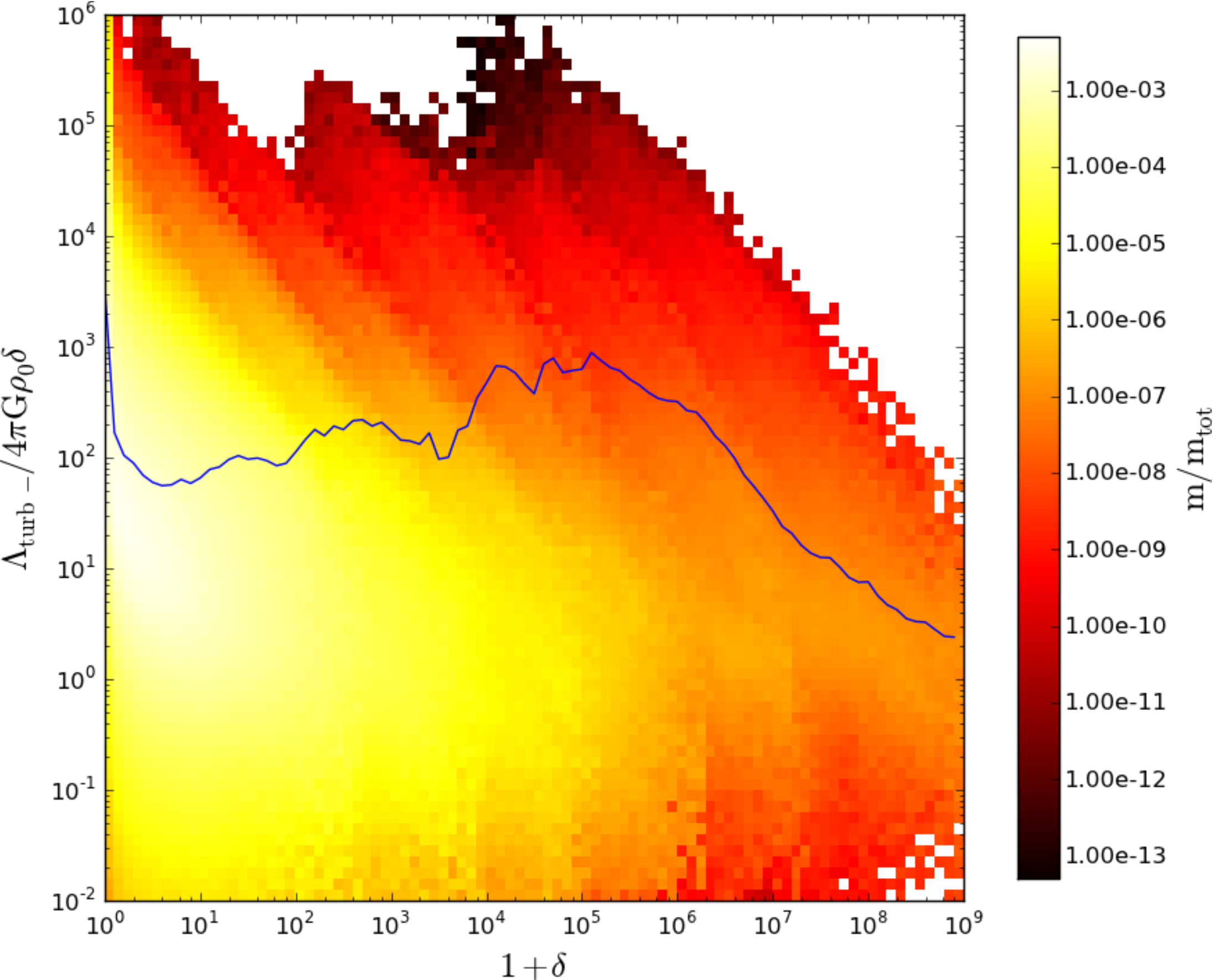}
\caption{Phase plots of the
	turbulent support relative to the gravitational compression rate (no magnetic field).}
\label{fig:hd_stab_turb_2d}
\end{figure*}

\begin{figure*}
\centering
  \includegraphics[width=0.48\linewidth]{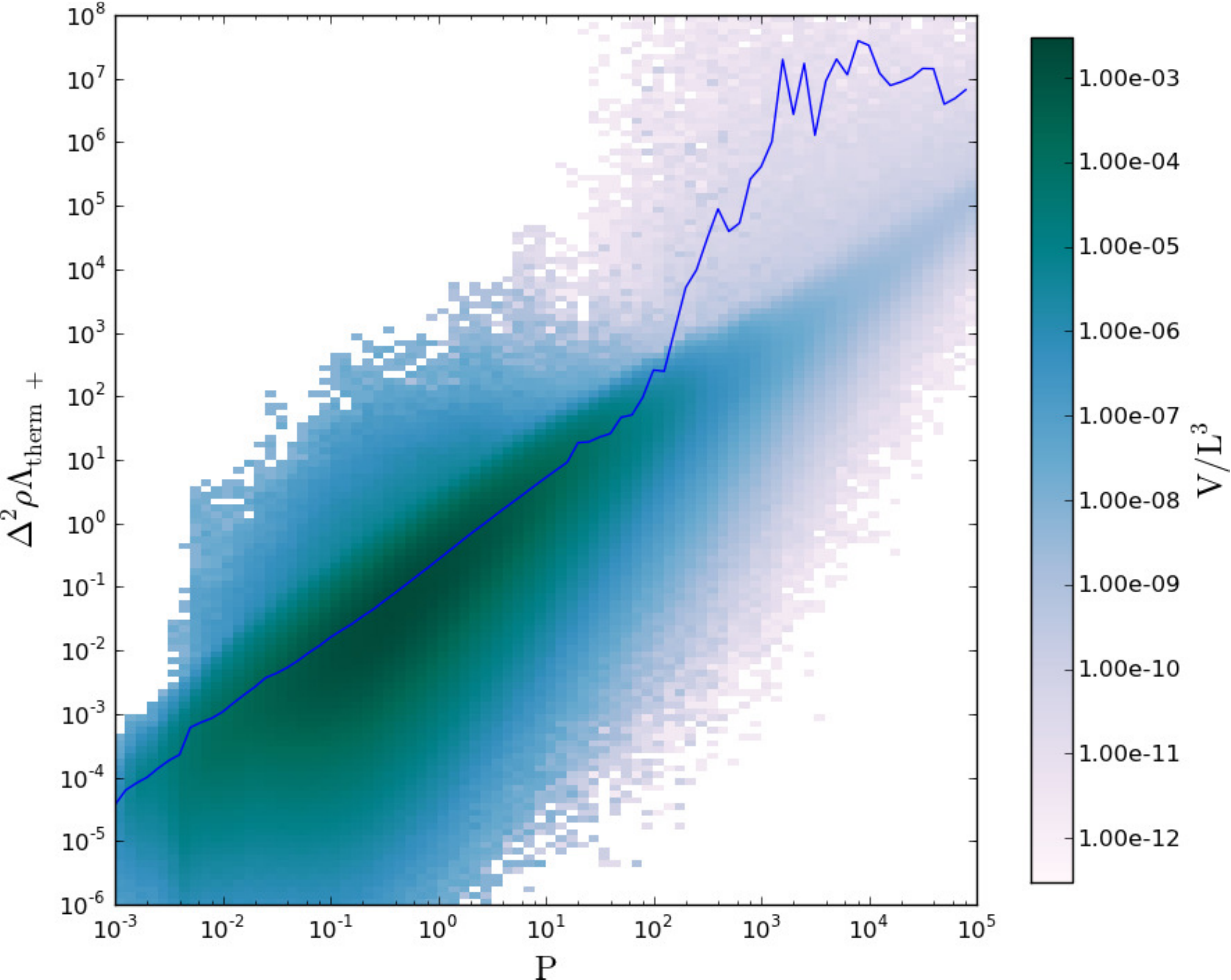}\quad
  \includegraphics[width=0.48\linewidth]{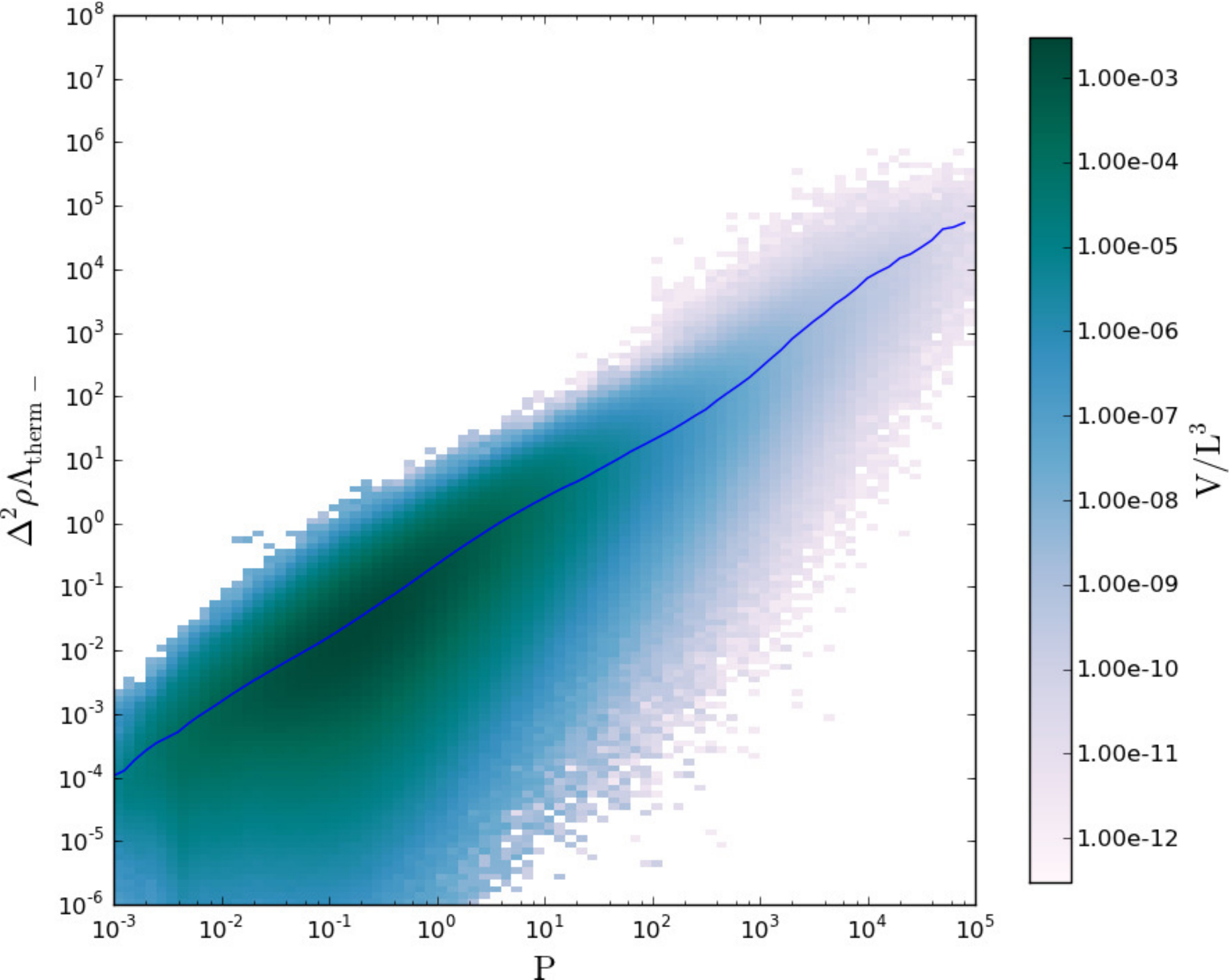}
\caption{Phase plots of normalized thermal support against the thermal pressure ($\beta_0=20.0$).}
\label{fig:b20_lambda_therm_2d}
\end{figure*}

\begin{figure*}
\centering
  \includegraphics[width=0.48\linewidth]{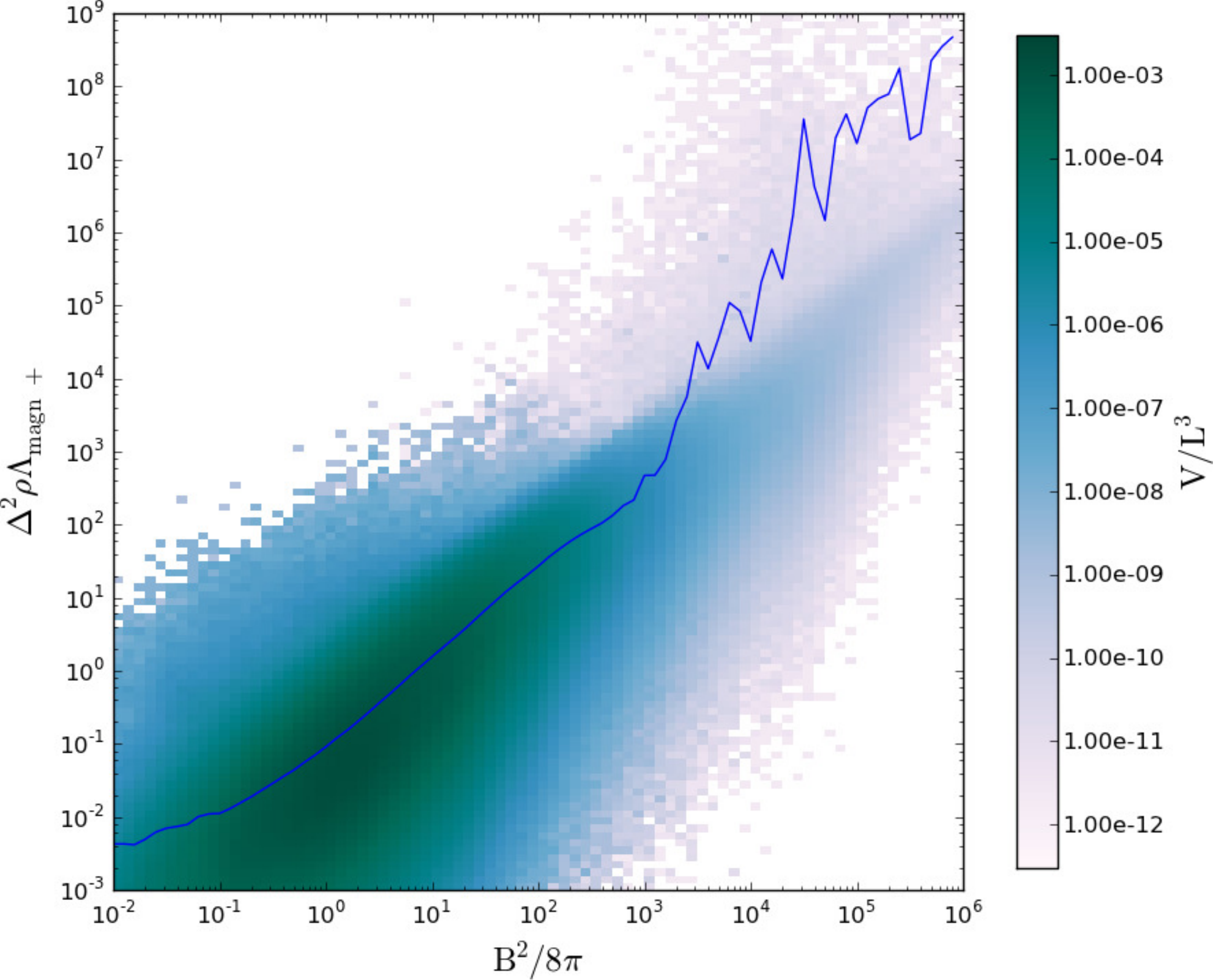}\quad
  \includegraphics[width=0.48\linewidth]{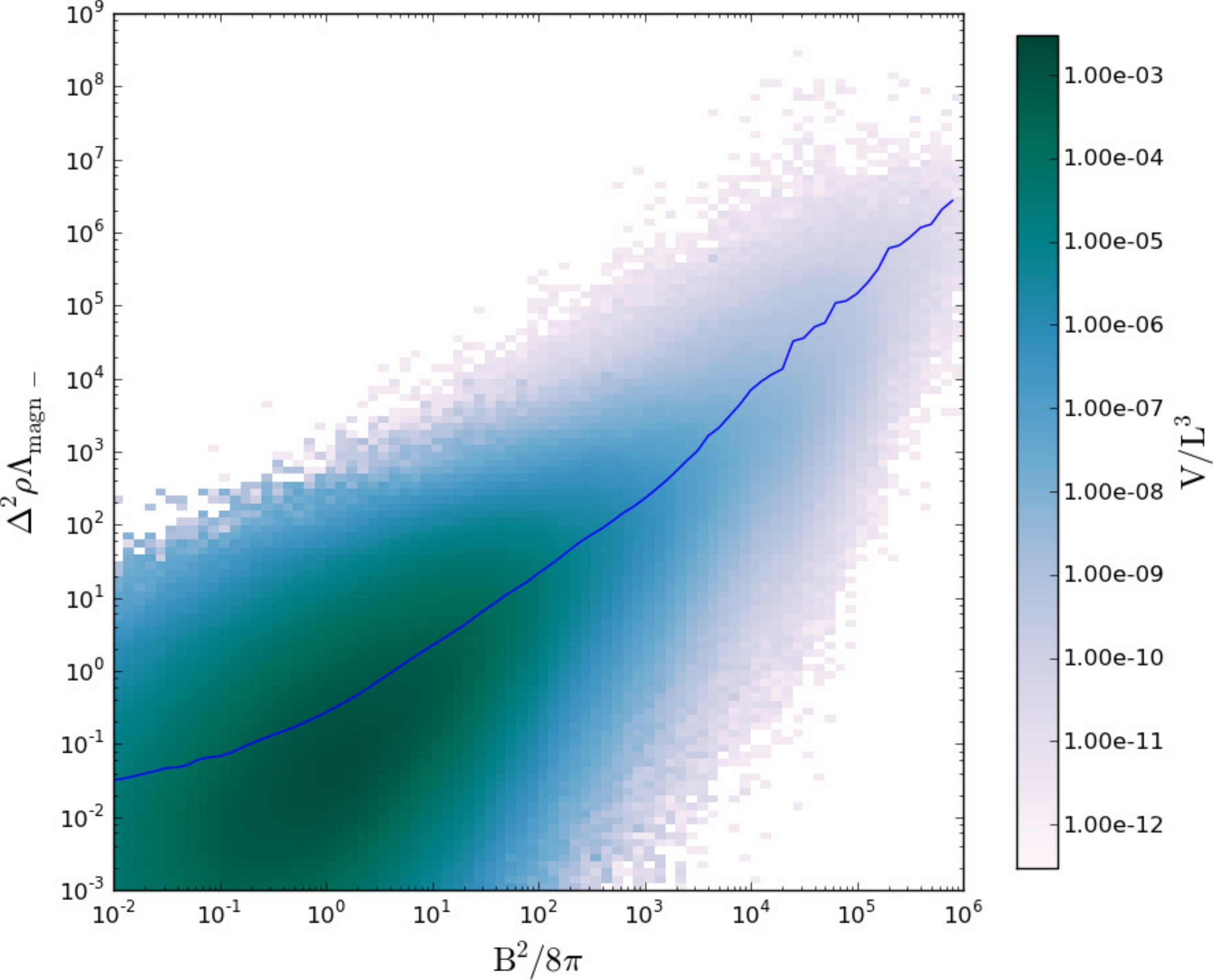}
\caption{Phase plots of normalized magnetic support against the magnetic pressure ($\beta_0=20.0$).}
\label{fig:b20_lambda_magn_2d}
\end{figure*}

\begin{figure*}
\centering
  \includegraphics[width=0.47\linewidth]{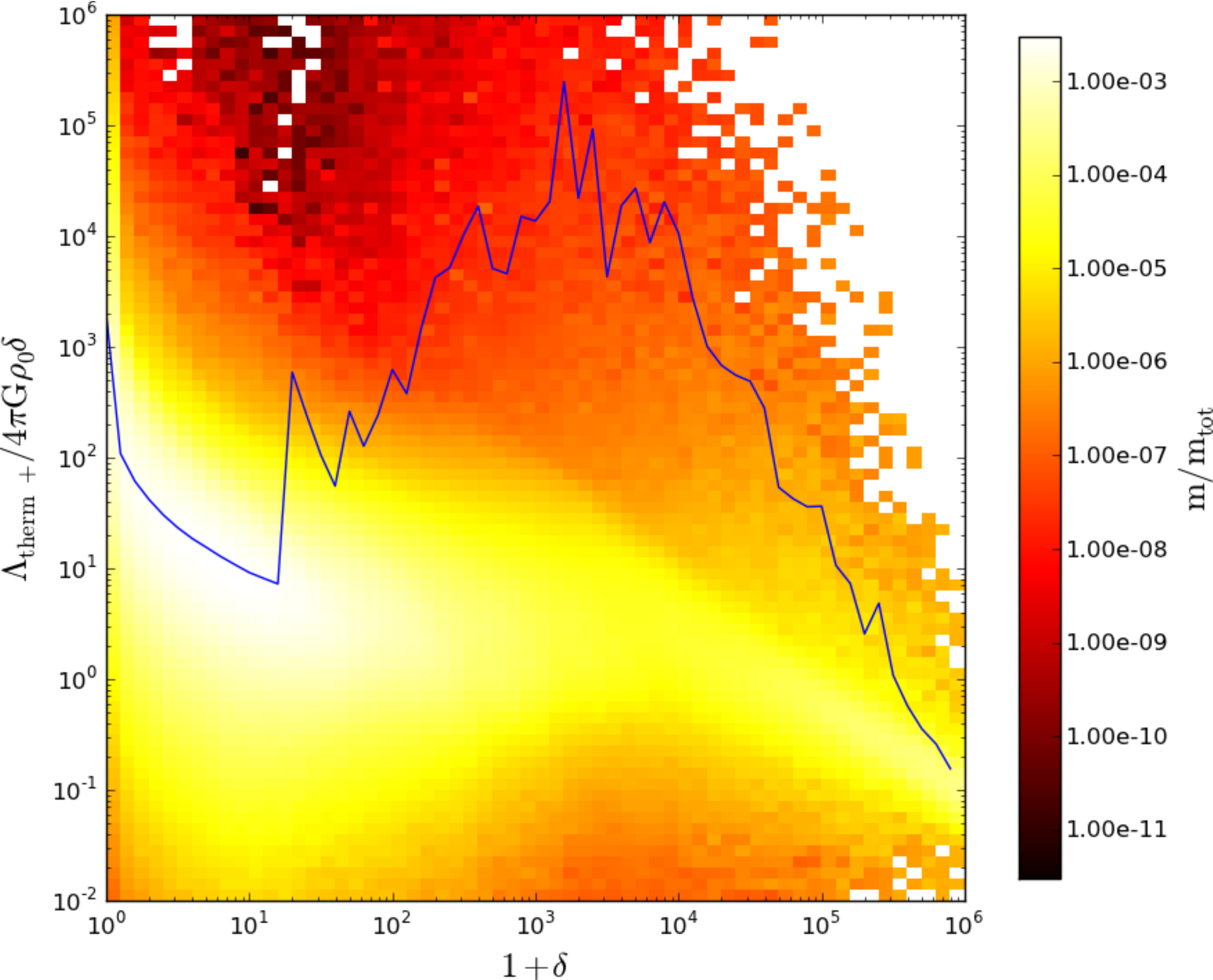}\quad
  \includegraphics[width=0.47\linewidth]{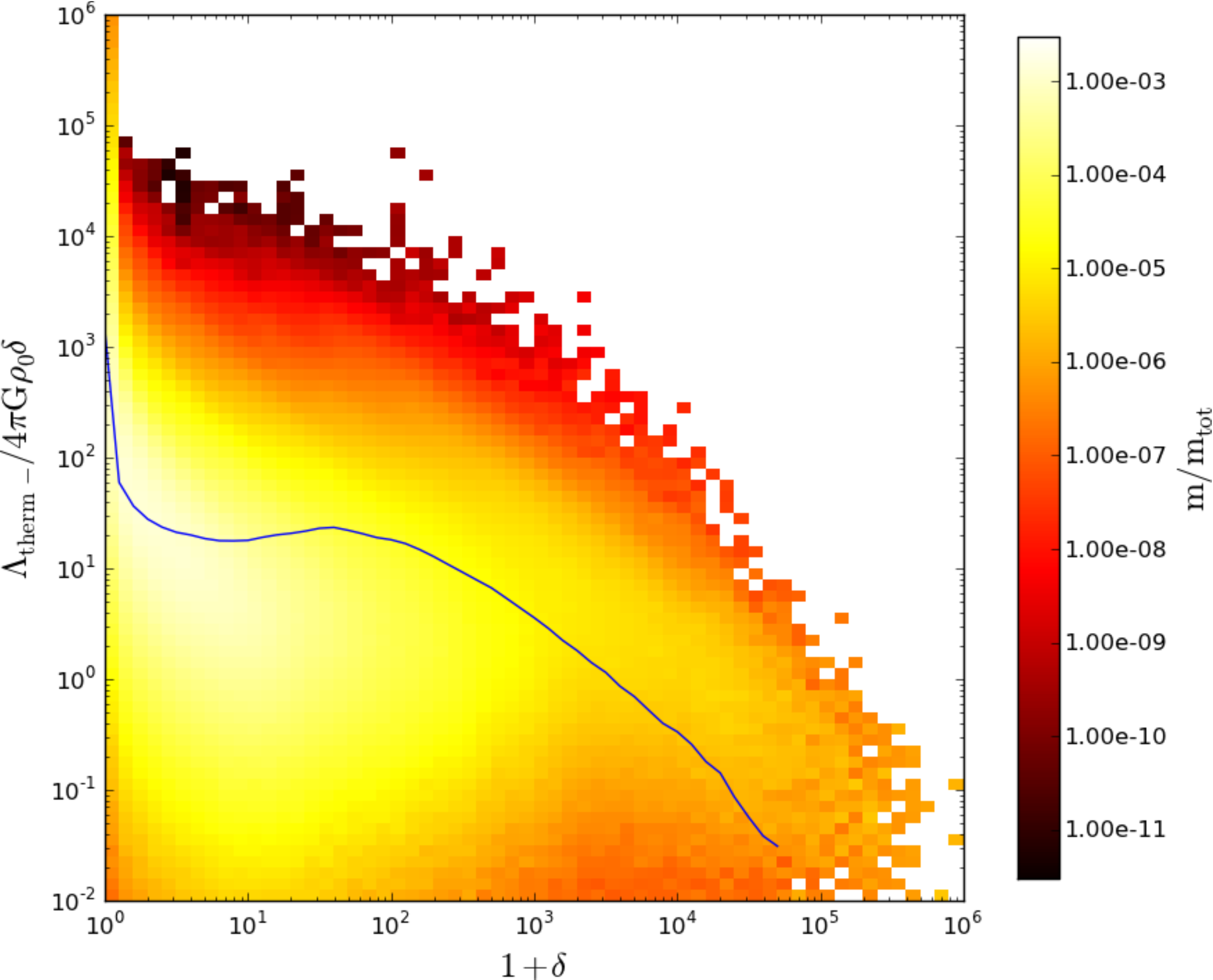}
\caption{Phase plots of the
	thermal support relative to the gravitational compression rate ($\beta_0=20.0$).}
	\label{fig:b20_stab_therm_2d}
\end{figure*}

\begin{figure*}
\centering
  \includegraphics[width=0.47\linewidth]{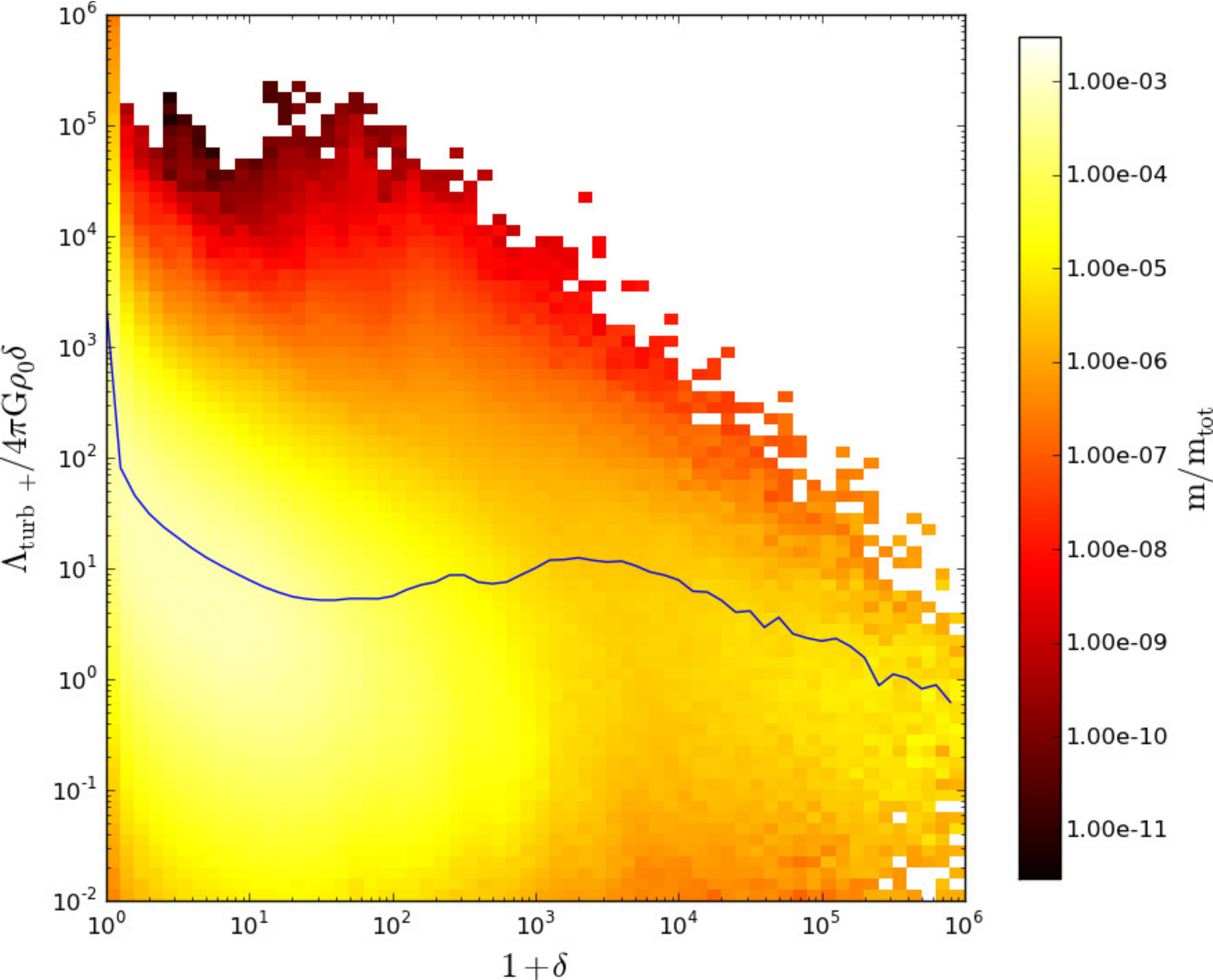}\quad
  \includegraphics[width=0.47\linewidth]{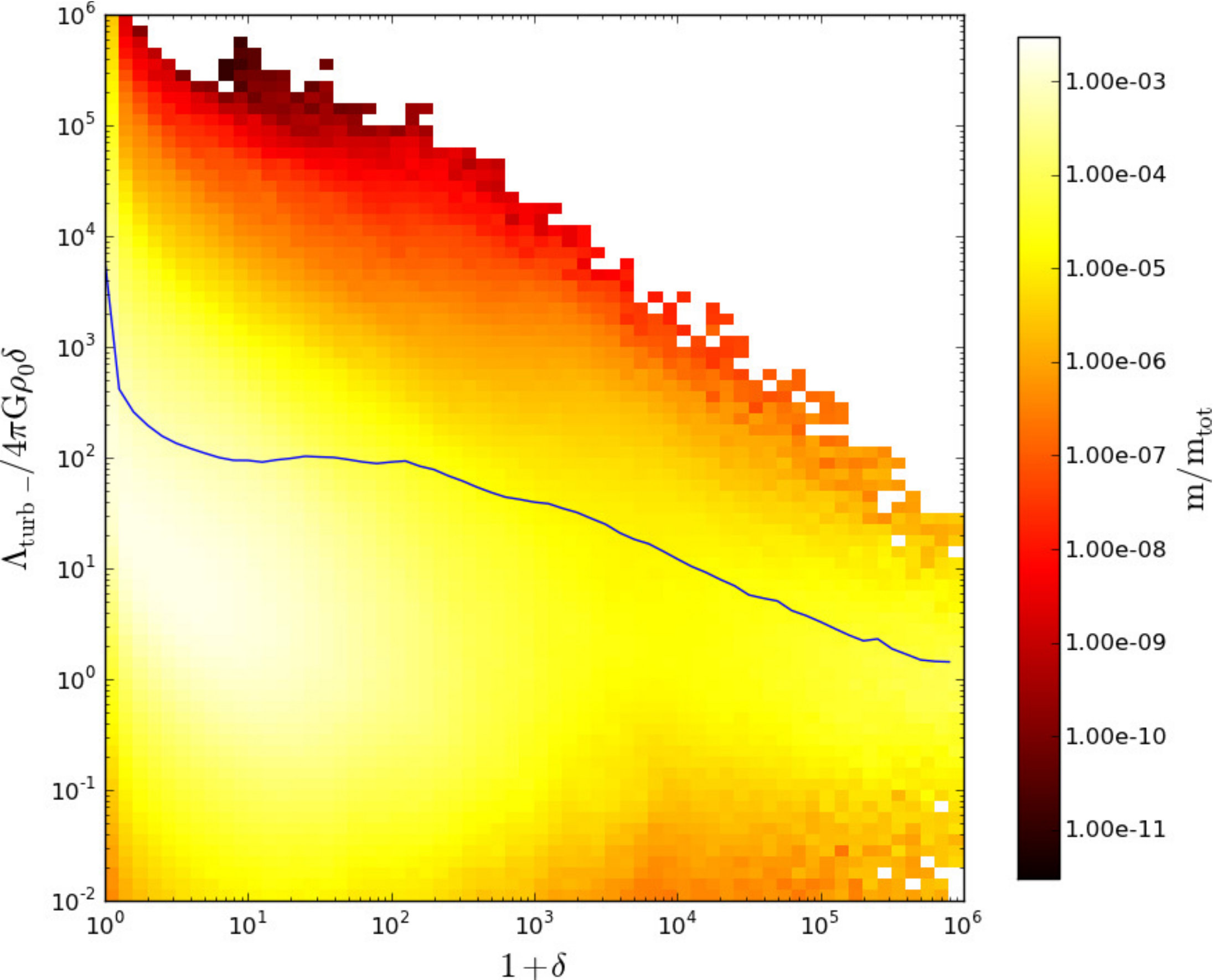}
\caption{Phase plots of the
	turbulent support relative to the gravitational compression rate ($\beta_0=20.0$).}
\label{fig:b20_stab_turb_2d}
\end{figure*}

\begin{figure*}
\centering
  \includegraphics[width=0.47\linewidth]{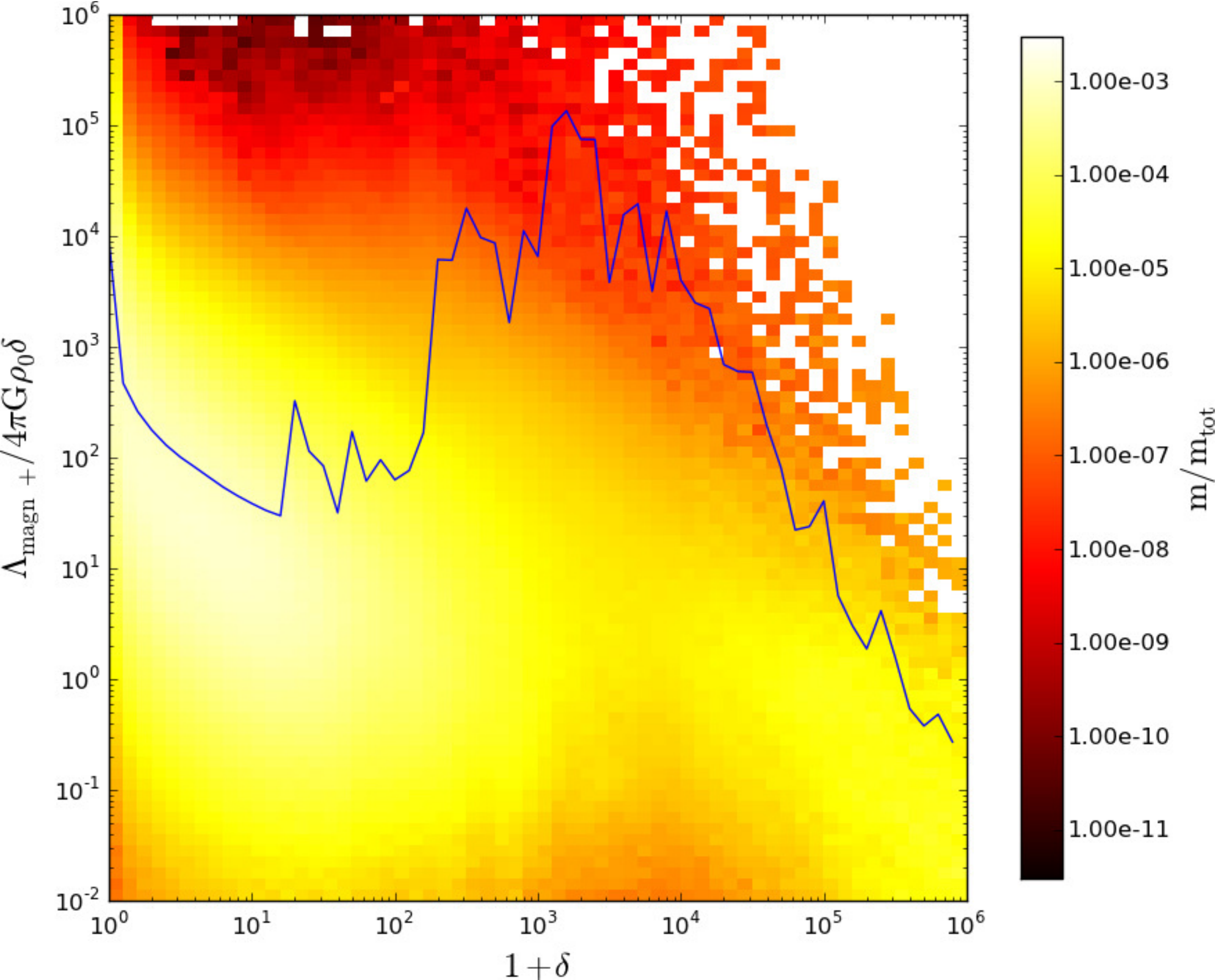}\quad
  \includegraphics[width=0.47\linewidth]{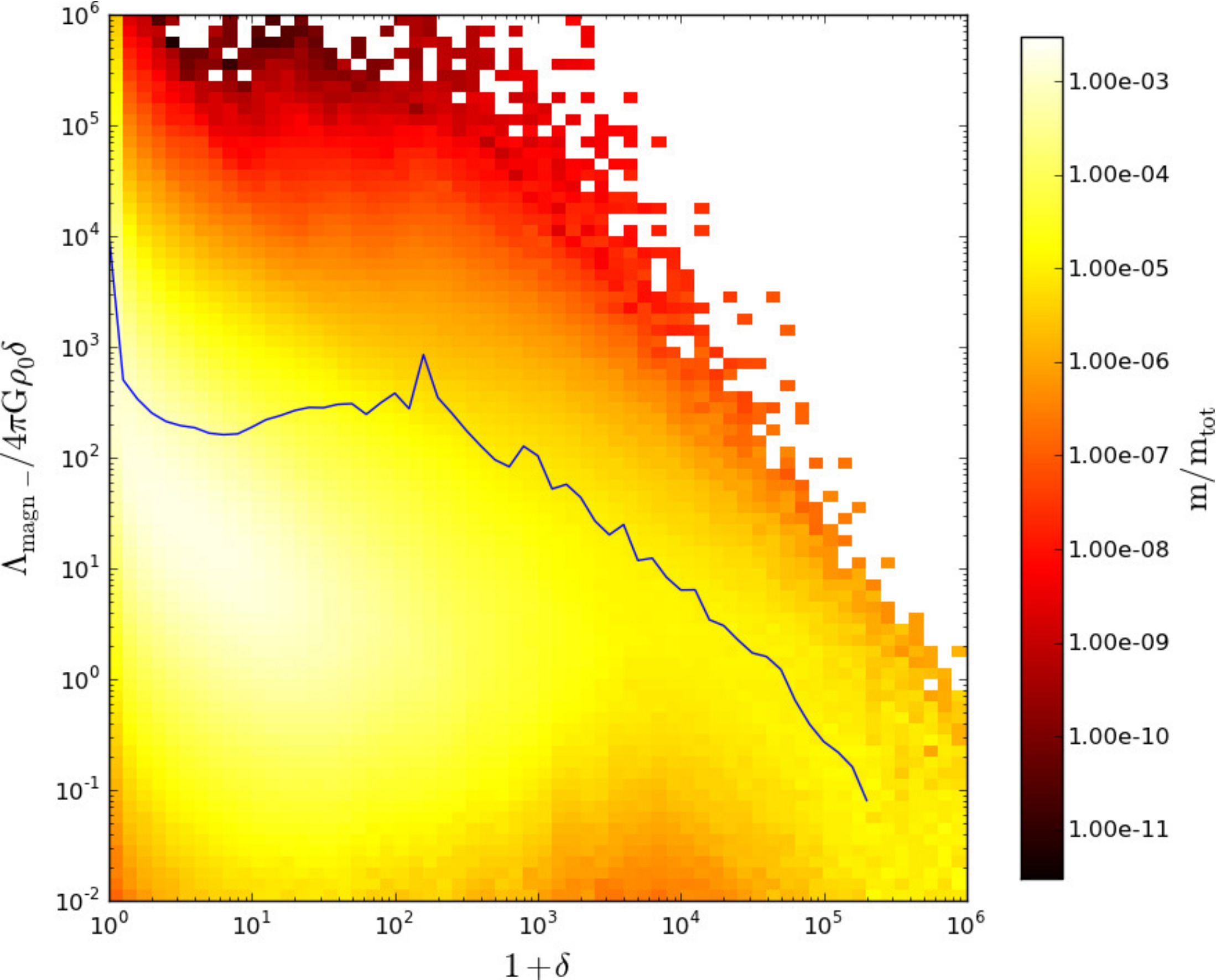}
\caption{Phase plots of the
	magnetic support relative to the gravitational compression rate ($\beta_0=20.0$).}
\label{fig:b20_stab_magn_2d}
\end{figure*}

\end{document}